\documentclass[a4paper,11pt]{article}
\pdfoutput=1 

\usepackage{jheppub} 

\usepackage[T1]{fontenc} 
\usepackage{subfigure} 

\usepackage{preprintcover}

\PreprintCoverPaperTitle{\boldmath Search for pair and single production of new heavy quarks that decay to a $Z$ boson and a third-generation quark in $pp$ collisions at $\sqrt{s}=8$~TeV with the ATLAS detector}

\PreprintIdNumber{CERN-PH-EP-2014-188} 

\PreprintCoverAbstract{A search is presented for the production of new heavy quarks that decay to a $Z$ boson and a third-generation Standard Model quark.  In the case of a new charge $+2/3$ quark ($T$), the decay targeted is $T \rightarrow Zt$, while the decay targeted for a new charge $-1/3$ quark ($B$) is \mbox{$B \rightarrow Zb$}.  The search is performed with a dataset corresponding to $20.3~\mathrm{fb}^{-1}$ of $pp$ collisions at \mbox{$\sqrt{s}=8$~TeV} recorded in 2012 with the ATLAS detector at the CERN Large Hadron Collider.  Selected events contain a high transverse momentum $Z$ boson candidate reconstructed from a pair of oppositely charged same-flavor leptons (electrons or muons), and are analyzed in two channels defined by the absence or presence of a third lepton.  Hadronic jets, in particular those with properties consistent with the decay of a $b$-hadron, are also required to be present in selected events.  Different requirements are made on the jet activity in the event in order to enhance the sensitivity to either heavy quark pair production mediated by the strong interaction, or single production mediated by the electroweak interaction.  No significant excess of events above the Standard Model expectation is observed, and lower limits are derived on the mass of vector-like $T$ and $B$ quarks under various branching ratio hypotheses, as well as upper limits on the magnitude of electroweak coupling parameters.}

\PreprintJournalName{JHEP}

\title{\boldmath Search for pair and single production of new heavy quarks that decay to a $Z$ boson and a third-generation quark in $pp$ collisions at $\sqrt{s}=8$~TeV with the ATLAS detector}

\collaboration{The ATLAS Collaboration}

\abstract{
A search is presented for the production of new heavy quarks that decay to a $Z$ boson and a third-generation Standard Model quark.  In the case of a new charge $+2/3$ quark ($T$), the decay targeted is $T \rightarrow Zt$, while the decay targeted for a new charge $-1/3$ quark ($B$) is \mbox{$B \rightarrow Zb$}.  The search is performed with a dataset corresponding to $20.3~\mathrm{fb}^{-1}$ of $pp$ collisions at \mbox{$\sqrt{s}=8$~TeV} recorded in 2012 with the ATLAS detector at the CERN Large Hadron Collider.  Selected events contain a high transverse momentum $Z$ boson candidate reconstructed from a pair of oppositely charged same-flavor leptons (electrons or muons), and are analyzed in two channels defined by the absence or presence of a third lepton.  Hadronic jets, in particular those with properties consistent with the decay of a $b$-hadron, are also required to be present in selected events.  Different requirements are made on the jet activity in the event in order to enhance the sensitivity to either heavy quark pair production mediated by the strong interaction, or single production mediated by the electroweak interaction.  No significant excess of events above the Standard Model expectation is observed, and lower limits are derived on the mass of vector-like $T$ and $B$ quarks under various branching ratio hypotheses, as well as upper limits on the magnitude of electroweak coupling parameters.
}

\begin{document} 
\maketitle
\flushbottom

\section{Introduction}

A cornerstone of the Standard Model (SM) is the formulation of the electroweak interactions as arising from a spontaneously broken gauge symmetry.  Experiments over the past four decades have confirmed this hypothesis with precision, most notably the LEP and SLC collider programs~\cite{PEWT1,PEWT2}.  However, the nature of the symmetry-breaking mechanism is not yet determined.  The ATLAS and CMS collaborations have reported observations~\cite{HiggsATLAS,HiggsCMS} of a new particle produced at the CERN Large Hadron Collider (LHC) possessing properties thus far consistent with those predicted for the SM Higgs boson.  The default electroweak symmetry-breaking mechanism, whereby a weak-isospin doublet of fundamental scalar fields obtains a vacuum expectation value, therefore remains a valid hypothesis.

Even with the existence of a Higgs boson confirmed, the SM cannot be considered a complete description of Nature.  For example, the theory does not explain the fermion generations and mass hierarchy, nor the origin of the matter--antimatter asymmetry in the universe.  Neither does it possess a viable dark matter particle, nor describe gravitational interactions.  The SM is therefore generally regarded as a low-energy approximation of a more fundamental theory with new degrees of freedom and symmetries that would become manifest at higher energy.  In fact, the SM violates a concept of naturalness~\cite{Naturalness} when extrapolated to energies above the electroweak scale, as fine tuning is required to compensate the quadratic mass-squared divergence of a fundamental scalar field.

Proposed models of physics beyond the SM typically address the naturalness problem by postulating a new symmetry.  For example, supersymmetry is a Bose--Fermi symmetry, and the new states related to the SM bosons and fermions by this symmetry introduce new interactions that cancel the quadratically divergent ones.  Alternatively, the symmetry could be a spontaneously broken global symmetry of the extended theory, with the Higgs boson emerging as a pseudo-Nambu--Goldstone boson~\cite{StrongEWSB}.  Examples of models that implement this idea are Little Higgs~\cite{LittleHiggs,LittleHiggsRev} and Composite Higgs~\cite{CompHiggs1,CompHiggs2} models.  The new states realizing the enhanced symmetry are generically strongly coupled resonances of some new confining dynamics.  These include vector-like quarks, defined as color-triplet spin-$1/2$ fermions whose left- and right-handed chiral components have the same transformation properties under the weak-isospin gauge group.  Such quarks could mix with like-charge SM quarks~\cite{VLQmixing,delAguila}, and the mixing of the SM top quark with a charge $+2/3$ vector-like quark could play a role in regulating the divergence of the Higgs mass-squared.  Hence, vector-like quarks emerge as a characteristic feature of several non-supersymmetric {\it natural models}~\cite{FermNatural}.

Search strategies for vector-like quarks have been outlined previously~\cite{ContinoServant,JA_TP,TP_guide,JA_new}.  Results of searches for chiral fourth-generation quarks apply, though interpreting the exclusions was difficult in the past when the quarks were assumed to decay entirely via the charged-current process.  The GIM mechanism~\cite{GIM} ceases to operate when vector-like quarks are added to the SM, thus allowing for tree-level neutral-current decays of such new heavy quarks~\cite{delAguila2}.  Some searches traditionally targeting chiral quarks, and hence the charged-current decay, have since provided vector-like quark interpretations~\cite{WbATLAS}.  Dedicated searches for neutral-current decay channels have also been made~\cite{ZtCMS,ZbATLAS}.  More recently, the CMS collaboration has published an inclusive search for a vector-like top quark~\cite{VLTCMS} that achieves commensurate sensitivity in the charged- and neutral-current decay modes, and sets lower mass limits ranging from 690~GeV to 780~GeV.  These previous searches assumed the pair-production mechanism is dominant, and the strategies were tailored accordingly.

This paper describes a search with ATLAS data collected in $pp$ collisions at $\sqrt{s}=8$~TeV for the production of charge $+2/3$ ($T$) and $-1/3$ ($B$) vector-like quarks that decay to a $Z$ boson and a third-generation quark ($T \rightarrow Zt$ and $B \rightarrow Zb$).  Selected events contain a high transverse momentum $Z$ boson candidate reconstructed from a pair of oppositely charged same-flavor leptons (electrons or muons), and are analyzed in two channels defined by the absence or presence of a third lepton.  Hadronic jets, in particular those likely to have contained a $b$-hadron, are also required.  Lastly, different requirements on the jet activity in the event are made to enhance the sensitivity to heavy quark pair production mediated by the strong interaction, or single production mediated by the electroweak interaction.

\section{ATLAS detector}

The ATLAS detector~\cite{Detector} identifies and measures the momentum and energy of particles created in proton--proton ($pp$) collisions at the LHC.  It has a cylindrical geometry, approximate $4\pi$ solid angle coverage, and consists of particle-tracking detectors, electromagnetic and hadronic calorimeters, and a muon spectrometer.   At small radii transverse to the beamline, the inner tracking system utilizes fine-granularity pixel and microstrip detectors designed to provide precision track impact parameter and secondary vertex measurements.  These silicon-based detectors cover the pseudorapidity range $|\eta|<2.5$~\footnote{ATLAS uses a right-handed coordinate system with its origin at the nominal interaction point (IP) in the center of the detector and the $z$-axis coinciding with the axis of the beam pipe. The $x$-axis points from the IP to the center of the LHC ring, and the $y$-axis points upward. Cylindrical coordinates $(r,\phi)$  are used in the transverse plane, $\phi$ being the azimuthal angle around the beam pipe. The pseudorapidity is defined in terms of the polar angle $\theta$ as $\eta = -\ln \tan(\theta/2)$. For the purpose of the fiducial selection, this is calculated relative to the geometric center of the detector; otherwise, it is relative to the reconstructed primary vertex of each event.}.  A gas-filled straw-tube tracker complements the silicon tracker at larger radii.  The tracking detectors are immersed in a 2~T axial magnetic field produced by a thin superconducting solenoid located in the same cryostat as the barrel electromagnetic (EM) calorimeter.  The EM calorimeters employ lead absorbers and utilize liquid argon as the active medium.  The barrel EM calorimeter covers $|\eta|<1.5$, and the end-cap EM calorimeters $1.4<|\eta|<3.2$.  Hadronic calorimetry in the region $|\eta|<1.7$ is achieved using steel absorbers and scintillator tiles as the active medium.  Liquid-argon calorimetry with copper absorbers is employed in the hadronic end-cap calorimeters, which cover the region $1.5<|\eta|<3.2$.  Forward liquid-argon calorimeters employing copper and tungsten absorbers cover the region $3.1<|\eta|<4.9$.  The muon spectrometer measures the deflection of muons with $|\eta|<2.7$ using multiple layers of high-precision tracking chambers located in a toroidal field of approximately 0.5~T and 1~T in the central and end-cap regions, respectively.  The muon spectrometer is also instrumented with separate trigger chambers covering $|\eta|<2.4$.  The first-level trigger system is implemented in custom electronics, using a subset of the detector information to reduce the event rate to a design value of 75~kHz, while the second and third levels use software algorithms running on PC farms to yield a recorded event rate of approximately 400~Hz.

\section{Reconstruction of physics objects}

The physics objects utilized in this search are electrons, muons, and hadronic jets, including jets that have been tagged for the presence of a $b$-hadron.  This section briefly summarizes the reconstruction methods and identification criteria applied to each object.

Electron candidates~\cite{EleRef} are reconstructed from energy deposits (in clusters of cells) in the EM calorimeter that are matched to corresponding reconstructed inner detector tracks.  The candidates are required to have a transverse energy, $E_{\rm T}$, greater than 25~GeV and \mbox{$|\eta_{\rm cluster}|<2.47$} (where $\eta_{\rm cluster}$ is the pseudorapidity of the cluster associated with the electron candidate).  Candidates in the transition region between the barrel and end-cap calorimeters, $1.37<|\eta_{\rm cluster}|<1.52$, are not considered.  The longitudinal impact parameter of the electron track with respect to the selected primary vertex of the event is required to be less than 2~mm.  Electron candidates used to reconstruct $Z$ boson candidates satisfy medium quality requirements~\cite{EleRef} on the EM cluster and associated track.  No additional requirements, for example on calorimeter energy or track isolation, are made.  Electron candidates not associated with $Z$ candidates are required to satisfy tighter identification requirements~\cite{EleRef} to suppress contributions from jets misidentified as electrons (``fakes'').  Further, these electrons are required to be isolated in order to reduce the contribution of non-prompt electrons produced from semi-leptonic $b$- and $c$-hadron decays inside jets.  A calorimeter isolation requirement is applied, based on the scalar sum of transverse energy in cells within a cone of radius $\Delta R \equiv \sqrt{(\Delta \eta)^{2}+(\Delta \phi)^{2}}< 0.2$ around the electron, as well as a track isolation requirement, based on the scalar sum of track transverse momenta within $\Delta R < 0.3$.  Both isolation requirements are chosen to be 90\%{} efficient for electrons from $W$ and $Z$ boson decays.

Muon candidates~\cite{MuonRef1,MuonRef2} are reconstructed from track segments in the various layers of the muon spectrometer and matched to corresponding inner detector tracks.  The final candidates are refitted using the complete track information from both detector systems.  A muon candidate is required to have transverse momentum, $p_{\rm T}$, above $25$~GeV and $|\eta|<2.5$. The hit pattern in the inner detector must be consistent with a well-reconstructed track, and the longitudinal impact parameter of the muon track with respect to the selected primary vertex of the event is required to be less than 2~mm.  Muons must also satisfy a $p_{\mathrm{T}}$-dependent track isolation requirement: the scalar sum of the track $p_{\mathrm{\rm T}}$ in a cone of variable radius $\Delta R < 10~\mathrm{GeV}/p^{\mu}_{\mathrm{T}}$ around the muon (excluding the muon itself) must be less than 5\% of the muon $p_{\mathrm{T}}$.

Jets are reconstructed using the anti-$k_{t}$ algorithm~\cite{AntiKt1, AntiKt2,AntiKt3} with a radius parameter $R=0.4$ from calibrated topological clusters built from energy deposits in the calorimeters.  Prior to jet finding, a local cluster calibration scheme~\cite{HadCal} is applied to correct the topological cluster energy for the effects of non-compensation, dead material, and out-of-cluster leakage.  The corrections are obtained from simulation of charged and neutral particles.  After energy calibration~\cite{JES, JES2012}, central jets are defined as those reconstructed with \mbox{$|\eta|<2.5$} and satisfying $p_{\rm T}>25$~GeV.  To reduce the contribution of central jets originating from secondary $pp$ interactions, a requirement is made on jets with $p_{\rm T}<50$~GeV and $|\eta|<2.4$ to ensure that at least 50\% of the scalar sum of track transverse momenta associated with the jet comes from tracks also compatible with originating from the primary vertex.  Forward jets, utilized in the search for the electroweak single production of vector-like quarks, are defined as those with $2.5 < |\eta| < 4.5$ and \mbox{$p_{\rm T} > 35$~GeV}.  During jet reconstruction, no distinction is made between identified electron and hadronic-jet energy deposits. Therefore, if any selected jet is within $\Delta R<0.2$ of a selected electron, the jet is discarded in order to avoid double-counting of electrons as jets. After this, any electrons or muons within $\Delta R<0.4$ of selected jets are discarded.

Central jets are identified as originating from the hadronization of a $b$-quark ($b$-tagging) using a multivariate discriminant that combines information from the impact parameters of displaced tracks as well as topological properties of secondary and tertiary decay vertices reconstructed within the jet~\cite{btag1,BTAG_NEW}.  The operating point used corresponds to a $b$-tagging efficiency of 70\%, as determined for $b$-tagged jets with $p_{\mathrm{T}}>20$~GeV and $|\eta|<2.5$ in simulated $t\bar{t}$ events, with light- and charm-quark rejection factors of approximately 130 and 5, respectively.

\section{Data sample and event preselection}

The data analyzed in this search were collected with the ATLAS detector between April and December 2012 during LHC $pp$ collisions at $\sqrt{s}=8$~TeV and correspond to an integrated luminosity of \mbox{$20.3\pm0.6~\mathrm{fb}^{-1}$}~\cite{Lumi}.  Events recorded by single-electron or single-muon triggers under stable beam conditions and for which all detector subsystems were operational are considered.  Single-lepton triggers with different $p_{\mathrm{T}}$ thresholds are combined to increase the overall efficiency.  The $p_{\mathrm{T}}$ thresholds are 24~GeV and 60~GeV for the electron triggers and 24~GeV and 36~GeV for the muon triggers.  The lower-threshold triggers include isolation requirements on the candidate leptons, resulting in inefficiencies at higher $p_{\mathrm{T}}$ that are recovered by the higher-$p_{\mathrm{T}}$ threshold triggers.  Events satisfying the trigger requirements must also have a reconstructed vertex with at least five associated tracks, consistent with the beam collision region in the $(x,y)$ plane.  If more than one such vertex is found, the primary vertex selected is the one with the largest sum of the squared transverse momenta of its associated tracks.  

Events selected for analysis contain at least one pair of same-flavor reconstructed leptons (electrons or muons) with opposite electric charge, and at least one reconstructed lepton in the event must match ($\Delta R<0.15$) a lepton reconstructed by the trigger system.  Reconstructed $Z$ boson candidates are formed if the invariant mass of a same-flavor opposite-charge lepton pair differs from the $Z$ boson mass by less than 10~GeV.   If more than one $Z$ boson candidate is reconstructed in an event, the one whose mass is closest to the $Z$ boson mass is considered.  Selected events are then separated into two categories defined by the absence or presence of a third electron or muon that is not associated with the $Z$ candidate, referred to as the dilepton and trilepton channels.  After preselection, $12.5\times10^{6}$ and $1.76\times10^{3}$ events are selected in the dilepton and trilepton channels, respectively.

\section{Signal modeling}

This section introduces the production mechanisms and decay properties of new heavy quarks, and describes how they are modeled in this analysis.

\subsection{Heavy quark pair production and vector-like quark decay modes}

\begin{figure}[tbp]
\begin{minipage}[b]{0.40\linewidth} 
\centering
\epsfxsize=1.13\textwidth\epsfbox{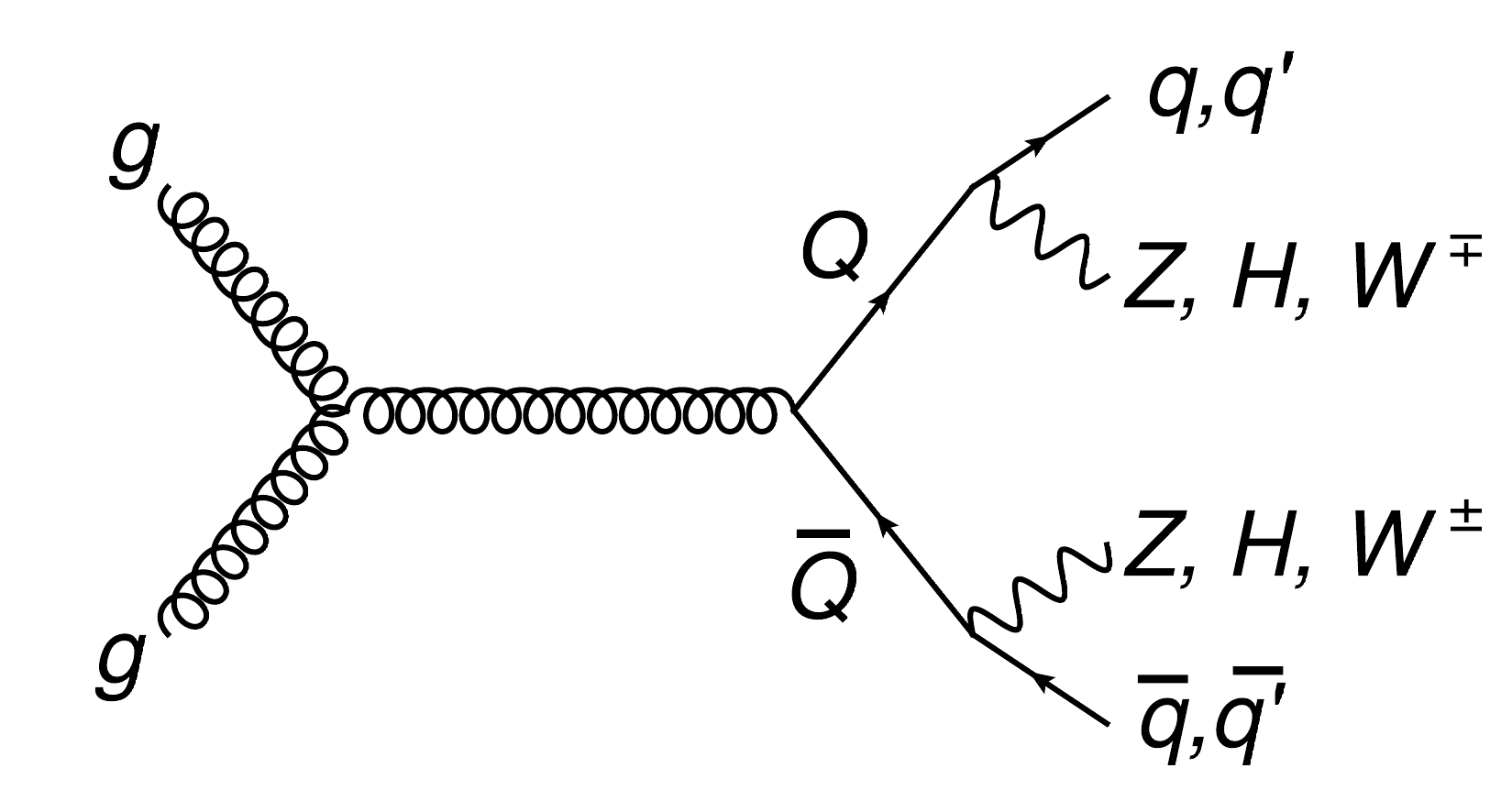}
\vspace{1.0in}
(a)
\end{minipage}
\hspace{0.04\linewidth}
\begin{minipage}[b]{0.55\linewidth} 
\centering
\epsfxsize=1.05\textwidth\epsfbox{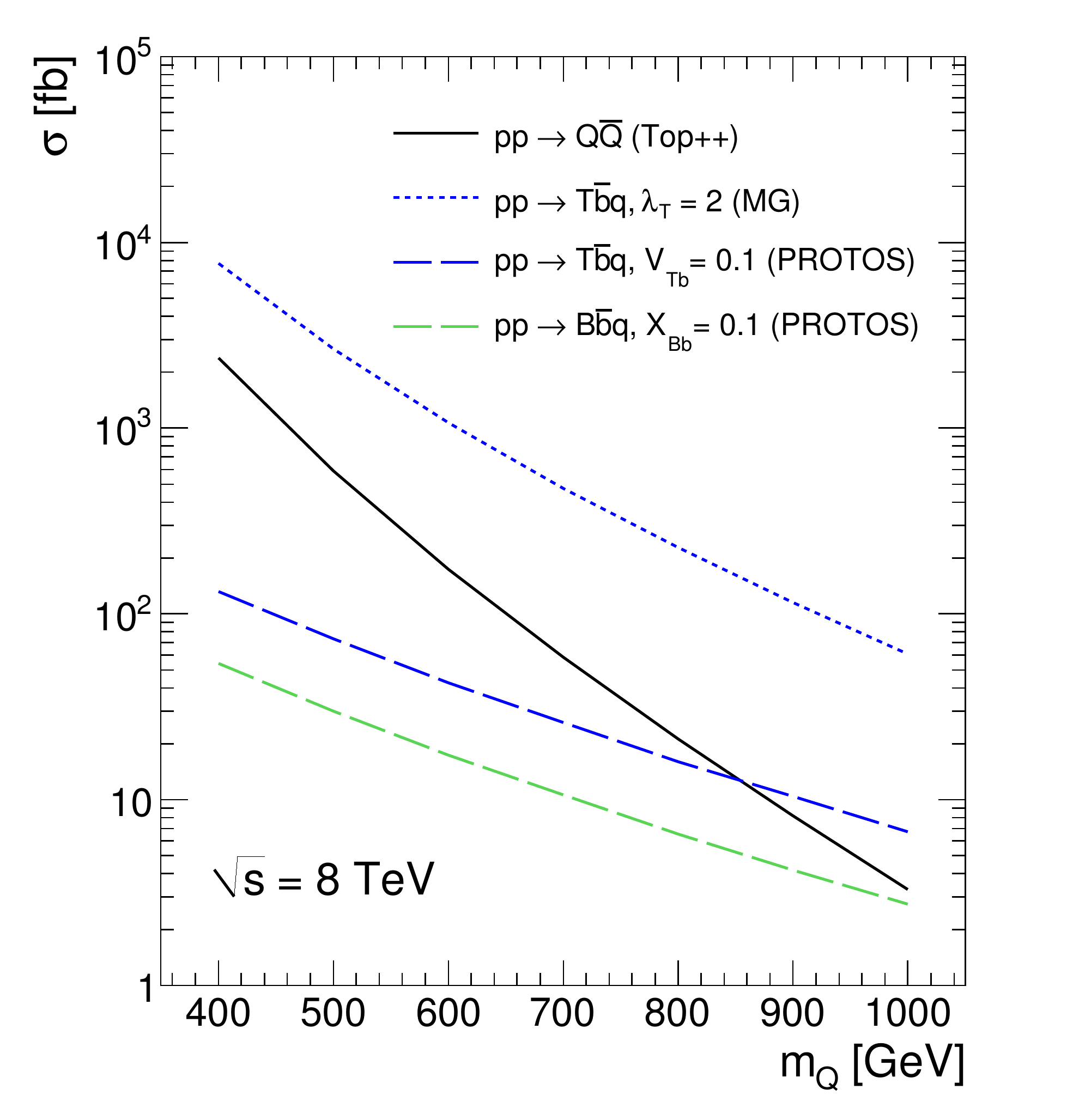}
(b)
\end{minipage}
\caption{A representative diagram (a) illustrating the pair production and decay modes of a vector-like quark ($Q=T,B$).  The $\sqrt{s}=8$~TeV LHC cross section as a function of the quark mass (b) for pair production, denoted by the solid line, as well as for the $T\bar{b}q$ and $B\bar{b}q$ single-production processes, denoted by dashed lines.  The pair-production cross section has been calculated with {\sc Top++}~\cite{toppp2}.  The single-production cross sections were calculated with {\sc protos}~\cite{PROTOS} and {\sc madgraph}~\cite{Madgraph} (MG) using different electroweak coupling parameters that are discussed in the text.}
\label{fig:xsec}
\end{figure}

One source of heavy quark production at the LHC is through pair production via the strong interaction, as illustrated in figure~\ref{fig:xsec}(a).  The cross section at $\sqrt{s}=8$~TeV as a function of the new quark mass is denoted by the solid line in figure~\ref{fig:xsec}(b).  The prediction was computed using {\sc Top++} v2.0~\cite{toppp1,toppp2}, a next-to-next-to-leading-order (NNLO) calculation in QCD including resummation of next-to-next-to-leading logarithm (NNLL) soft gluon terms, using the MSTW 2008 NNLO~\cite{MSTW,MSTW2} set of parton distribution functions (PDFs).  It is independent of the charge of the new heavy quark.  The cross-section prediction ranges from 2.4~pb for a quark mass of 400~GeV to 3.3~fb for a quark mass of 1000~GeV, with an uncertainty that increases from 8\% to 14\% over this mass range.  The PDF and $\alpha_{s}$ uncertainties dominate over the scale uncertainties, and were evaluated according to the PDF4LHC recommendations~\cite{PD4LHC}.

\begin{figure}[tbp]
\begin{minipage}[b]{0.5\linewidth} 
\centering
\epsfxsize=0.99\textwidth\epsfbox{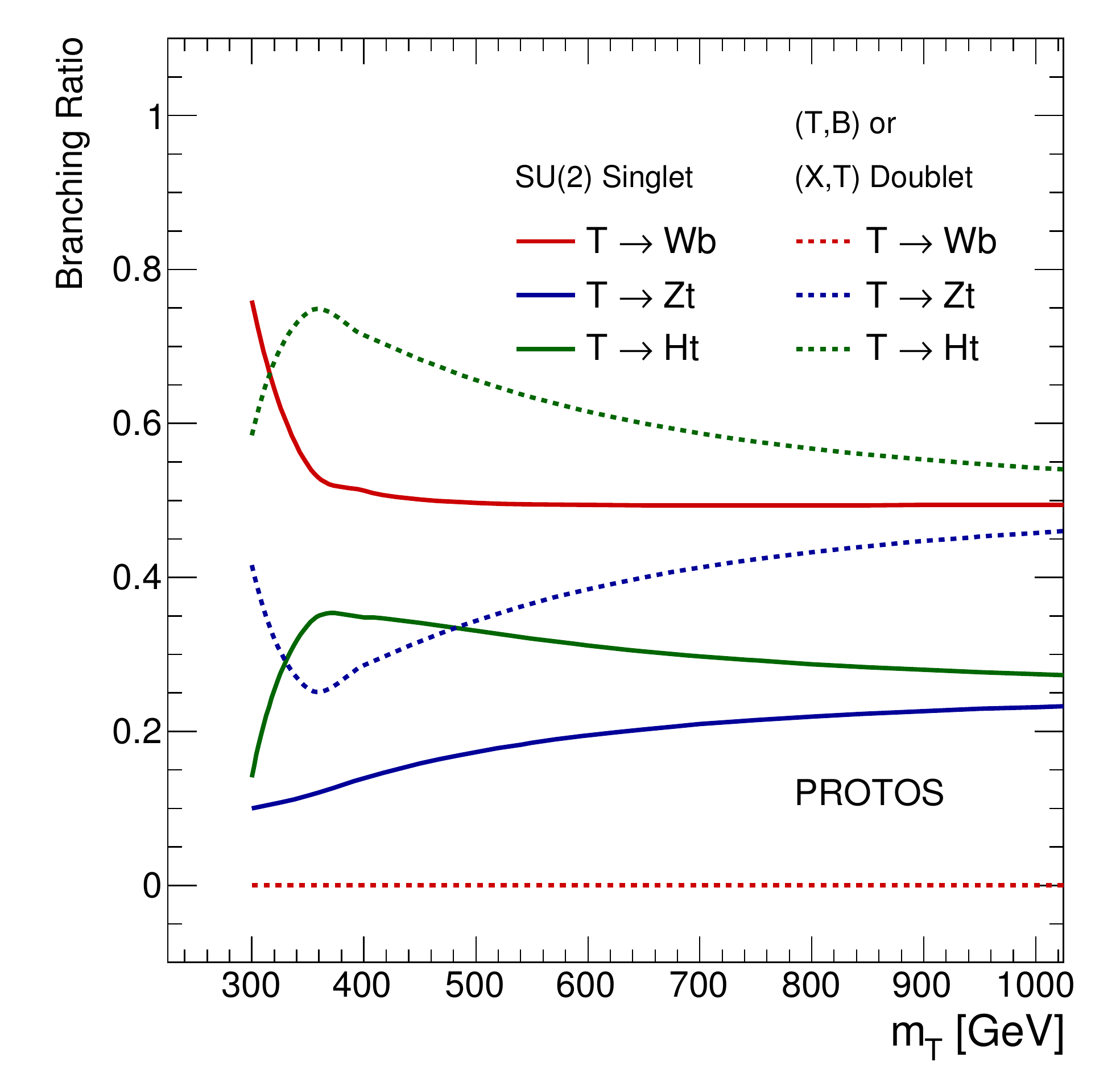}
(a)
\end{minipage}
\begin{minipage}[b]{0.5\linewidth} 
\centering
\epsfxsize=0.99\textwidth\epsfbox{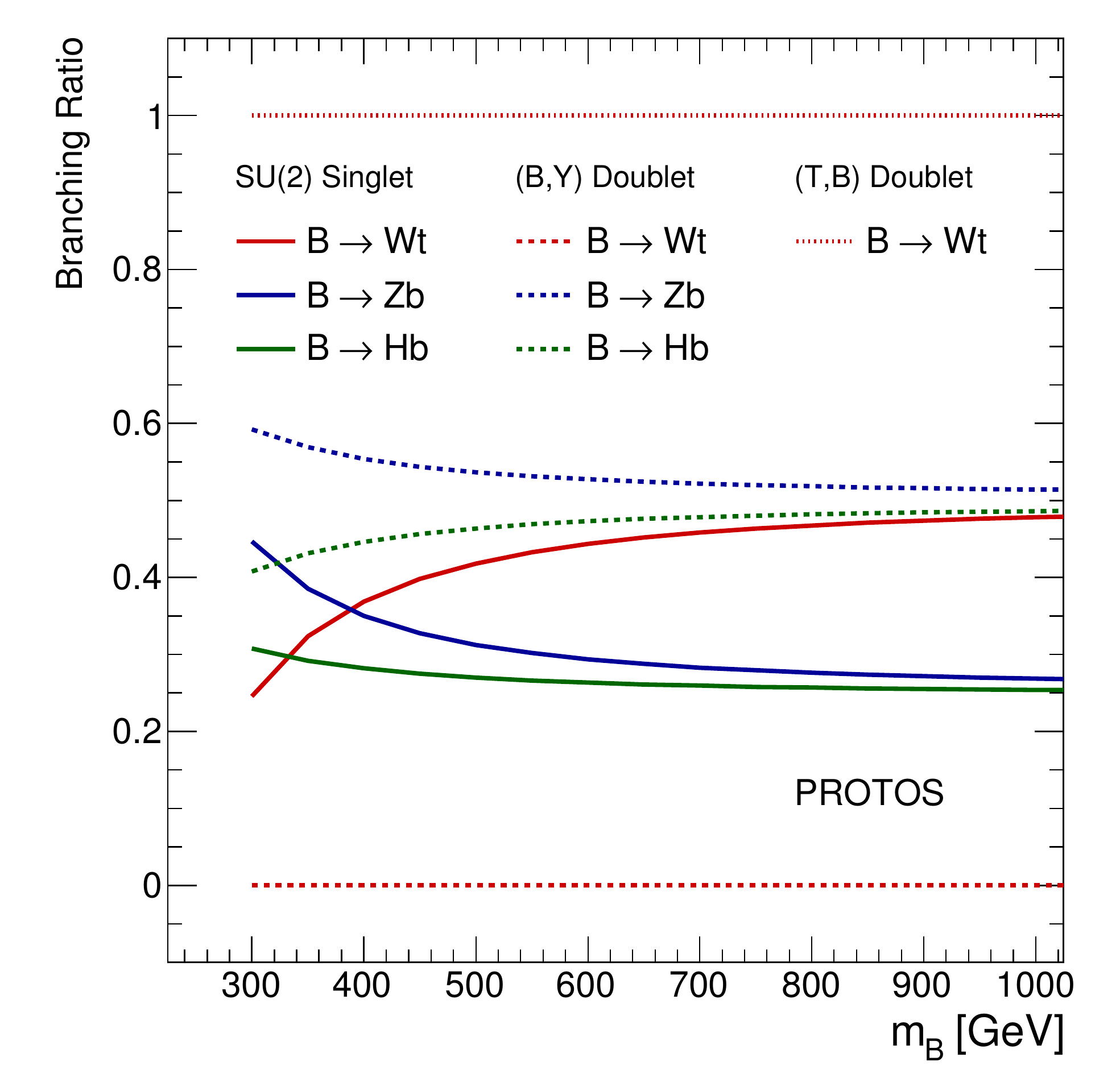}
(b)
\end{minipage}
\caption{Vector-like $T$ quark branching ratios (a) to the $Wb$, $Zt$, and $Ht$ decay modes as a function of the $T$ quark mass, computed with {\sc protos}~\cite{PROTOS} for an $SU(2)$ singlet and two types of doublets.  Likewise, vector-like $B$ quark branching ratios (b) to the $Wt$, $Zb$, and $Hb$ decay modes for a singlet and two types of doublets.  The $X$ quark in an $(X,T)$ doublet has charge $+5/3$, and the $Y$ quark in a $(B,Y)$ doublet has charge $-4/3$.}
\label{fig:BRs}
\end{figure}

The final-state topology depends on the decay modes of the new quarks.  Unlike chiral quarks, which only decay at tree level in the charged-current decay mode, vector-like quarks may decay at tree level to a $W$, $Z$, or $H$ boson plus an SM quark.  Additionally, vector-like quarks are often assumed to couple preferentially to third-generation SM quarks~\cite{VLQmixing,JA_mixing}, particularly in the context of naturalness arguments.  Thus, figure~\ref{fig:xsec}(a) depicts a $T$ or a $B$ vector-like quark, represented by $Q$, decaying to either an SM $t$ or $b$ quark, represented by $q$ or $q'$, and a $Z$, $H$, or $W$ boson.  The branching ratios of a $T$ quark as a function of its mass, as computed by {\sc protos}~v2.2~\cite{PROTOS,JA_TP}, are shown in figure~\ref{fig:BRs}(a).\footnote{The branching ratios in figure~\ref{fig:BRs} are valid for small mixing between the new heavy quark and the third-generation quark.  For example, using the mass eigenstate basis notation of refs.~\cite{JA_TP,JA_new,RealisticCHM}, and the relations in appendix~A of ref.~\cite{JA_new}, $V_{Tb} \approx X_{tT}$ in the limit of small mixing, and hence these mixing parameters cancel when computing branching ratios using the width expressions in eq.~(22) of ref.~\cite{JA_TP}.}  A weak-isospin ($SU(2)$) singlet $T$ quark hypothesis is depicted, as well as a $T$ that is part of an $SU(2)$ doublet.  The doublet prediction is valid for an $(X,T)$ doublet, where the charge of the $X$ quark is $+5/3$, as well as a $(T,B)$ doublet when a mixing assumption of $V_{Tb} \ll V_{tB}$ is made~\cite{JA_TP}.  The charged-current mode, $BR(T \rightarrow Wb)$, is absent in the doublet cases.  Similarly, figure~\ref{fig:BRs}(b) shows the branching ratio of a $B$ quark as a function of its mass for the singlet and doublet hypotheses.  In the case of a $(T,B)$ doublet, $BR(B \rightarrow Wt)=1$.  Branching ratio values are also shown in figure~\ref{fig:BRs}(b) for a $(B,Y)$ doublet, where the charge of the $Y$ quark is $-4/3$. The charged-current mode, $BR(B \rightarrow Wt)$, is absent in the $(B,Y)$ doublet case.

Monte Carlo (MC) simulated samples of leading-order (LO) pair-production events were generated for the $T\bar{T}$ and $B\bar{B}$ hypotheses with {\sc protos}~v2.2 interfaced with {\sc pythia}~\cite{Pythia} v6.421 for parton shower and fragmentation, and using the MSTW 2008 LO~\cite{MSTW} set of PDFs.  These samples are normalized using the {\sc Top++} cross-section predictions.  The vector-like quarks decay with a branching ratio of $1/3$ to each of the three modes ($W,Z,H$).  Arbitrary sets of branching ratios consistent with the three modes summing to unity are obtained by reweighting the samples using particle-level information.  An SM Higgs boson with a mass of 125~GeV is assumed.  The primary set of samples spans quark masses between 350~GeV and 850~GeV in steps of 50~GeV and implement $SU(2)$ singlet couplings.  Additional samples were produced at two mass points (350~GeV and 600~GeV) using $SU(2)$ doublet couplings in order to confirm that kinematic differences arising from the different chirality of singlet and doublet couplings are negligible in this analysis. The above samples were passed through a simulation of the ATLAS detector~\cite{FullSim} that employs a fast simulation of the response of the calorimeters~\cite{AFII_new}. Additional samples with quark masses of 400~GeV, 600 GeV, and 800~GeV were also produced using the standard {\sc geant}~v4~\cite{GEANT4} based simulation of all the detector components, to test the agreement.

\subsection{Electroweak single production}

\begin{figure*}[tbp]
\centering
\subfigure[]{\includegraphics[width=0.44\textwidth]{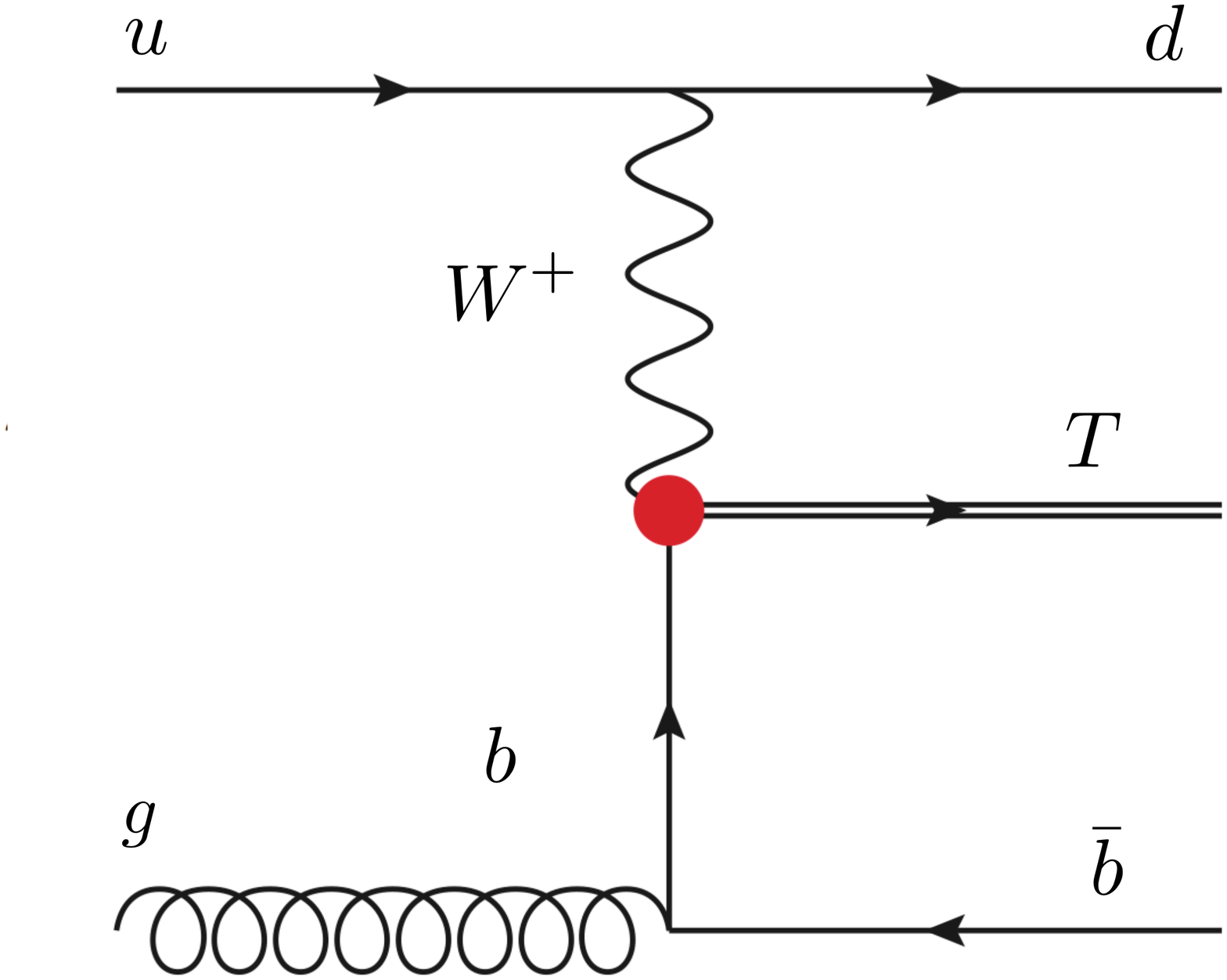}\label{fig:Tbq}}
\subfigure[]{\includegraphics[width=0.44\textwidth]{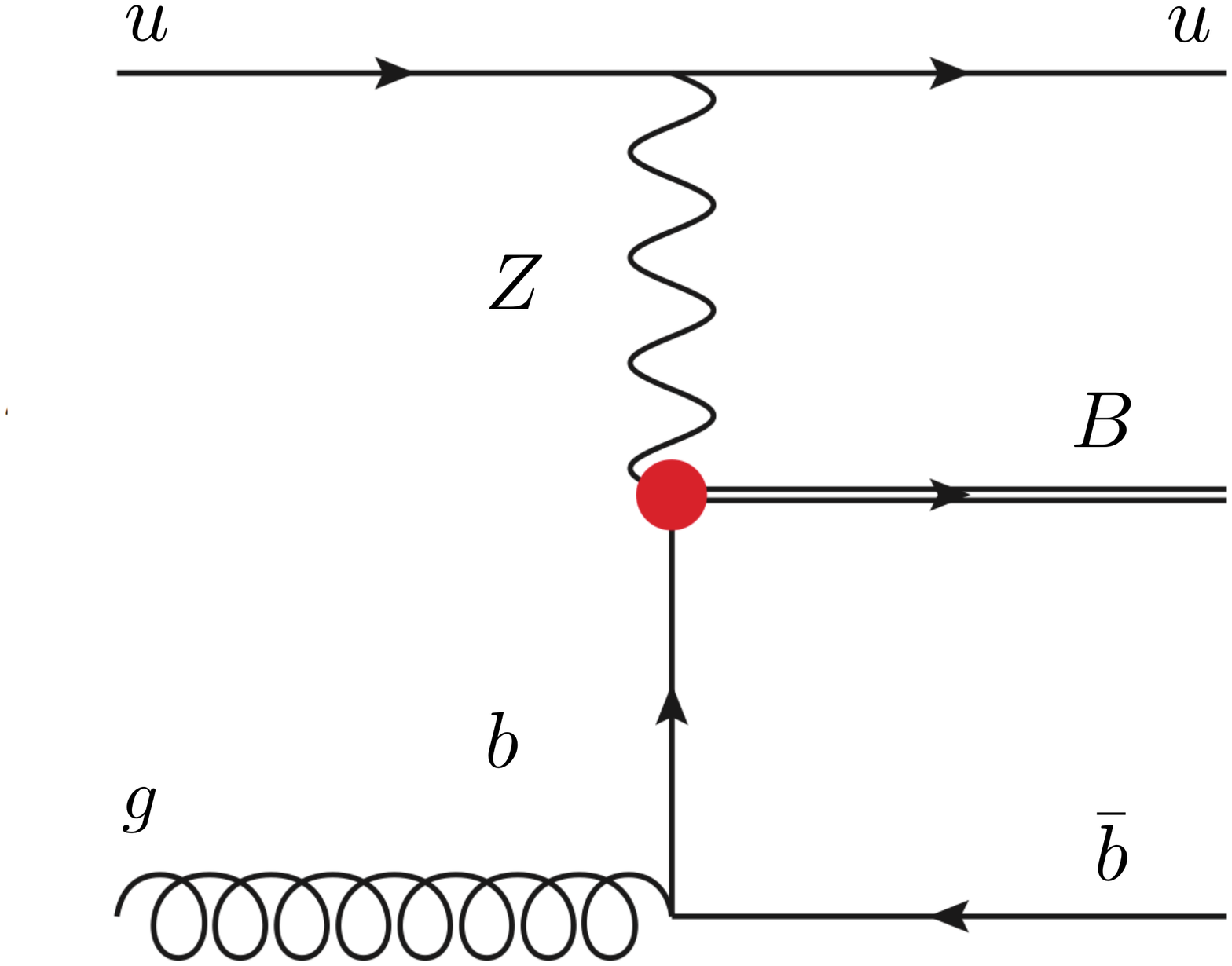}\label{fig:Bbq}}
\caption{Representative diagrams illustrating the $t$-channel electroweak single production of (a) a $T$ quark via the $T\bar{b}q$ process and (b) a $B$ quark via the $B\bar{b}q$ process.}
\label{fig:singleprod_diagrams}
\end{figure*}

Another way to produce heavy quarks is singly via the electroweak interaction.  The $t$-channel process provides the largest contribution, as is also the case for SM single-top production at the LHC.  Figures~\ref{fig:singleprod_diagrams}(a,b) illustrate the $t$-channel $2 \rightarrow 3$ process producing a vector-like $T$ or $B$ quark, respectively, in association with a $b$-quark\footnote{The $t$-channel production in association with a top quark is also possible, but the cross section is over an order of magnitude smaller for the same heavy quark mass, and for the same mixing parameter value.} and a light-generation quark.  Cross sections as a function of the heavy quark mass are also shown in figure~\ref{fig:xsec}(b) for the $T\bar{b}q$ and $B\bar{b}q$ processes, with the long-dashed lines indicating the prediction using {\sc protos} with mixing parameter values~\cite{JA_TP,JA_new} of $V_{Tb}=0.1$ and $X_{bB}=0.1$, respectively.  These reference values were chosen to reflect the magnitude of indirect upper bounds on mixing~\cite{JA_new,RealisticCHM} from precision electroweak data when assuming a single vector-like multiplet is present in the low-energy theory.  No kinematic requirements are placed on the $b$-quark or the light-flavor quark produced in association with the heavy quark.  The single-production cross sections scale quadratically with the mixing parameter.

The indirect constraints on the mixing parameters may be relaxed if several multiplets are present in the low-energy spectrum, as would be the case in realistic composite Higgs models~\cite{RealisticCHM}.  Several authors have emphasized the importance of the single-production mechanism in this context~\cite{TP_guide,RealisticCHM,CH_Natascia}, in particular, that it could represent a more favorable discovery mode than the pair-production mechanism.  Figure~\ref{fig:xsec}(b) shows the predicted $T\bar{b}q$ cross section in a specific composite Higgs model~\cite{CH_Natascia} that was implemented in {\sc madgraph} v5~\cite{Madgraph} and provided by the authors of the model.  In this model, the $WTb$ vertex is parameterized by the variable $\lambda_{T}$, which is related to the Yukawa coupling in the composite sector and the degree of compositeness of the third-generation SM quarks.\footnote{The notation of ref.~\cite{CH_Natascia} follows that adopted in ref.~\cite{ContinoServant}, and uses the weak eigenstate basis.  For small values of $\lambda_{T}$, or large heavy quark masses, $V_{Tb} \approx (\lambda_{T} v)/(\sqrt{2}M_{T})$, with $v=246$~GeV.  See footnote~2 of ref.~\cite{ContinoServant} for more details.}  The prediction shown corresponds to $\lambda_{T}=2$, and values between 1 and 5 were considered in ref.~\cite{CH_Natascia}. 

Fast-simulation samples for the $T\bar{b}q$ process were produced for the $T$ singlet of Ref.~\cite{CH_Natascia} with {\sc madgraph}.  Samples were generated for $T$ masses between 400~GeV and 1050~GeV in 50~GeV steps setting $\lambda_{T}=2$.  In addition, samples were generated for $\lambda_{T}$ between 1 and 5 in integer steps at the 700~GeV mass point, in order to study the dependence of the experimental acceptance and the sensitivity to large $T$ widths.  Particle-level $T\bar{b}q$ samples were also produced with {\sc protos} for several mass and $V_{Tb}$ values to check the degree of consistency between the two generators in the kinematic distributions of relevance to this analysis.  Fully simulated samples for the $B\bar{b}q$ process were produced with {\sc protos} for $SU(2)$ singlet $B$ quarks with masses between 400~GeV and 1200~GeV and $X_{bB}=0.1$.  Particle-level $B\bar{b}q$ process samples were also produced for different mixing values, and for a $B$ in a $(B,Y)$ doublet.  The $B\bar{b}q$ process is absent in some composite Higgs models~\cite{TP_guide,CH_Natascia}.  This is not a generic prediction, however, and the $B\bar{b}q$ process may be relevant in the context of a $(B,Y)$ doublet and corresponding improvements to electroweak fits~\cite{JA_new}.

\section{Background modeling}

The SM backgrounds in this analysis are predicted primarily with simulated samples normalized to next-to-leading order, or higher, cross-section calculations.  Unless stated otherwise, all samples for SM processes are passed through a full detector simulation.  Two leading-order multi-parton event generators, {\sc alpgen}~\cite{Alpgen} and {\sc sherpa}~\cite{Sherpa}, were carefully compared at each stage of the dilepton channel analysis to provide a robust characterization of the dominant $Z+{\rm jets}$ background.  The cross-section normalization of both is set by the NNLO prediction calculated with the {\sc dynnlo} program~\cite{DYNNLO}.  

The {\sc alpgen} $Z+{\rm jets}$ samples were produced using v2.13 with the CTEQ6L1~\cite{CTEQ} PDF set and interfaced to {\sc pythia} v6.426 for parton-shower and hadronization.  Separate inclusive \mbox{$Z+{\rm jets}$} and dedicated $Z+c\bar{c}+{\rm jets}$ and $Z+b\bar{b}+{\rm jets}$ samples were simulated.  Heavy-flavor quarks in the former arise from the parton shower, while in the latter they can be produced directly in the matrix element.   To avoid double-counting of partonic configurations generated by both the matrix element and the parton shower, a parton--jet matching scheme~\cite{MLM} is employed in the generation of the samples.  Likewise, to remove double-counting when combining the inclusive and dedicated heavy-flavor samples, another algorithm is employed based on the angular separation between heavy quarks ($q_{h}=c,b$).  The matrix-element prediction is used if $\Delta R(q_{h},\bar{q}_{h})>0.4$, and the parton-shower prediction is used otherwise.

The {\sc sherpa} $Z+{\rm jets}$ samples were produced using v1.4.1 with the CT10~\cite{CT10} PDF set, and generated setting the charm and bottom quarks to be massive.  Filters are used to divide the samples into events containing a bottom hadron, events without a bottom hadron but containing a charm hadron, and events with neither a charm nor a bottom hadron.  In this paper, the $Z+{\rm bottom~jet(s)}$ background category corresponds to the bottom hadron filtered samples, while the $Z+{\rm light~jets}$ category combines the two other samples without a bottom hadron\footnote{Further, these categories are referred to more concisely as $Z$+bottom and $Z$+light in tables~\ref{table:yields_dilepton_1} and \ref{table:yields_dilepton_2}.}.  To increase the statistical precision of the prediction at large values of the $Z$ boson transverse momentum, $p_{\rm T}(Z)$, each hadron filtered sample was produced in different $p_{\rm T}(Z)$ intervals: inclusive, $70-140$~GeV, $140-280$~GeV, $280-500$~GeV, and greater than $500$~GeV.  The first three samples are reconstructed with a fast detector simulation while the latter two use full detector simulation.  As a result of the higher statistical precision in the final stages of selection, these {\sc sherpa} samples constitute the default $Z+{\rm jets}$ prediction.

The dominant source of background events in the early selection stages of the trilepton channel analysis arise from $Z$ bosons produced in association with $W$ bosons.  The diboson processes ($WZ$, $ZZ$, and $WW$) are generated with {\sc sherpa}, and normalized to NLO cross-section predictions obtained with {\sc mcfm}~\cite{xsec_DB}.  In the final selection stages of the trilepton analysis, an important source of background events arise from $Z$ bosons produced in association with a pair of top quarks.  The $t\bar{t}+V$ processes, where $V=W,Z$, are modeled with {\sc madgraph}~\cite{Madgraph}, using {\sc pythia} for parton shower and hadronization.  These samples are also normalized to NLO cross-section predictions~\cite{xsec_TTV}.

Processes that do not contain a $Z$ boson constitute subleading background contributions.  Simulated $t\bar{t}$ events are produced using {\sc powheg}~\cite{Powheg,Powheg2,Powheg3,Powheg4} for the matrix element with the CT10 PDF set.  Parton shower and hadronization are performed with {\sc pythia} v6.421.  The $t\bar{t}$ cross section is determined by the {\sc Top++} prediction, computed as in the signal hypothesis, but setting the top quark mass to 172.5~GeV.  Samples generated with {\sc mc@nlo}~\cite{MCatNLO,MCatNLO2} interfaced to {\sc herwig}~6.520.2~\cite{Herwig1,Herwig2,Herwig} are used to estimate the $Wt$ and $s$-channel single-top processes, while {\sc AcerMC}~\cite{Acer} interfaced to {\sc pythia} is used to estimate the $t$-channel process.  The single-top processes are normalized to NLO cross-section predictions~\cite{xsec_ST}.

Events that enter the selected $Z$ candidate sample as a result of a fake or non-prompt lepton satisfying the lepton selection criteria are estimated with data, using samples obtained by relaxing or inverting certain lepton identification requirements.  Such contributions are found to be less than $5\%$ of the total background in the early stages of event selection and negligible in the final stages.

\section{Search strategies}
\label{sec:strategy}

This section outlines the search strategies.  The single- and pair-production signal hypotheses are targeted in both the dilepton and trilepton channels.  A common set of event selection requirements are made first, and a small number of specific requirements are added to enhance the sensitivity of the dilepton and trilepton channels to the single- or pair-production hypotheses.  Table~\ref{table:selection} summarizes the selection criteria for reference.

\begin{figure*}[tbp]
\centering
\subfigure[]{\includegraphics[width=0.48\textwidth]{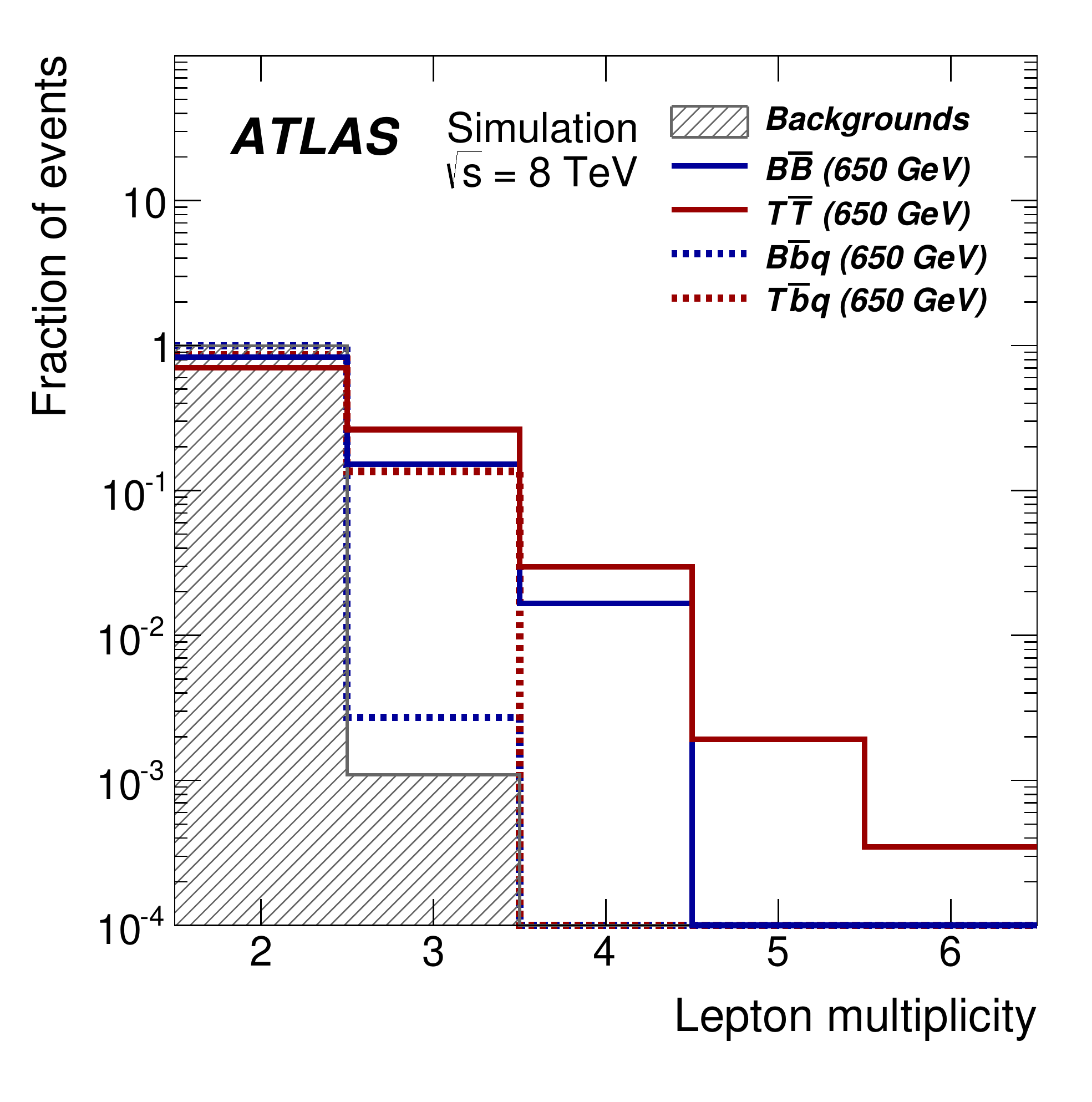}\label{fig:shape_nlep}}
\subfigure[]{\includegraphics[width=0.48\textwidth]{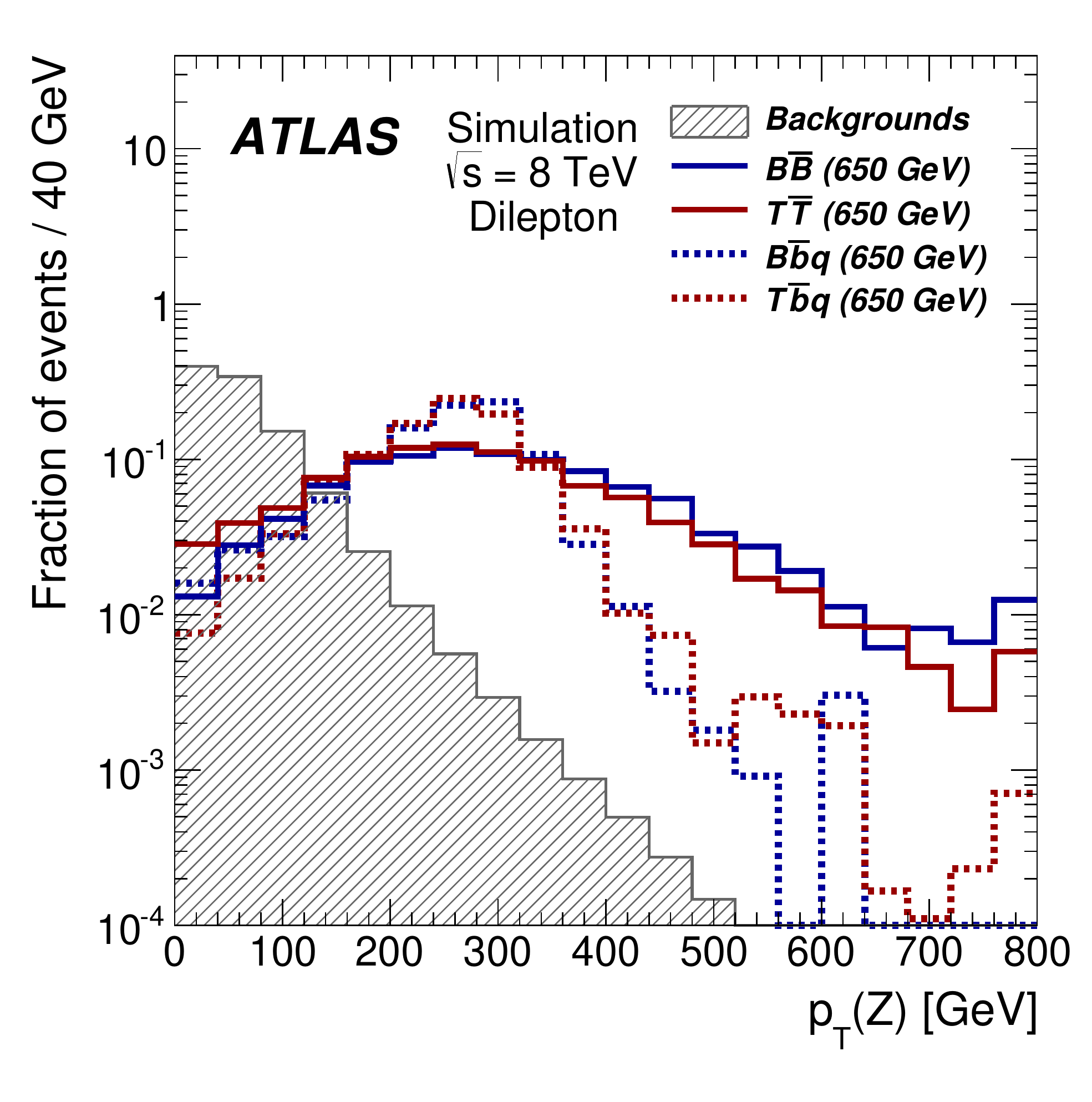}\label{fig:shape_ptz}}
\subfigure[]{\includegraphics[width=0.48\textwidth]{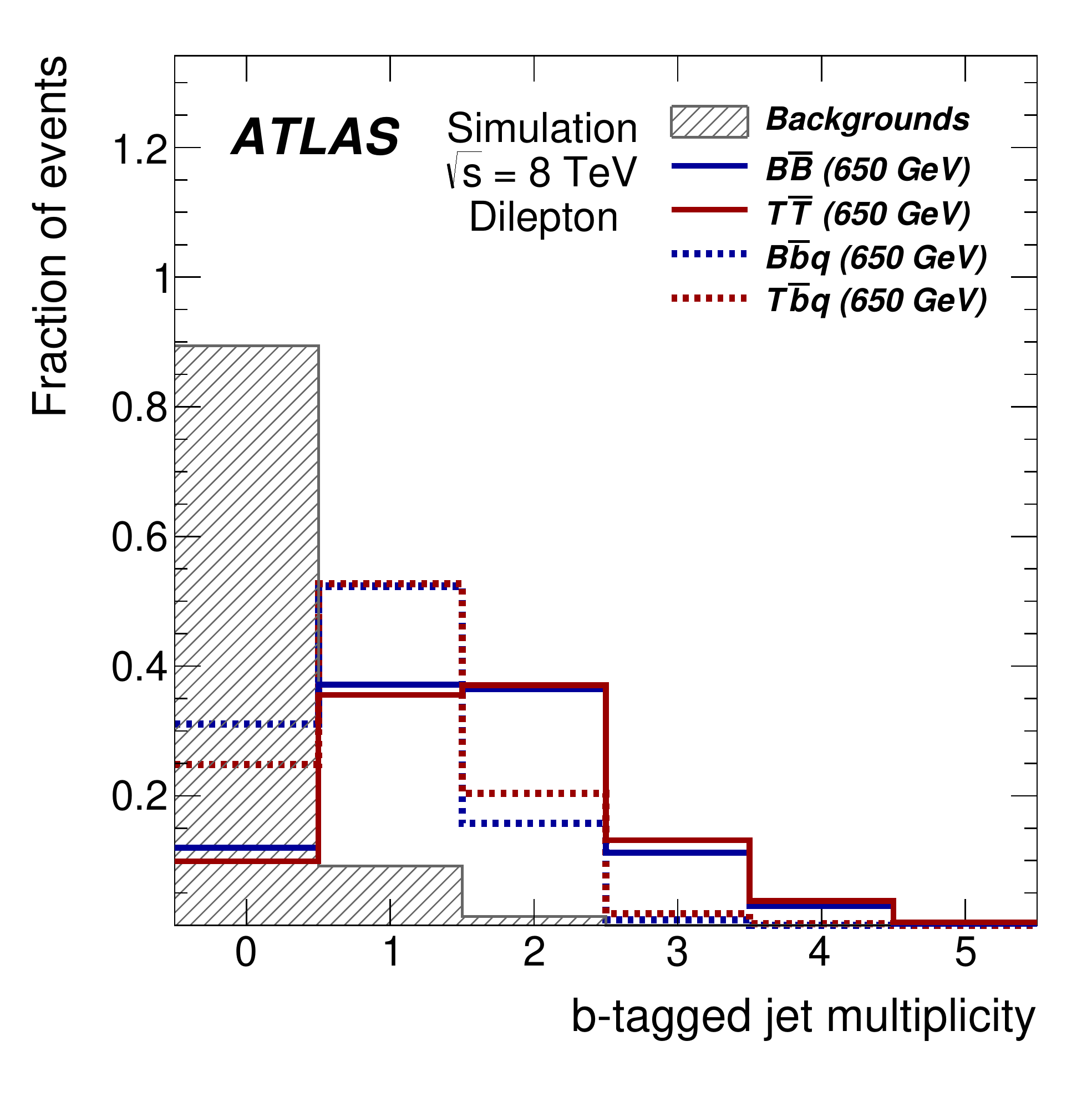}\label{fig:shape_nbtag}}
\subfigure[]{\includegraphics[width=0.48\textwidth]{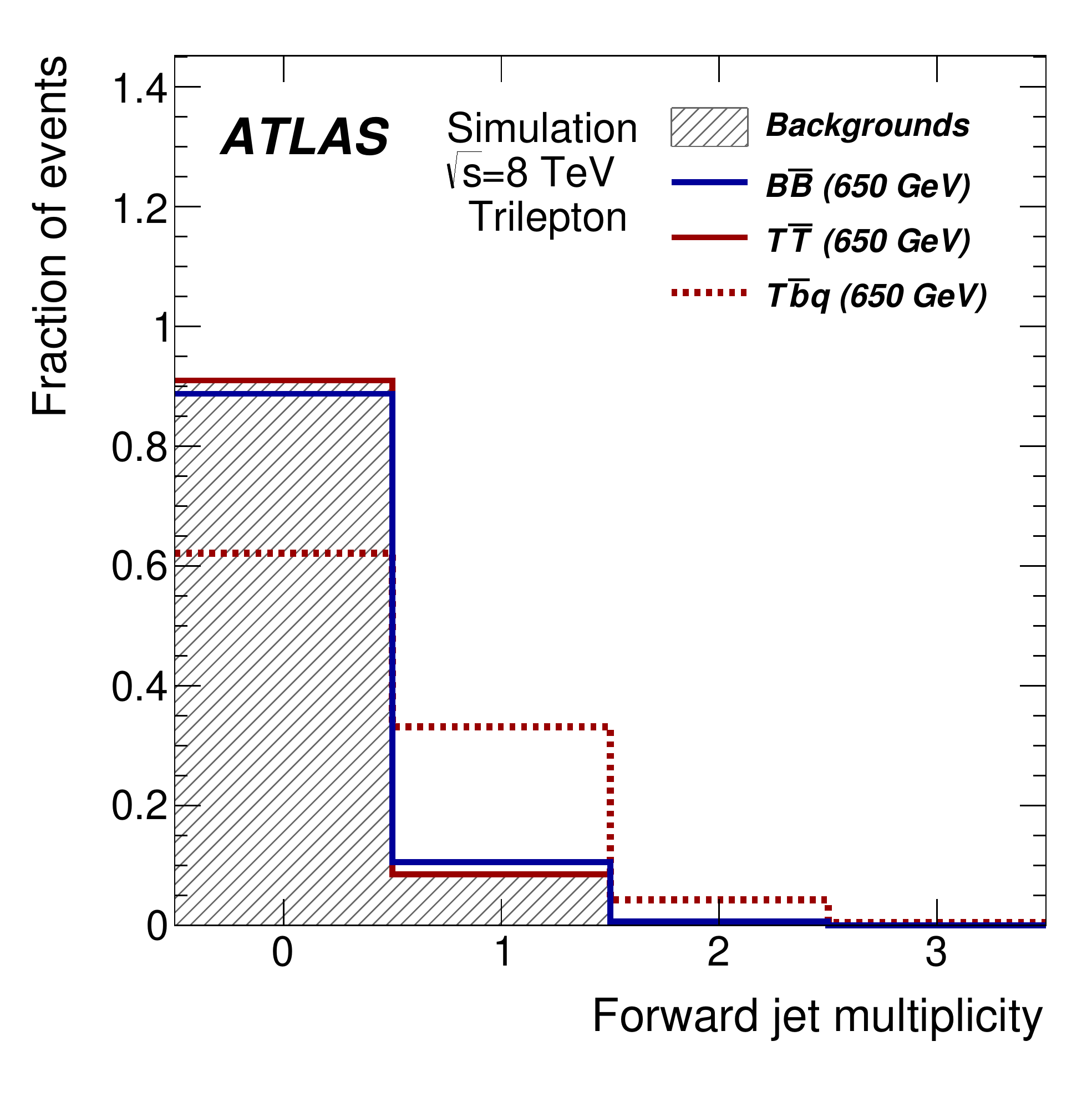}\label{fig:shape_njjet}}
\caption{Unit-normalized distributions of signal-sensitive variables employed in this analysis.  The filled histograms correspond to SM backgrounds.  Unfilled histograms correspond to signal, with solid (dashed) lines representing pair (single) production of $SU(2)$ singlet $T$ and $B$ quarks with a mass of 650~GeV.  The rightmost bin in each histogram contains overflow events.  Panel (a) shows the lepton multiplicity distribution after a $Z+\geq2$ central jets selection.  Panel (b) shows the $p_{\rm T}(Z)$ distribution, and (c) the $b$-tagged jet multiplicity distribution, for dilepton channel events.  Panel (d) shows the forward-jet multiplicity distribution in trilepton events.}
\label{fig:discvar1}
\end{figure*}

Figure~\ref{fig:discvar1} presents unit-normalized distributions of simulated signal and background events in several discriminating variables employed in the event selection.  The reference signals shown correspond to the single and pair production of $SU(2)$ singlet $T$ and $B$ quarks with a mass of 650~GeV.  Figure~\ref{fig:discvar1}(a) presents the lepton multiplicity distribution after selecting events with a $Z$ boson candidate and at least two central jets.  The shapes of the signal and background distributions motivate separate criteria for events with exactly two leptons, and those with  three or more, with the strategy for the former focused on background rejection, and the strategy for the latter focused on maintaining signal efficiency.  The only signal hypothesis not expected to produce events with a third isolated lepton is the $B(\rightarrow Zb)\bar{b}q$ process.  The other three processes are capable of producing, in addition to the $Z$ boson, a $W$ boson that decays to leptons.  The $W$ boson could arise from a top quark decay, or directly from the other heavy quark decay in the case of the pair-production signal.

\begin{table}[tbp]
\begin{center}
\begin{tabular}{|c||c||c||c|}
\hline
\multicolumn{4}{|c|}{\textbf{Event selection}}                                                \\\hline\hline
\multicolumn{4}{|c|}{$Z$ boson candidate preselection}                                                \\
\multicolumn{4}{|c|}{$\geq 2$ central jets}                                                     \\
\multicolumn{4}{|c|}{$p_{\rm T}(Z)\geq 150$~GeV}                                                  \\ \hline
\multicolumn{2}{|c||}{\textbf{Dilepton channel}}                 & \multicolumn{2}{c|}{\textbf{Trilepton channel}}      \\ \hline
\multicolumn{2}{|c||}{$=2$ leptons}                 & \multicolumn{2}{c|}{$\geq 3$ leptons}      \\ \hline
\multicolumn{2}{|c||}{$\geq2$ $b$-tagged jets}        & \multicolumn{2}{c|}{$\geq 1$ $b$-tagged jet} \\ \hline
\textbf{Pair production}               & \textbf{Single production} & \textbf{Pair production}     & \textbf{Single production}    \\ \hline
$H_{\rm T}({\rm jets}) \geq 600$~GeV & $\geq 1$ fwd. jet & --                  & $\geq 1$ fwd. jet    \\ \hline\hline
\multicolumn{4}{|c|}{\textbf{Final discriminant}}                                                \\\hline
\multicolumn{2}{|c||}{$m(Zb)$}        & \multicolumn{2}{c|}{$H_{\rm T}$(jets+leptons)} \\ \hline
\end{tabular}
\end{center}
 \caption{Summary of the event selection criteria.  Preselected $Z$ boson candidate events are divided into dilepton and trilepton categories.  The requirements on the number of central jets and the $Z$ candidate transverse momentum are common to both channels, and for the pair- and single-production hypotheses.  Other requirements are specific to a lepton channel or the targeted production mechanism.  The last row lists the final discriminant used for hypothesis testing.}
\label{table:selection}
\end{table}

At least two central jets are required in both lepton channels, and when testing both production mechanism hypotheses.  The requirement is over $95\%$ efficient for the pair-production signals, and over $70\%$ efficient for the single-production signals, while suppressing the backgrounds by a factor of 20 and 5 in the dilepton and trilepton channels, respectively.  A second common requirement is on the minimum transverse momentum of the $Z$ boson candidate: $p_{\rm T}(Z)>150$~GeV.  Figure~\ref{fig:discvar1}(b) presents the $p_{\rm T}(Z)$ distribution in signal and background dilepton channel events after the $Z+\geq 2$ central jets selection.

Figure~\ref{fig:discvar1}(c) presents the $b$-tagged jet multiplicity distribution, also after the $Z+\geq2$ central jets selection in the dilepton channel.  Pair-production signal events are expected to yield at least two $b$-jets, whether produced directly from a heavy quark decay, the decay of a top quark, or the decay of a Higgs boson.  Single-production signal events also yield two $b$-jets, but the one arising from the $b$-quark produced in association is less often in the acceptance for $b$-tagging.  In order to effectively suppress the large $Z+{\rm jets}$ background, dilepton channel events are required to contain at least two $b$-tagged jets when testing both the single- and pair-production hypotheses.  A requirement of at least one $b$-tagged jet sufficiently balances signal efficiency and background rejection in the trilepton channel.

\begin{figure*}[tbp]
\centering
\subfigure[]{\includegraphics[width=0.48\textwidth]{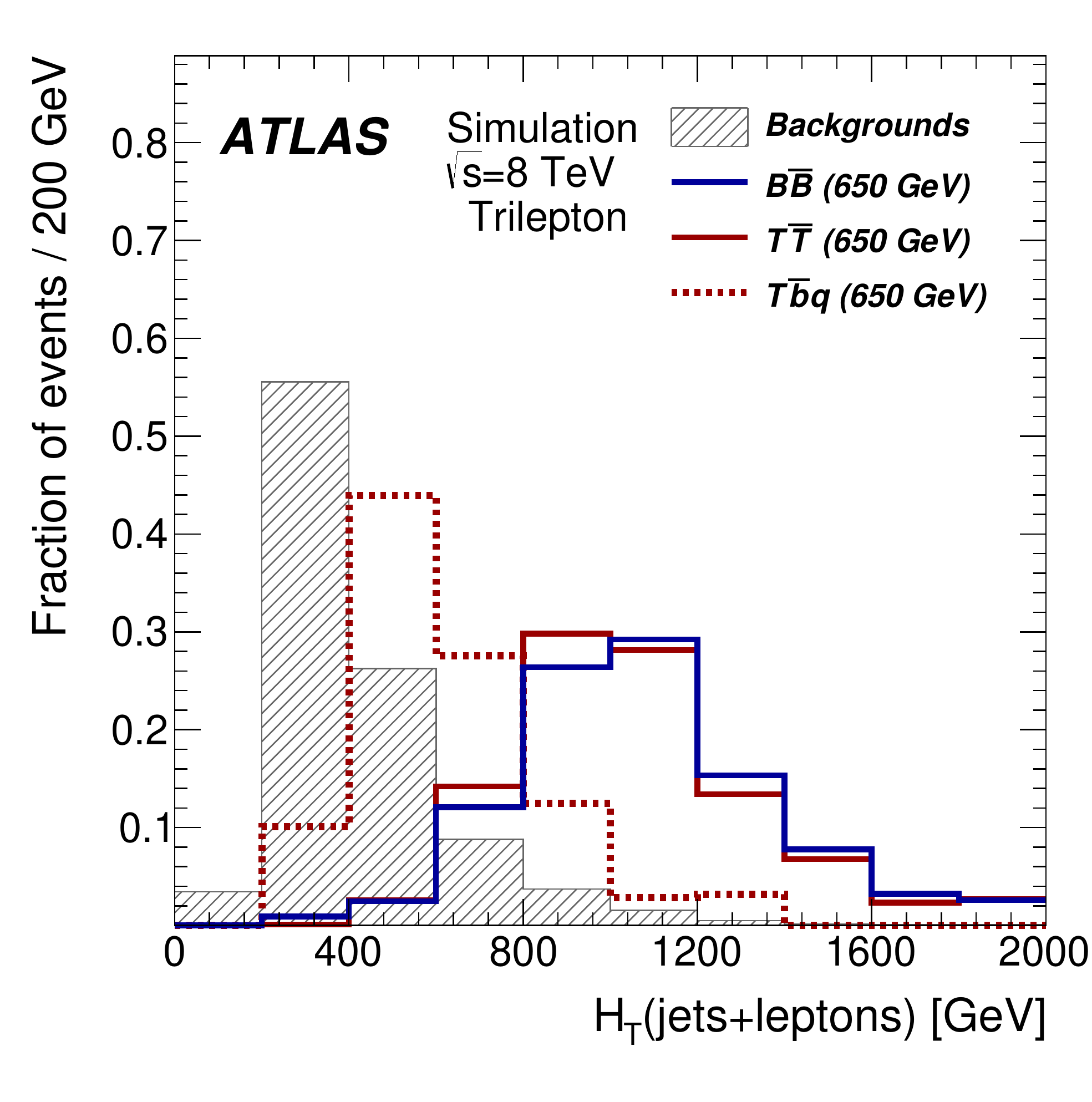}\label{fig:shape_ht3}}
\subfigure[]{\includegraphics[width=0.48\textwidth]{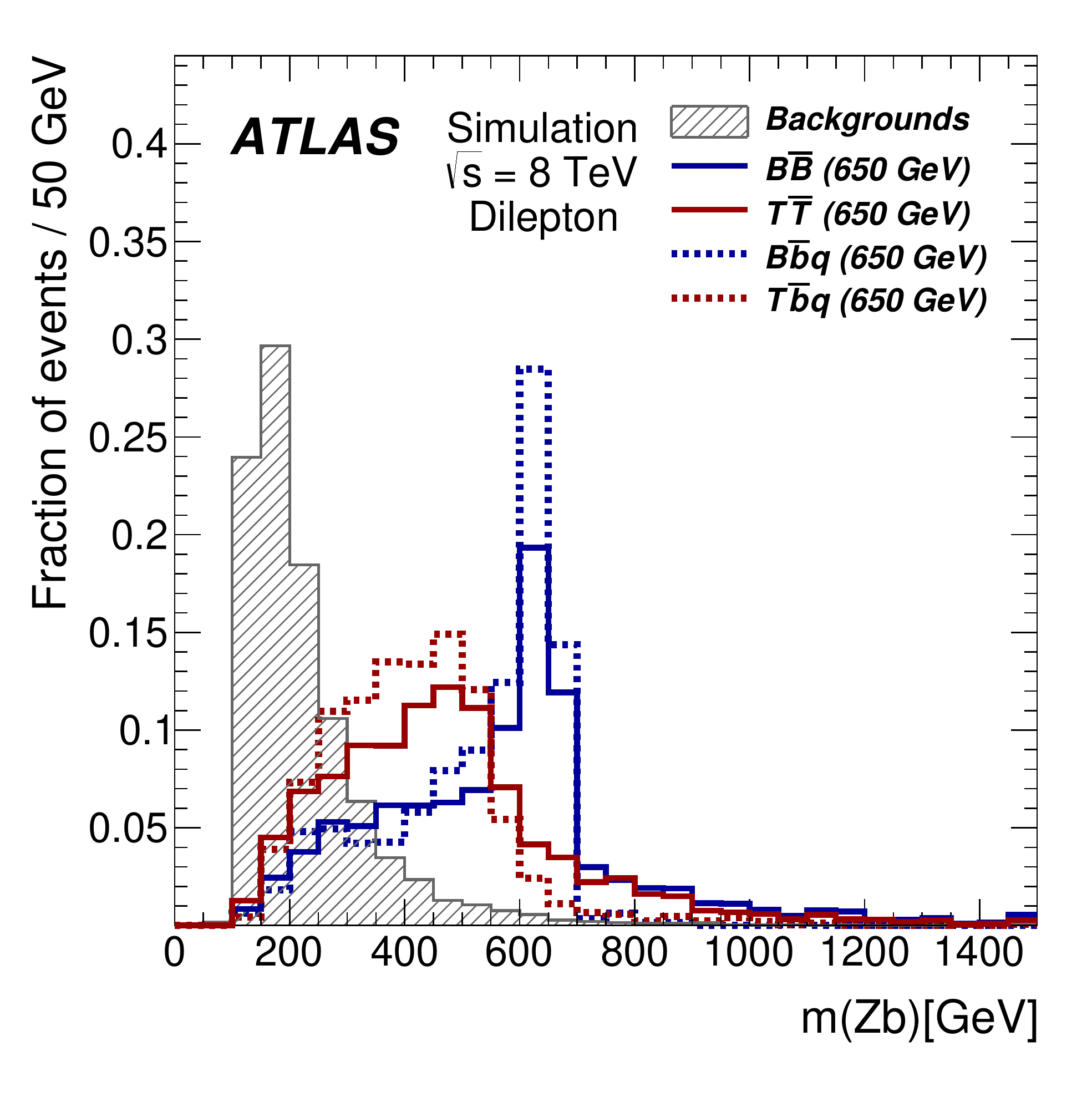}\label{fig:shape_mzb}}
\caption{Unit-normalized distributions of the discriminating variables used for hypothesis testing, shown at the $Z+\geq2$ central jets selection stage: (a) $H_{\rm T}({\rm jets+leptons})$ in the trilepton channel, and (b) the $m(Zb)$ distribution in the dilepton channel.  The filled histograms correspond to SM backgrounds.  
Unfilled histograms correspond to signal, with solid (dashed) lines representing pair (single) production of $SU(2)$ singlet $T$ and $B$ quarks with a mass of 650~GeV.  The rightmost bin in each histogram contains overflow events.}
\label{fig:discvar2}
\end{figure*}

Signal events from pair production often produce several energetic jets.  The scalar sum of the transverse momentum of all central jets in the event, $H_{\rm T}({\rm jets})$, is a powerful variable to further reduce the background in the dilepton channel.  Selected events in this channel are required to satisfy $H_{\rm T}({\rm jets})>600$~GeV when testing the pair-production hypotheses.  The transverse momentum of leptons is not included, as the same information is effectively utilized in the $p_{\rm T}(Z)$ requirement, and it is advantageous to study the jet activity separately.  In the trilepton channel, however, the lepton transverse momenta are used in the variable $H_{\rm T}({\rm jets+leptons})$ to include, in particular, the discriminating power of the transverse momentum of the third lepton.  Figure~\ref{fig:discvar2}(a) shows the $H_{\rm T}({\rm jets+leptons})$ distribution in trilepton events with at least two central jets.  A minimum-value requirement on this variable is not imposed, but rather the full shape is used as the final discriminant for hypothesis testing.  The variable provides good separation between the background and pair-production signals.  Although the separation is not as powerful for the single-production signals, the variable becomes increasingly effective for higher quark masses.

The associated light-flavor quark produced in the electroweak single production of heavy quarks gives rise to an energetic forward jet.  Figure~\ref{fig:discvar1}(d) presents the forward-jet multiplicity distribution in trilepton channel events after all requirements are made to select events for the pair-production hypotheses.  The presence of a forward jet is an additional requirement when testing the single-production hypotheses.

The invariant mass of the $Z$ boson candidate and highest-$p_{\rm T}$ $b$-tagged jet, $m(Zb)$, is used as the final discriminant in the dilepton channel, and is shown in figure~\ref{fig:discvar2}(b).  The distribution is strongly peaked at the heavy quark mass in the case of a $B$ quark.  The distribution peaks at a lower value and is wider in the case of a $T$ quark;  both features are consequences of the $W$ boson that is not included in the mass reconstruction.  The $H_{\rm T}({\rm jets})$ requirement is removed and the forward-jet requirement is added when testing the single-production hypotheses in the dilepton channel.

\section{Comparison of the data to the predictions}

Section~\ref{sec:strategy} motivated the selection criteria that are applied in the dilepton and trilepton channel analyses and when considering the single- and pair-production hypotheses.  This section presents the comparison of the data to the predictions.  Section~\ref{subsec:dileptonpair} presents the dilepton channel analysis, and focuses on the pair-production hypotheses.  Section~\ref{subsec:trileptonpair} presents the trilepton channel analysis, also focusing on the pair-production hypotheses.  Section~\ref{subsec:single} shows the results of both channels under the modified selection criteria used to test the single-production hypotheses.

\subsection{Dilepton channel analysis targeting the pair-production hypotheses}
\label{subsec:dileptonpair}

The preselected sample of $Z$ boson candidate events with exactly two leptons comprises $12.5\times10^{6}$ events ($5.5\times10^{6}$ and $7.0\times10^{6}$ events in the $ee$ and $\mu\mu$ channels, respectively).  These yields are consistent with the predictions within uncertainties, which at this stage of the analysis are less than $5\%$ and dominated by the Drell--Yan cross section and acceptance, luminosity, and lepton reconstruction uncertainties.  The predicted distributions of several kinematic variables are observed to agree well with the data, and the sample is then restricted to the subset of events with at least two central jets.  This sample comprises $501\times10^{3}$ events, and is also found to be well described by the {\sc sherpa} and {\sc alpgen} predictions within the uncertainties, now also including those associated with jet reconstruction.

\begin{figure*}[tbp]
\centering
\subfigure[]{\includegraphics[width=0.48\textwidth]{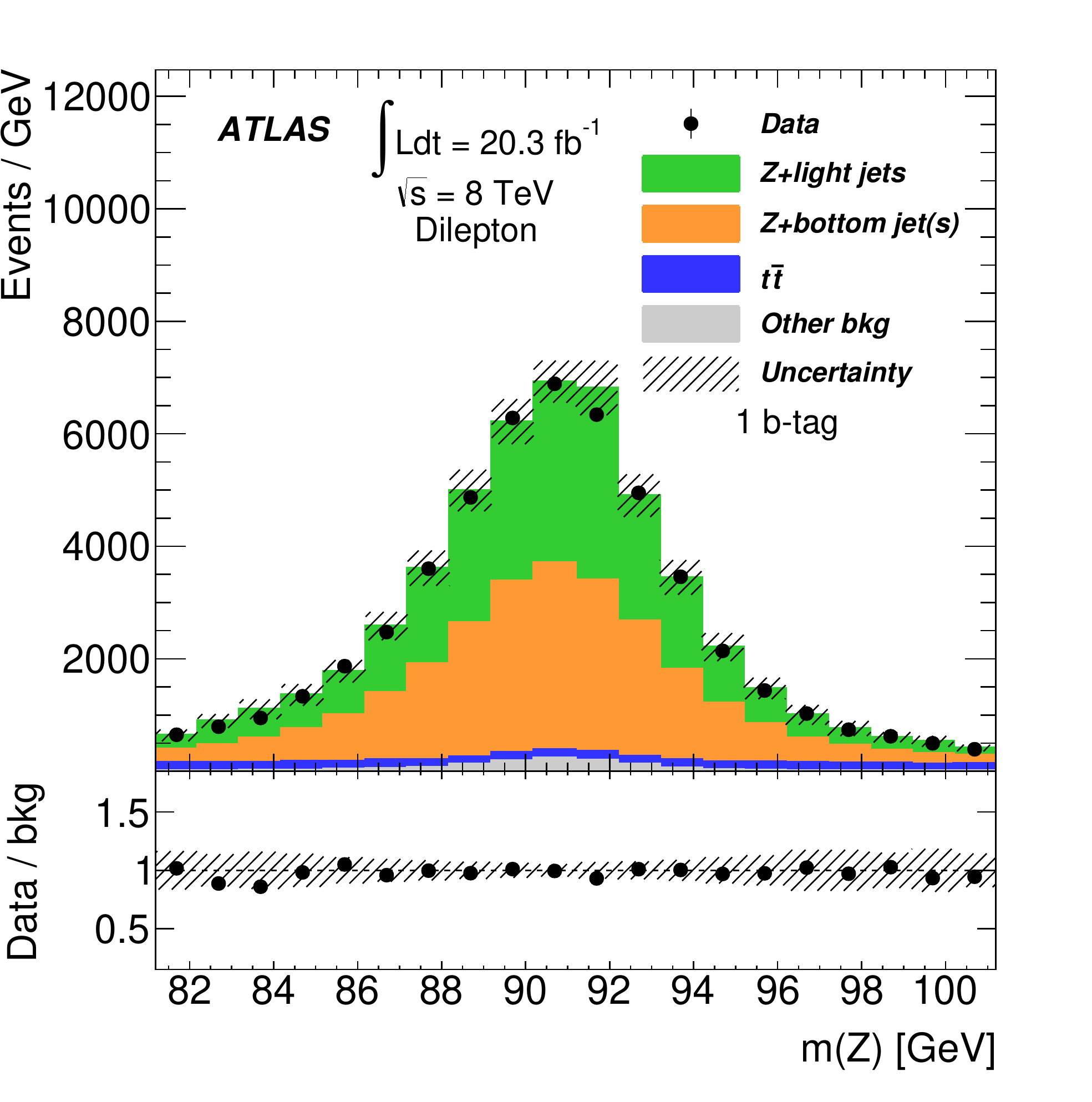}\label{fig:dilepton_zmass_1}}
\subfigure[]{\includegraphics[width=0.48\textwidth]{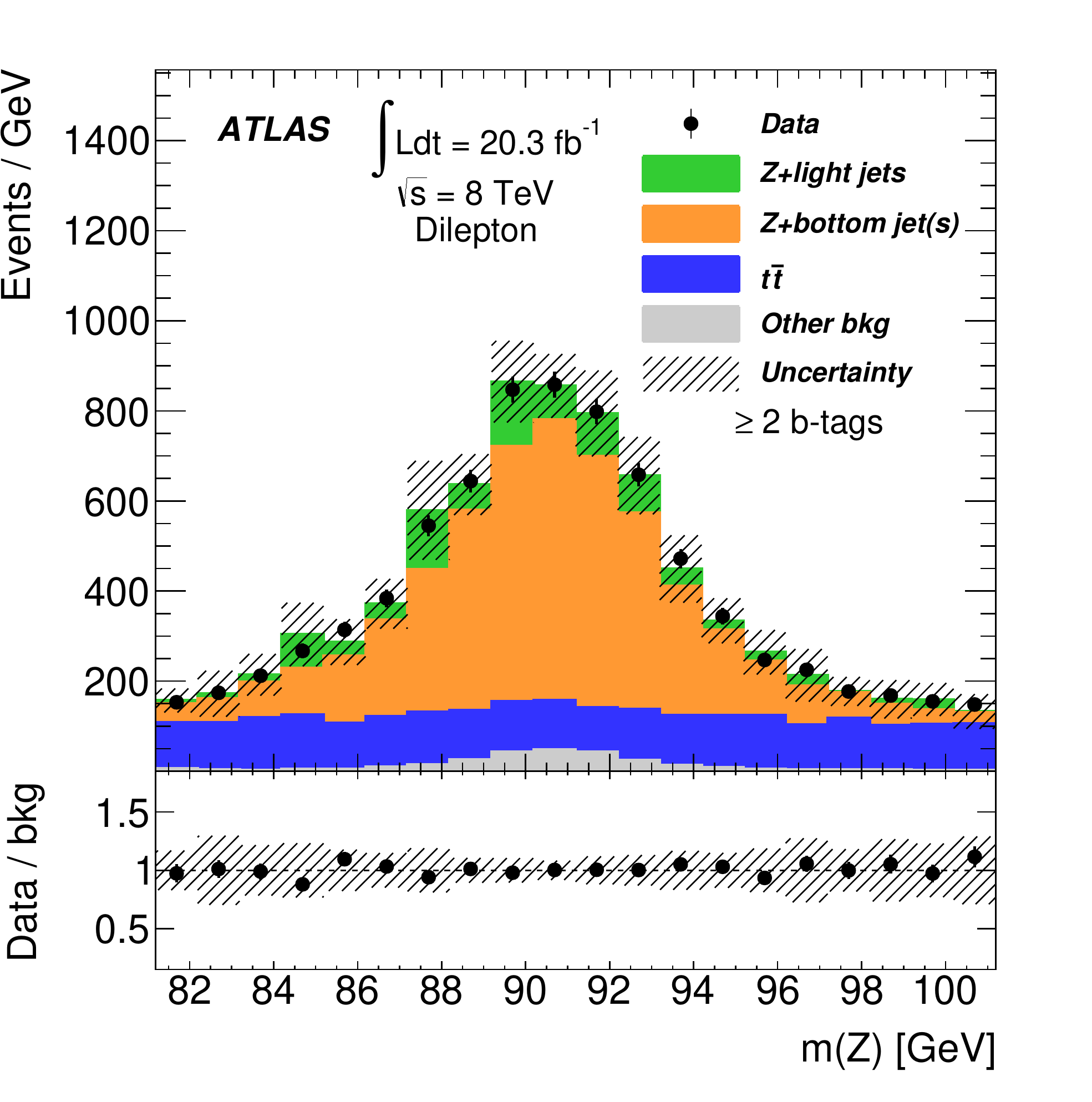}\label{fig:dilepton_zmass_2}}
\subfigure[]{\includegraphics[width=0.48\textwidth]{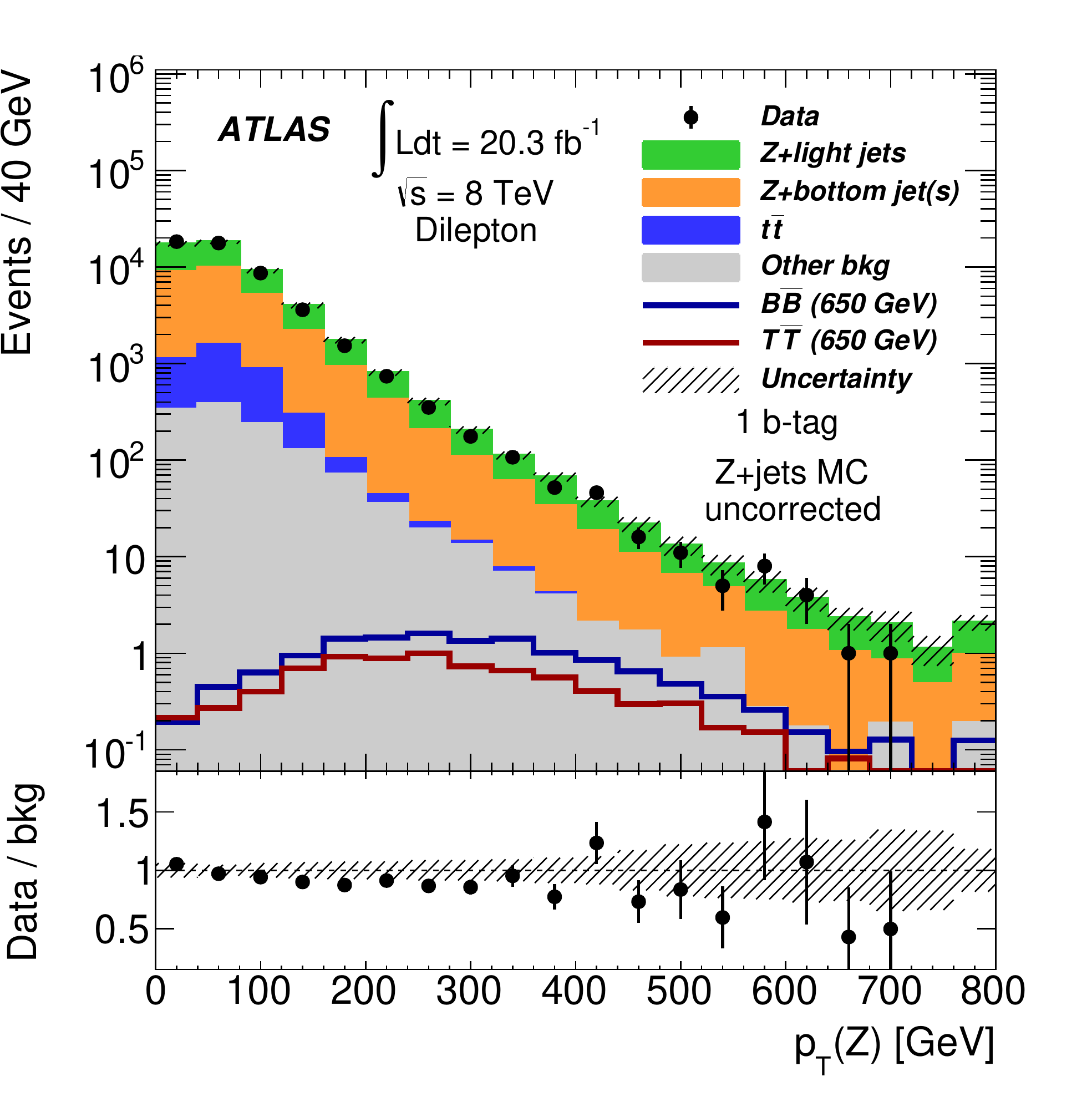}\label{fig:dilepton_zpt_1}}
\subfigure[]{\includegraphics[width=0.48\textwidth]{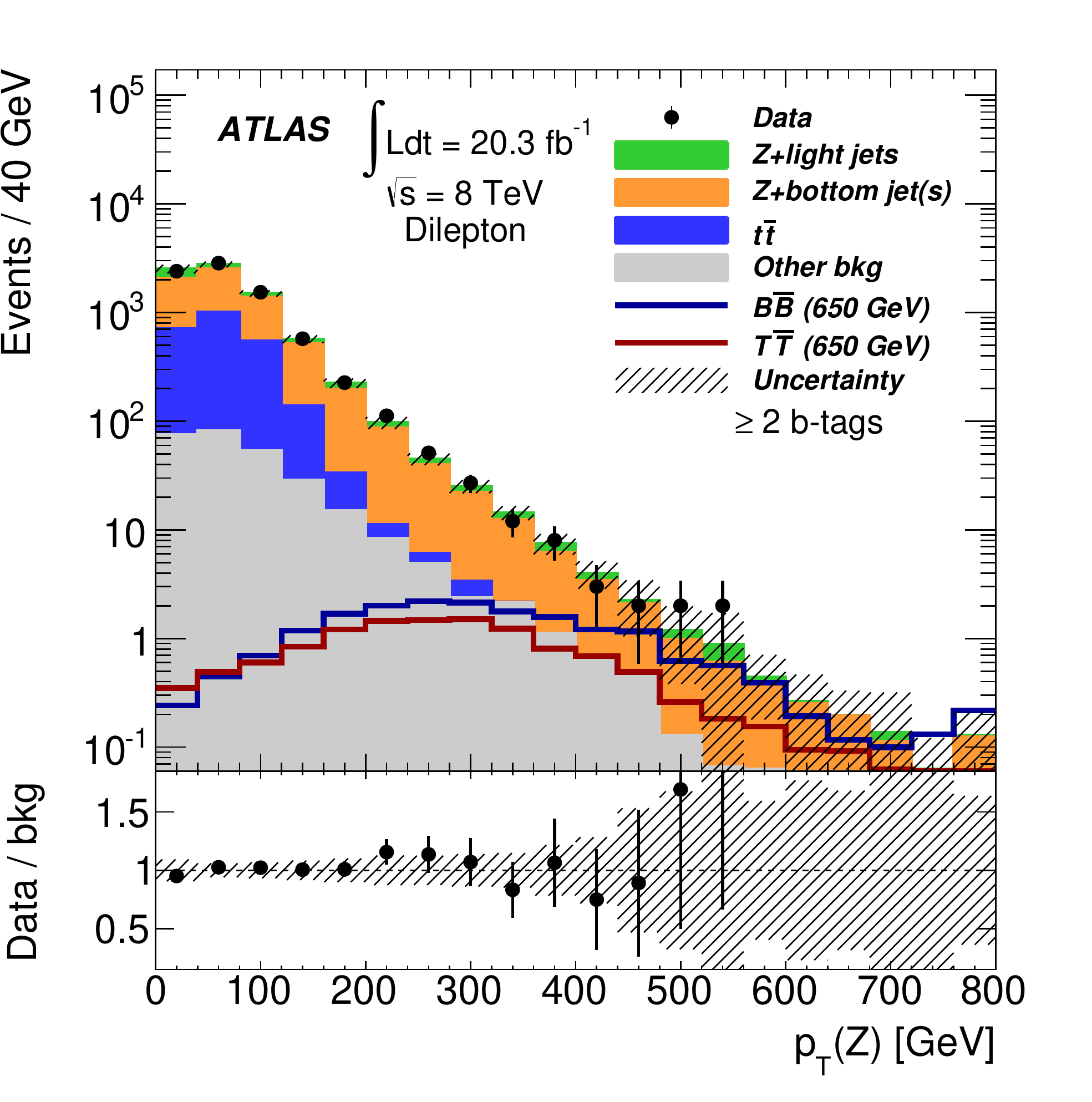}\label{fig:dilepton_zpt_2}}
\caption{The distribution of the $Z$ boson candidate mass, $m(Z)$, in dilepton channel events with $\geq2$ central jets and (a) $N_{\rm tag}=1$ or (b) $N_{\rm tag}\geq2$.  Panels (c) and (d) show the distribution of the transverse momentum, $p_{\rm T}(Z)$, under the same selection criteria.  Panel (c) presents the $Z+$~jets prediction before the $p_{\rm T}(Z)$ spectrum correction described in the text is applied, while panel (d) is shown with it applied.  Reference signals are displayed for $B\bar{B}$ and $T\bar{T}$ production assuming $SU(2)$ singlet quarks with a mass of 650~GeV.  The hatched bands in the upper and lower panels represent the total background uncertainty.}
\label{fig:dilepton_zmasspt}
\end{figure*}

Events passing the $Z+\geq2$~central jets selection are then separated according to the number of $b$-tagged jets in the event ($N_{\rm tag}$).  Figure~\ref{fig:dilepton_zmasspt}(a) shows the $Z$ candidate mass distribution using the {\sc sherpa} $Z+{\rm jets}$ prediction in the control region consisting of events with $N_{\rm tag}=1$.  Table~\ref{table:yields_dilepton_1} presents the corresponding event yield.  Figure~\ref{fig:dilepton_zmasspt}(b) shows the $Z$ candidate mass distribution in the signal region consisting of events with $N_{\rm tag}\geq2$.  Table~\ref{table:yields_dilepton_2} presents the corresponding yield.  Differences in the predicted yields are observed in both the $N_{\rm tag}=1$ and $N_{\rm tag}\geq2$ categories when using {\sc alpgen} in place of {\sc sherpa}.  While the predictions using {\sc sherpa} are consistent with the data within the experimental uncertainties (5--8\%), those with {\sc alpgen} are systematically low by $20\%$ and $15\%$ in the $N_{\rm tag}=1$ and $N_{\rm tag}\geq2$ categories, respectively.  Agreement between data and the prediction outside the 10~GeV mass window, particularly in events with $N_{\rm tag}\geq2$ where $t\bar{t}$ events are predicted to contribute significantly, indicates that {\sc alpgen} underestimates the $Z+{\rm jets}$ contribution in events with $b$-tagged jets.  Therefore, scaling factors for the $Z+{\rm jets}$ prediction are derived at this stage such that the total background prediction matches the data yields in the signal-depleted region defined by $p_{\rm T}(Z)<100$~GeV.  The procedure is performed separately for events with $N_{\rm tag} =1$ and $N_{\rm tag}\geq2$, and is repeated when evaluating the impact of systematic uncertainties.  It is also applied to the {\sc sherpa} prediction, though not necessary a priori, in order that the same data-driven correction methods are applied to both generators.

\begin{table}[tbq]
\begin{center}
\begin{tabular}{cccc}\hline\hline
 & $Z+\geq2$~jets {\footnotesize ($N_{\rm tag} = 1$)} & $p_{\rm T} (Z) > 150$ GeV & $H_{\rm T}(\rm{jets})>600$ GeV\\\hline
$Z+$light {\footnotesize (no $p_{\rm T}$ corr.)} & $24000\pm1500$ & $1940\pm190$ & $104.6\pm8.6$\\
$Z+$light {\footnotesize ($p_{\rm T}$ corr.)} & $23600\pm1500$ & $1700\pm150$ & $89\pm12$\\\hline
$Z+$bottom {\footnotesize (no $p_{\rm T}$ corr.)} & $24100\pm1700$ & $1970\pm240$ & $82.5\pm8.0$\\
$Z+$bottom {\footnotesize ($p_{\rm T}$ corr.)} & $23600\pm1700$ & $1730\pm160$ & $71\pm11$\\\hline
$t\bar t$ & $2850\pm230$ & $68\pm11$ & $8.0\pm2.9$\\
Other SM & $1250\pm370$ & $180\pm60$ & $17.9\pm5.7$\\\hline
Total SM {\footnotesize (no $p_{\rm T}$ corr.)} & $52200\pm2300$ & $4150\pm310$ & $213\pm13$\\
Total SM {\footnotesize ($p_{\rm T}$ corr.)} & $51300\pm2300$ & $3690\pm230$ & $186\pm16$\\\hline\hline
Data & $51291$ & $3652$ & $171$\\\hline\hline
$B\bar B$~($m_B =$ 650 GeV) & $13.6\pm1.0$ & $11.7\pm0.9$ & $9.6\pm0.8$\\
$T\bar T$~($m_T = $ 650 GeV) & $7.9\pm0.5$ & $6.5\pm0.5$ & $5.2\pm0.5$\\\hline\hline
\end{tabular}
\end{center}
\caption{Predicted and observed number of events in the dilepton channel after selecting a $Z$ boson candidate and at least two central jets, exactly one of which is $b$-tagged.  The number of events further satisfying $p_{\rm T}(Z)>150$~GeV is listed next, followed by the number satisfying, in addition, $H_{\rm T}({\rm jets})>600$~GeV.  The $Z+$jets predictions, as well as the total background prediction, are shown before and after the $p_{\rm T}(Z)$ spectrum correction described in the text.  Reference $B\bar{B}$ and $T\bar{T}$ signal yields are provided for $m_{B/T}=650$~GeV and $SU(2)$ singlet branching ratios.  The uncertainties on the predicted yields include statistical and systematic sources.}
\label{table:yields_dilepton_1}          
\end{table} 

\begin{table}[h!]
\begin{center}
\begin{tabular}{cccc}
\hline\hline
 & $Z+\geq2$~jets {\footnotesize($N_{\rm tag} \geq2$)} & $p_{\rm T} (Z) \geq 150$ GeV & $H_{\rm T}(\rm{jets}) \geq 600$ GeV \\ \hline\hline
$Z+$light & ~$900\pm210$ &  $63\pm14$ & $4.0\pm1.3$ \\
$Z+$bottom & $4420\pm300$ & $382\pm49$ & $19.3\pm3.6$ \\
$t\bar t$ & $2190\pm230$ & $33.0\pm8.0$ & $4.6\pm1.5$ \\
Other SM & $270\pm70$ & $42\pm11$ & $4.0\pm1.1$ \\ \hline
Total SM & $7780\pm440$ & $519\pm53$ & $32.0\pm4.2$ \\ \hline\hline
Data & $7790$ & $542$ & $31$ \\ \hline\hline
$B\bar B$($m_B =$ 650 GeV) & $18.7\pm1.5$ & $16.5\pm1.4$ & $14.2\pm1.3$ \\
$T\bar T$($m_T =$ 650 GeV) & $12.1\pm0.8$ & $10.0\pm0.7$ & $8.6\pm0.7$ \\ \hline\hline
\end{tabular}
\end{center}
\caption{Predicted and observed number of events in the dilepton channel after selecting a $Z$ boson candidate and at least two central jets, at least two of which are $b$-tagged.  The number of events further satisfying $p_{\rm T}(Z)>150$~GeV is listed next, followed by the number satisfying, in addition, $H_{\rm T}({\rm jets})>600$~GeV.  Reference $B\bar{B}$ and $T\bar{T}$ signal yields are provided for $m_{B/T}=650$~GeV and $SU(2)$ singlet branching ratios. The uncertainties on the predicted yields include statistical and systematic sources.}
\label{table:yields_dilepton_2}
\end{table} 

\begin{figure*}[tbp]
\centering
\subfigure[]{\includegraphics[width=0.48\textwidth]{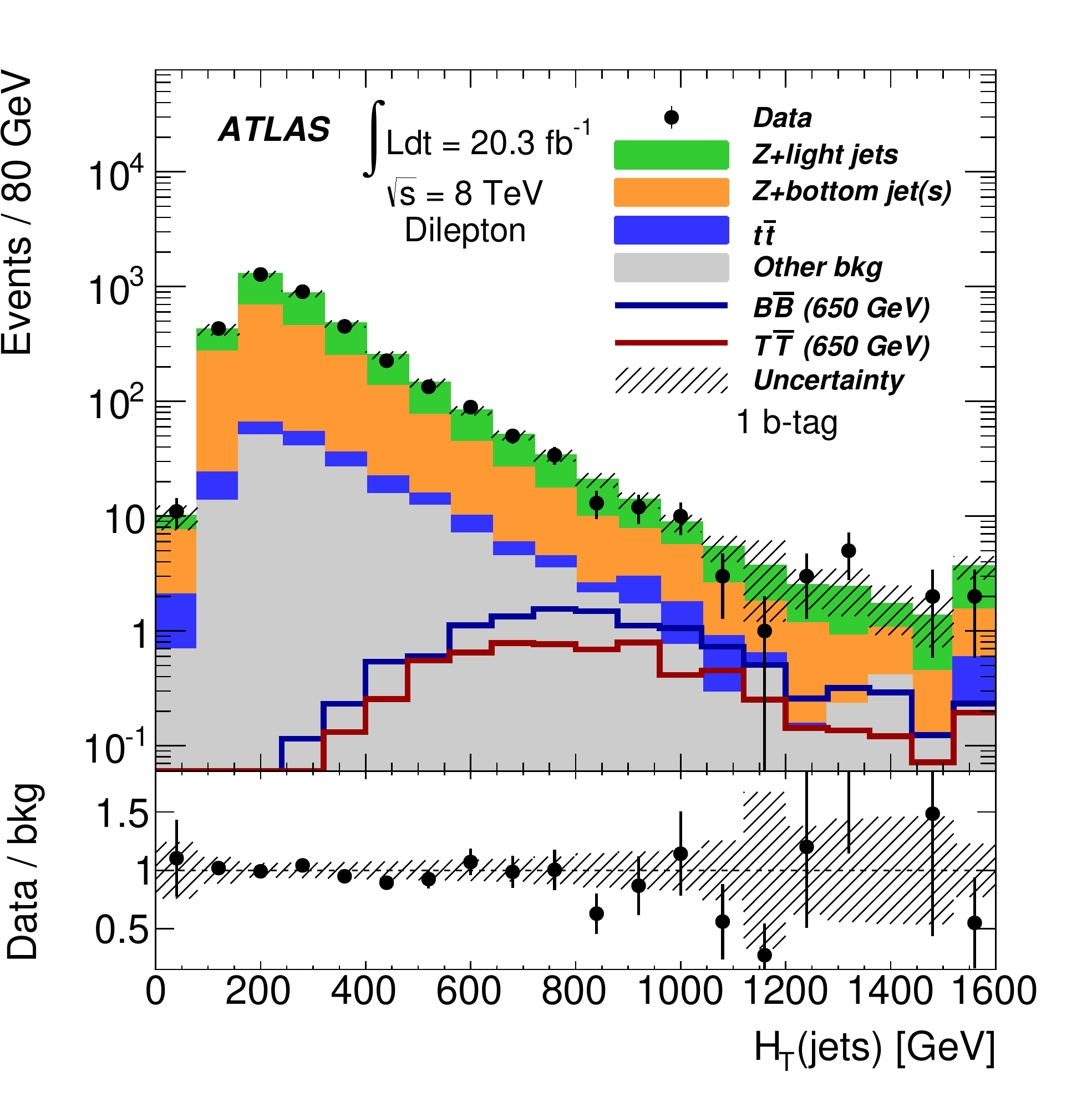}\label{fig:dilepton_ht_1}}
\subfigure[]{\includegraphics[width=0.48\textwidth]{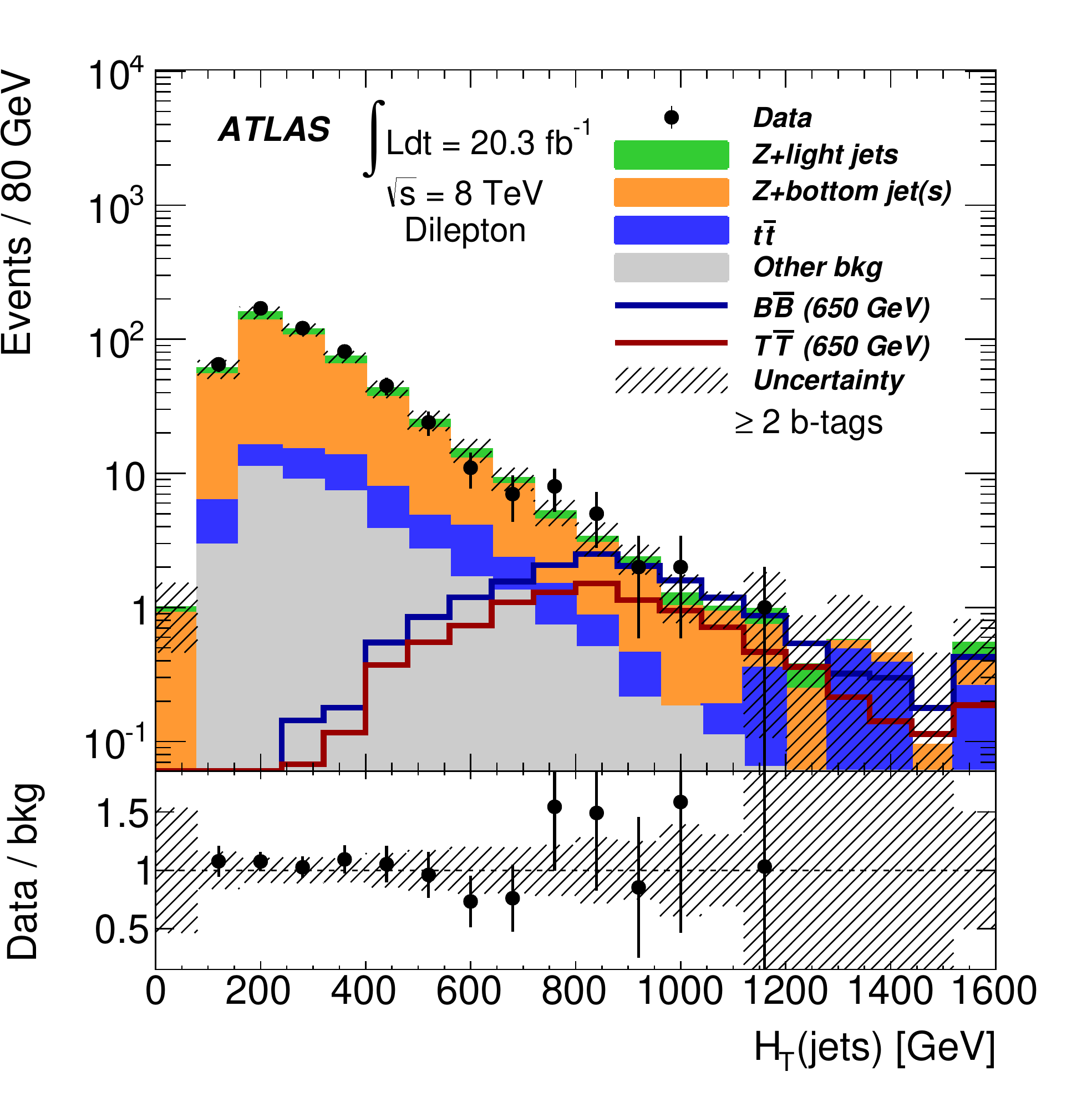}\label{fig:dilepton_ht_2}}
\subfigure[]{\includegraphics[width=0.48\textwidth]{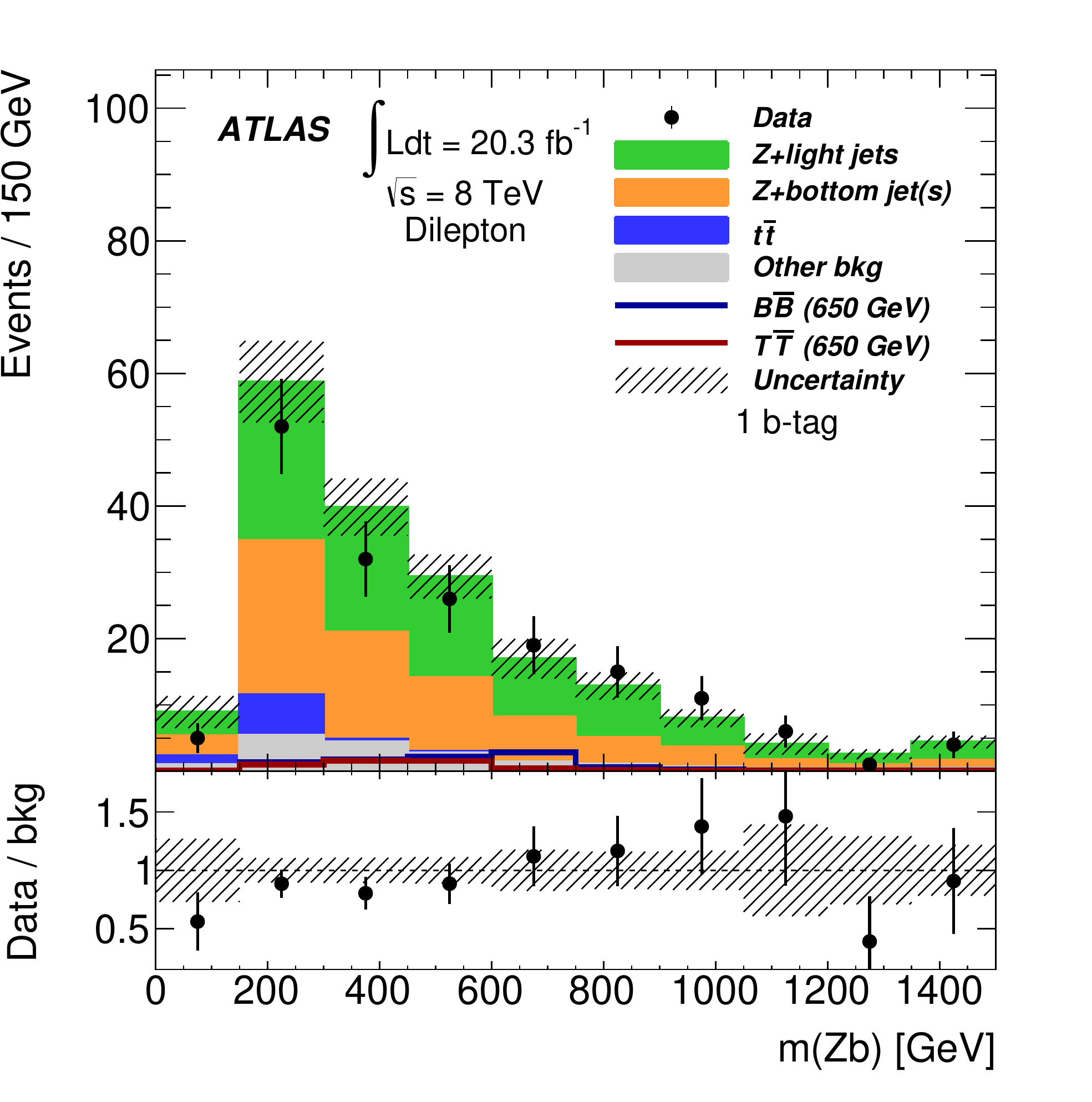}\label{fig:dilepton_mzb_1}}
\subfigure[]{\includegraphics[width=0.48\textwidth]{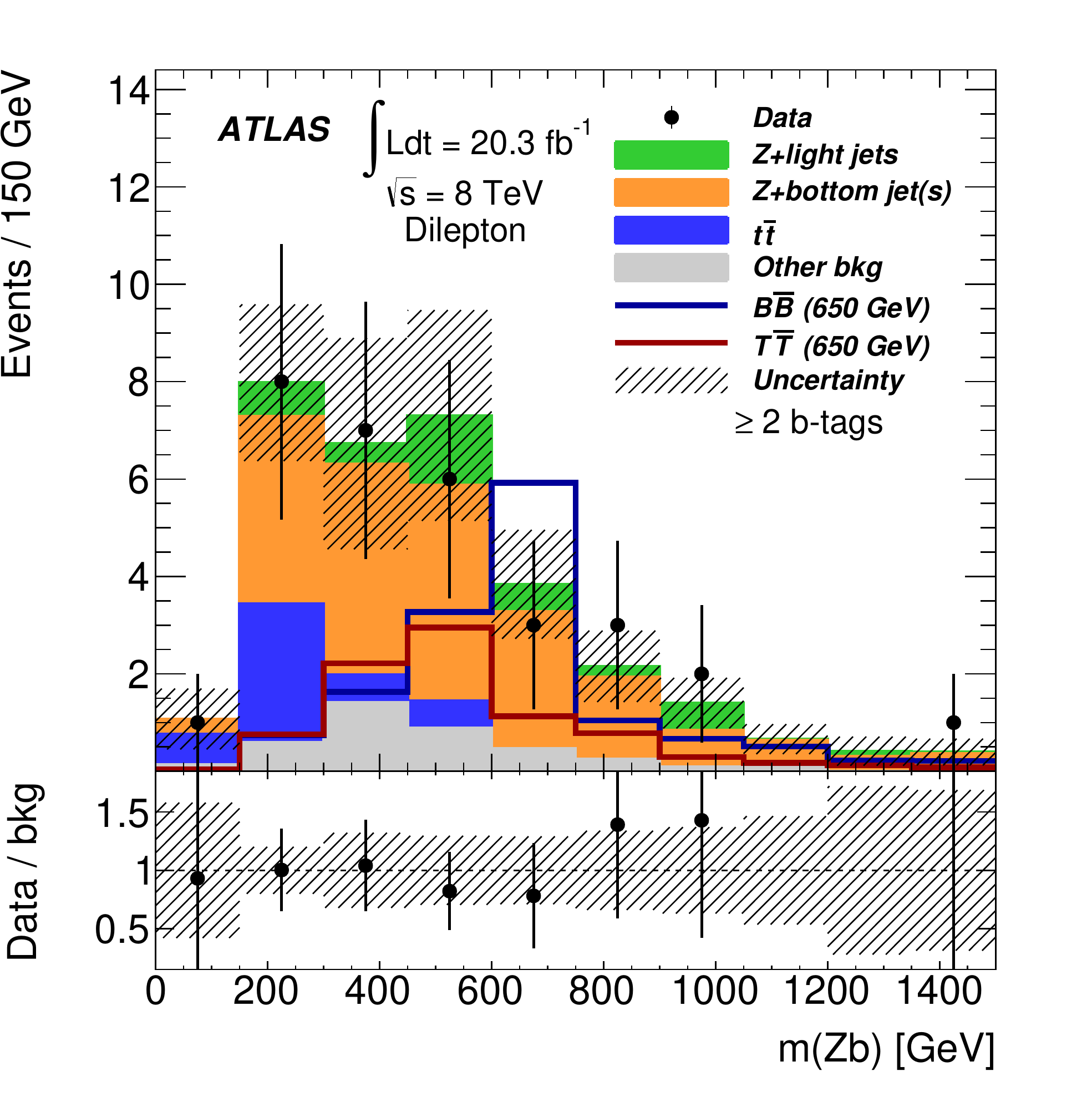}\label{fig:dilepton_mzb_2}}
\caption{The $H_{\rm T}({\rm jets})$ distribution after requiring $p_{\rm T}(Z)>150$~GeV in dilepton channel events with (a) $N_{\rm tag}=1$, or (b) $N_{\rm tag} \geq 2$.  The final $m(Zb)$ distribution after requiring \mbox{$p_{\rm T}(Z)>150$~GeV} and  $H_{\rm T}({\rm jets})>600$~GeV in events with (c) $N_{\rm tag}=1$, or (d) $N_{\rm tag} \geq 2$.}
\label{fig:dilepton_htmzb}
\end{figure*}

Figure~\ref{fig:dilepton_zmasspt}(c) shows the $Z$ boson candidate transverse momentum distribution in events with $N_{\rm tag}=1$, again using {\sc sherpa} to model the $Z+{\rm jets}$ processes.  The expected background shows a trend to increasingly overestimate the data with increasing $p_{\rm T}(Z)$.  This bias would result in a $14\%$ overestimate of the number of $N_{\rm tag}=1$ events passing the $p_{\rm T}(Z)>150$~GeV requirement, compared with the $8\%$ experimental uncertainty.  The trend is likewise observed in the $N_{\rm tag}=0$ control region, and also to a similar degree when using the {\sc alpgen} samples.  In order to mitigate this bias, a $Z+{\rm jets}$ reweighting function is derived by fitting a third-degree polynomial to the residuals defined by $w^{i} \equiv [(N^{\rm data}-N^{\rm pred}_{\rm non~Z+jets})/N^{\rm pred}_{\rm Z+jets}]^{i}$, where $N^{\rm data}$, $N^{\rm pred}_{\rm non~Z+jets}$, and $N^{\rm pred}_{\rm Z+jets}$, denote the number of data, predicted non $Z+{\rm jets}$ background, and predicted $Z+{\rm jets}$ background events, respectively, in the $i^{\rm th}$ bin of the $p_{\rm T}(Z)$ distribution shown in figure~\ref{fig:dilepton_zmasspt}(c).  The degree of the polynomial is chosen to accurately fit the trend while avoiding higher-order terms that could fit statistical fluctuations.  The fit is also performed separately in the dielectron and dimuon channels, and consistent results are obtained.  Table~\ref{table:yields_dilepton_1} presents the predicted yields in the control region with and without this correction applied.  Figure~\ref{fig:dilepton_zmasspt}(d) shows the $p_{\rm T}(Z)$ distribution in the $N_{\rm tag}\geq2$ signal region after the correction has been applied.  The correction results in a $9\%$ ($7\%$) decrease in the predicted number of events satisfying $p_{\rm T}(Z)>150$~GeV when using {\sc sherpa} ({\sc alpgen}).

Figures~\ref{fig:dilepton_htmzb}(a,b) present the $H_{\rm T}({\rm jets})$ distributions in the $N_{\rm tag}=1$ and $N_{\rm tag}\geq2$ categories, respectively, after applying the $p_{\rm T}(Z)$ spectrum correction and requiring $p_{\rm T}(Z)>150$~GeV.  The distributions are well modeled, and the final $H_{\rm T}({\rm jets})>600$~GeV requirement for testing the pair-production hypotheses is made.  Figures~\ref{fig:dilepton_htmzb}(c,d) present the resulting $m(Zb)$ distributions.  The final predicted background yields using {\sc sherpa} are listed in table~\ref{table:yields_dilepton_1} and table~\ref{table:yields_dilepton_2}, and are consistent with predictions using {\sc alpgen} within the $10\%$ statistical uncertainty on the latter.  The tables also present the predicted signal yields for the pair-production of $SU(2)$ singlet $B$ and $T$ quarks with a mass of 650~GeV.

\subsection{Trilepton channel analysis targeting the pair-production hypotheses}
\label{subsec:trileptonpair}

\begin{figure*}[tbp]
\centering
\subfigure[]{\includegraphics[width=0.49\textwidth]{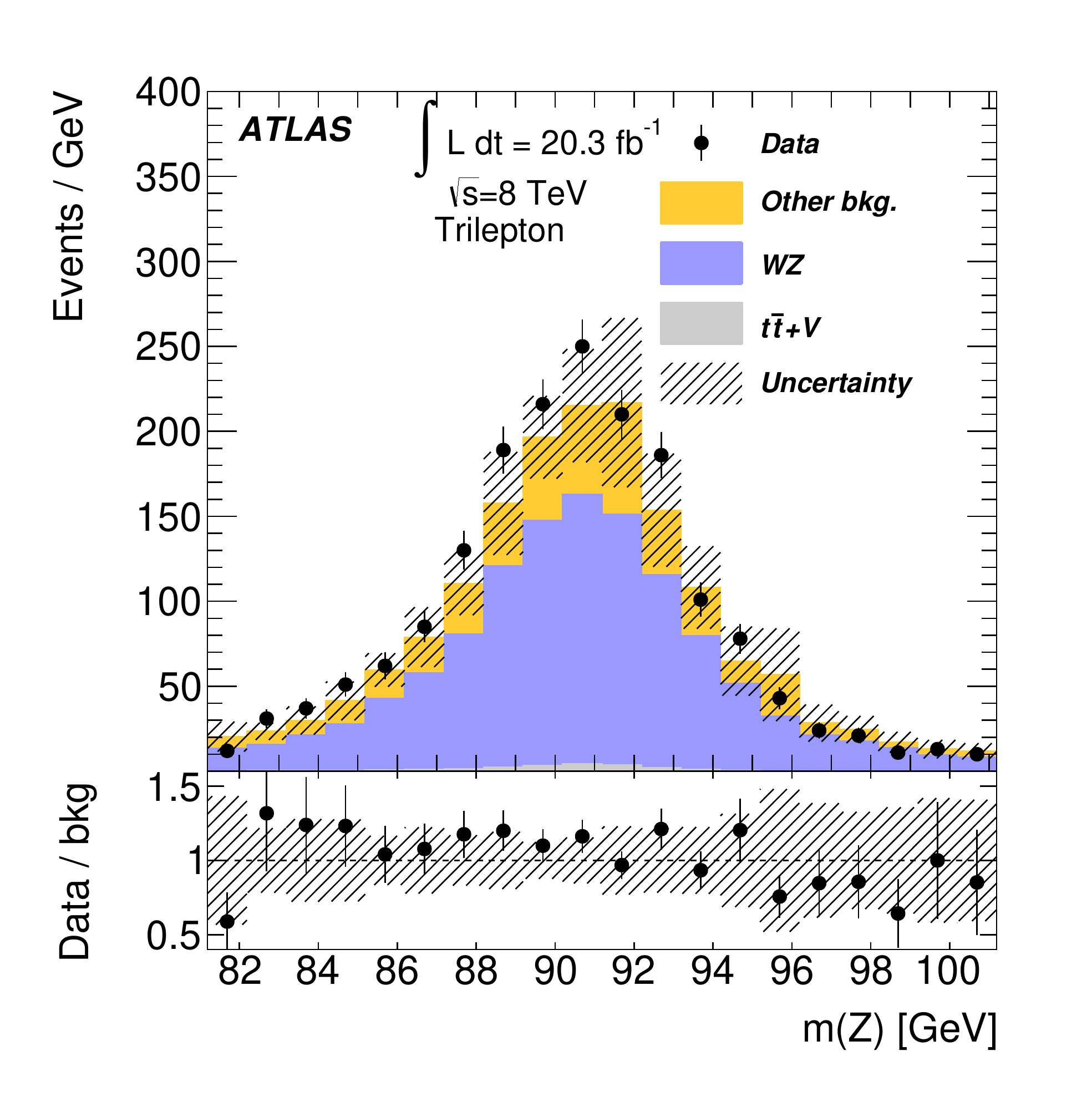}\label{fig:trilepton_1}}
\subfigure[]{\includegraphics[width=0.49\textwidth]{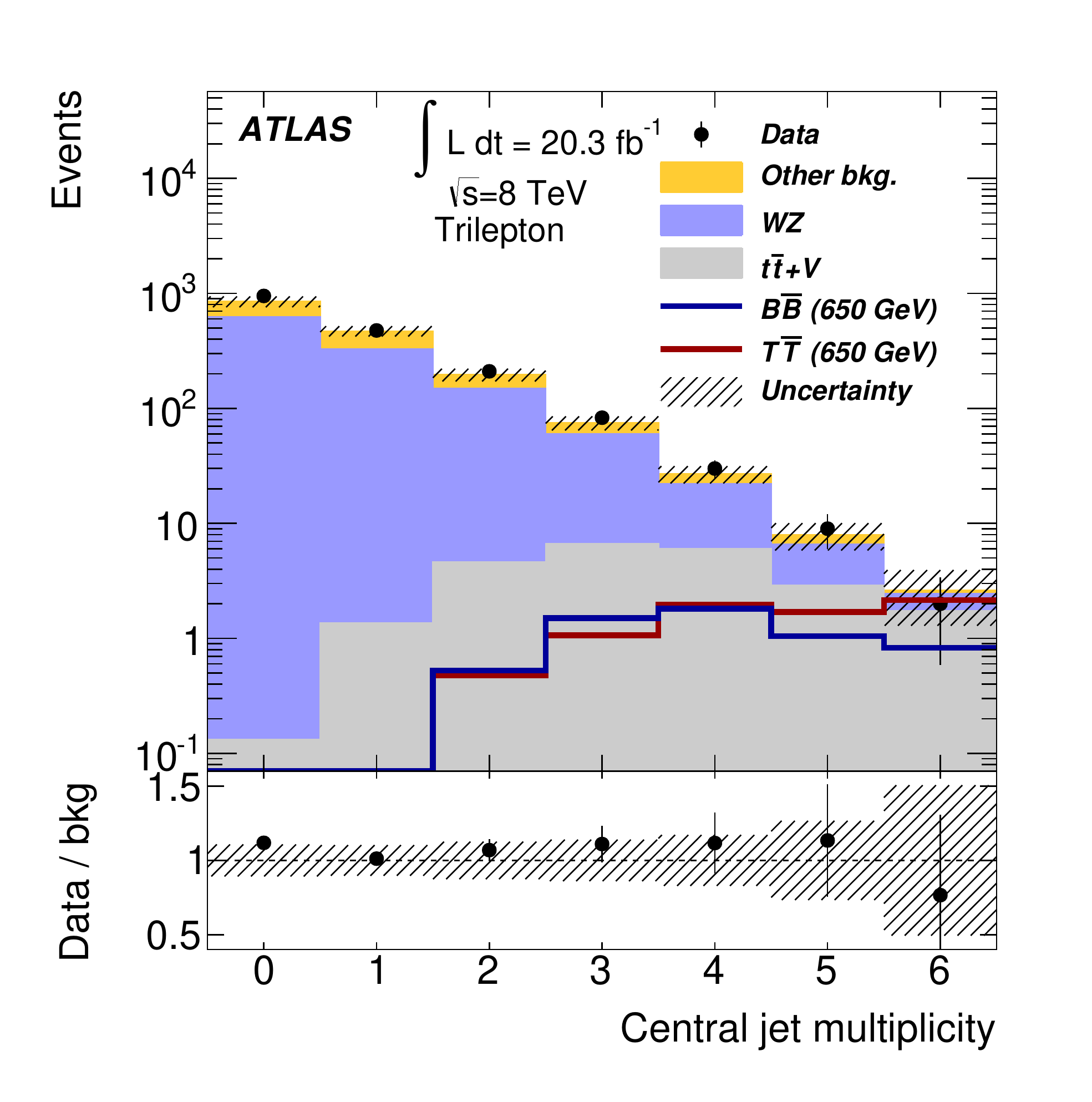}\label{fig:trilepton_2}}
\subfigure[]{\includegraphics[width=0.49\textwidth]{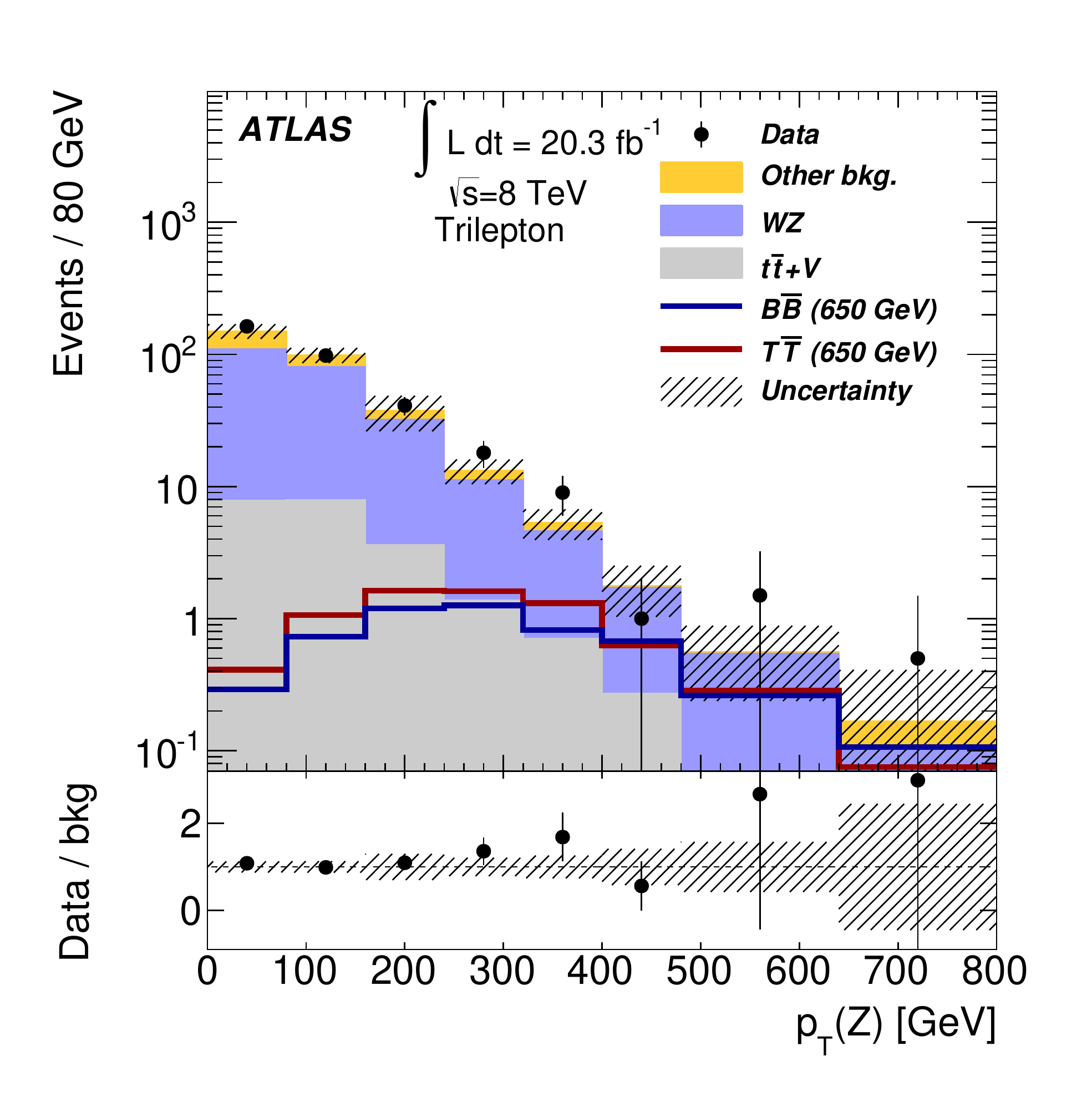}\label{fig:trilepton_3}}
\subfigure[]{\includegraphics[width=0.49\textwidth]{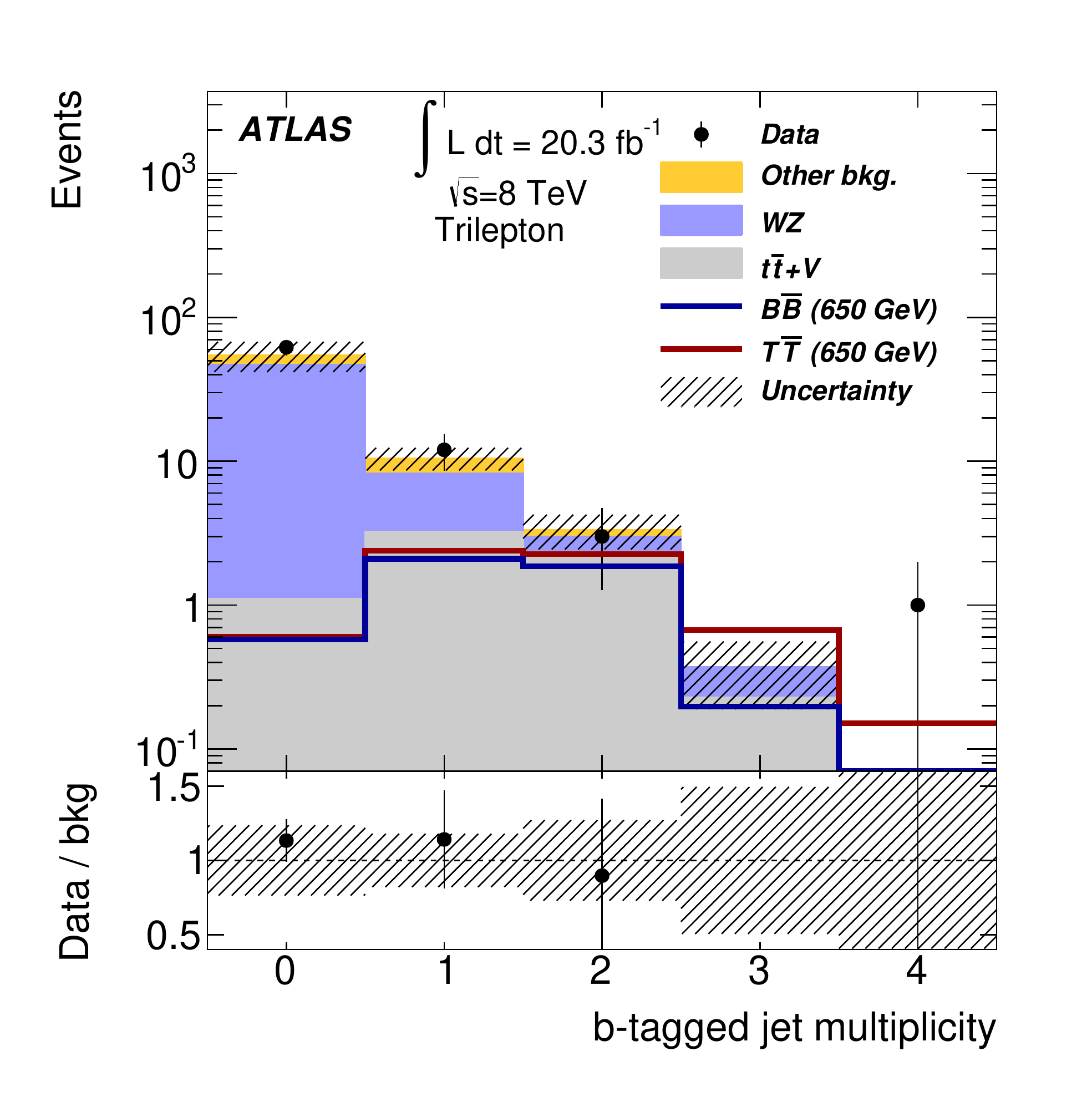}\label{fig:trilepton_4}}
\caption{The distributions of the $Z$ boson candidate mass (a), $m(Z)$, and central-jet multiplicity (b), in trilepton channel events.  The distribution of the $Z$ candidate transverse momentum (c), $p_{\rm T}(Z)$, after requiring $\geq2$ central jets.  The $b$-tagged jet multiplicity distribution (d) after requiring $\geq2$ central jets and $p_{\rm T}(Z)>150$~GeV.}
\label{fig:trilepton1}
\end{figure*}

The trilepton analysis selects events with a $Z$ boson candidate and a third isolated lepton, yielding a total of 1760 events in data.  The $Z$ boson candidate is reconstructed in the $ee$ ($\mu\mu$)  channel in 760 (1000) of these events, and the third lepton is an electron (muon) in 768 (992) of these events.  Figure~\ref{fig:trilepton1}(a) presents the $Z$ candidate mass distribution after the inclusive trilepton channel selection.  Events from $WZ$ processes constitute approximately 70\% of the predicted background.  The leading contributions to the remaining background are predicted to arise from $ZZ$ processes, with smaller contributions from $Z+{\rm jets}$, $t\bar{t}$, and $t\bar{t}+V$ processes.  Figure~\ref{fig:trilepton1}(b) presents the central-jet multiplicity, also after the inclusive trilepton channel selection.  Events with at least two central jets are considered further, and figure~\ref{fig:trilepton1}(c) shows the $Z$ candidate transverse momentum distribution after this requirement is made.  The data are well modeled by the background prediction, and the subset of events with $p_{\rm T}(Z)>150$~GeV are then selected.  The $b$-tagged jet multiplicity distribution is shown in figure~\ref{fig:trilepton1}(d) following the $p_{\rm T}(Z)$ requirement.  Events without a $b$-tagged jet are predicted to arise mostly from $WZ$ processes, while a similar number of $WZ$ and $t\bar{t}+V$ events are predicted to populate the background in events with at least one $b$-tagged jet.  At least one $b$-tagged jet is predicted to be present in a high fraction of pair-production signal events.

\begin{figure*}[tbp]
\centering
\subfigure[]{\includegraphics[width=0.49\textwidth]{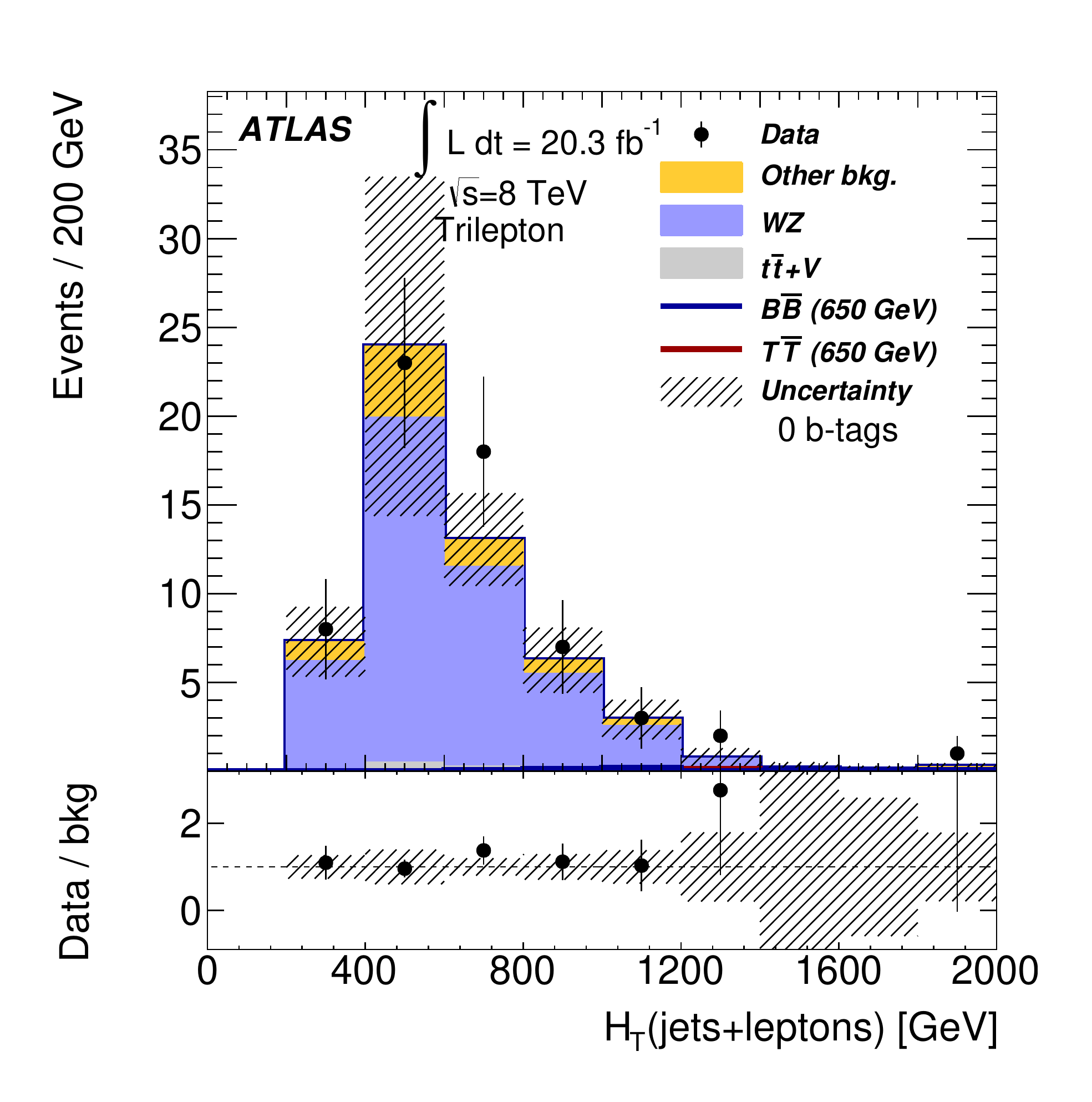}\label{fig:trilepton_5}}
\subfigure[]{\includegraphics[width=0.49\textwidth]{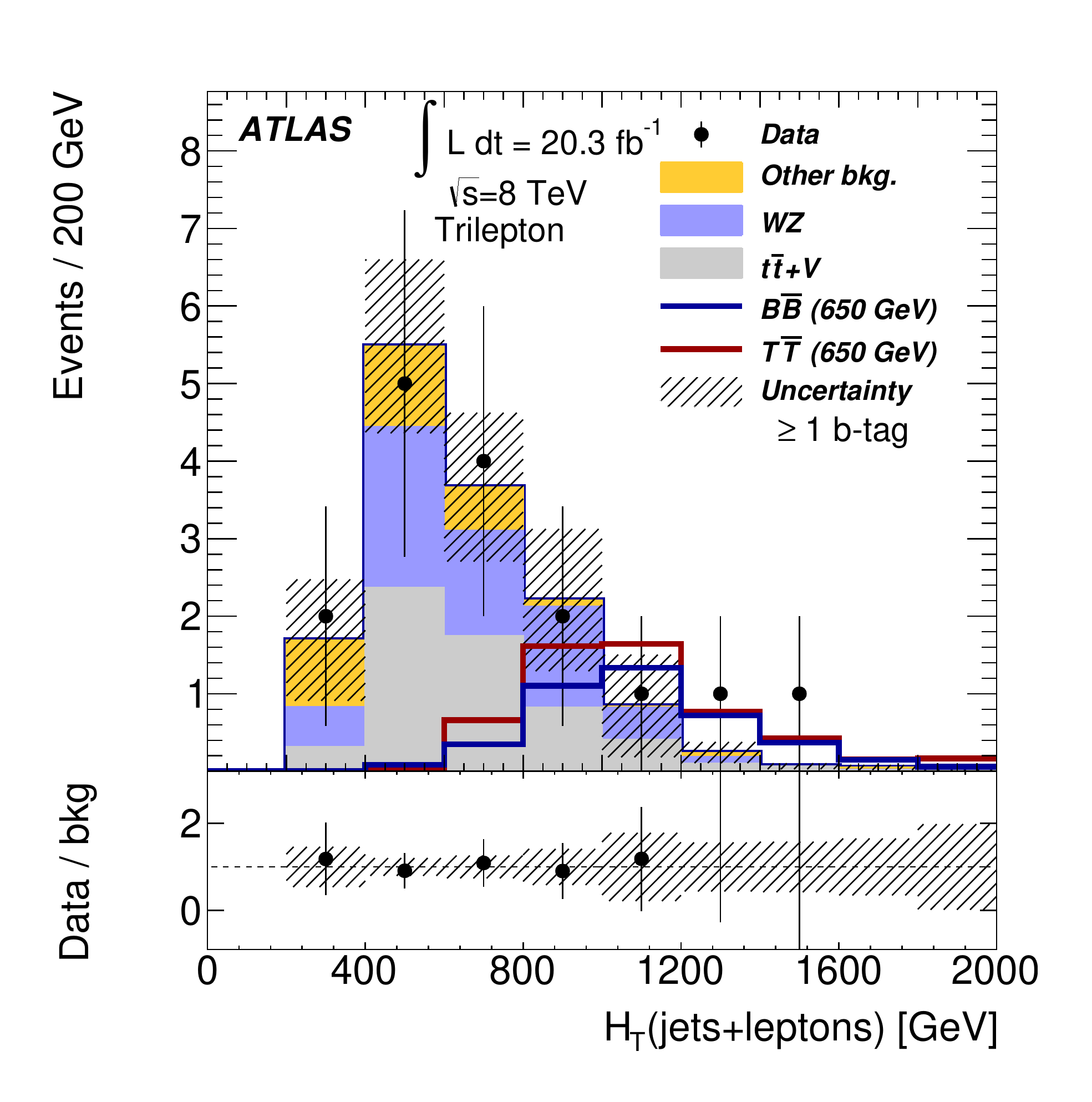}\label{fig:trilepton_6}}
\caption{The $H_{\rm T}({\rm jets+leptons})$ distribution in trilepton channels events with $\geq2$ central jets, $p_{\rm T}(Z)>150$~GeV, and (a) $N_{\rm tag}=0$, or (b) $N_{\rm tag}\geq1$.}
\label{fig:trilepton2}
\end{figure*}

Figure~\ref{fig:trilepton2}(a) shows the $H_{\rm T}({\rm jets+leptons})$ variable in the $N_{\rm tag}=0$ control region.  The distribution is well modeled by the background prediction.  Figure~\ref{fig:trilepton2}(b) presents the $H_{\rm T}({\rm jets+leptons})$ discriminant in the signal region consisting of events with $N_{\rm tag} \geq 1$.  Table~\ref{table:yields_trilepton1} presents the observed and predicted yields at each stage of the trilepton channel event selection.  In addition, the table lists the predicted signal yields for the pair production of $SU(2)$ singlet $B$ and $T$ quarks with a mass of 650~GeV.

\begin{table}[tbp]
\begin{center}
\begin{tabular}{ccccc}\hline\hline
                         	& Trilepton ch. 		& $\geq2$~central jets  	& $p_{\rm T} (Z) > 150$ GeV & $N_{\rm tag}\geq1$  \\ \hline
$WZ$ & $1170\pm130$ & $219\pm 32$ & $51.5\pm8.9$ & $5.8\pm1.4$ \\ 
$t\bar{t}+X$ & $23.5\pm6.7$ & $22.0\pm6.3 $ & $7.0\pm 2.1$ & $5.8\pm1.8$ \\ 
Other SM & $435\pm50$ & $67\pm13$ & $10.4\pm9.2$ & $2.6\pm1.3$ \\ 
\hline 
Total SM & $1630\pm170$ & $309\pm 39$ & $69\pm14$ & $14.3\pm 2.6$ \\ 
\hline 
\hline 
Data & 1760 & 334 & 78 & 16 \\ 
\hline 
\hline 
$B\bar{B}$ ($m_B$=650 GeV) & $5.8\pm 0.4$ & $5.7\pm 0.4$ & $4.99\pm 0.33$ & $4.17\pm 0.30$ \\ 
$T\bar{T}$ ($m_T$=650 GeV) & $7.4\pm 0.5$ & $7.4\pm 0.5$ & $6.7\pm 0.5$ & $5.5\pm 0.4$  	\\\hline\hline
\end{tabular}
\end{center}
\caption{Predicted and observed number of events in the trilepton channel, starting on the left with the selection stage of a $Z$ boson candidate plus a third isolated lepton,  followed by the yields after the additional requirements outlined in the text.  The final column represents the signal region for testing the pair production hypotheses.  Reference $B\bar{B}$ and $T\bar{T}$ signal yields are provided for $m_{B/T}=650$~GeV and $SU(2)$ singlet branching ratios.  The uncertainties on the predicted yields include statistical and systematic sources.}
\label{table:yields_trilepton1}          
\end{table} 

\subsection{Modified selection criteria to target the single-production hypotheses}
\label{subsec:single}

A characteristic feature of the signal events that produce a single heavy quark via the electroweak interaction is the presence of an energetic forward light-flavor jet that is produced in association.  Such a jet is required when testing the single-production hypotheses, in addition to the requirements discussed in sections~\ref{subsec:dileptonpair} and \ref{subsec:trileptonpair} in the context of the pair-production hypotheses.  In the dilepton channel, the $H_{\rm T}({\rm jets})$ requirement is removed, as it is not efficient for the single-production signals.

\begin{figure*}[tbp]
\centering
\subfigure[]{\includegraphics[width=0.48\textwidth]{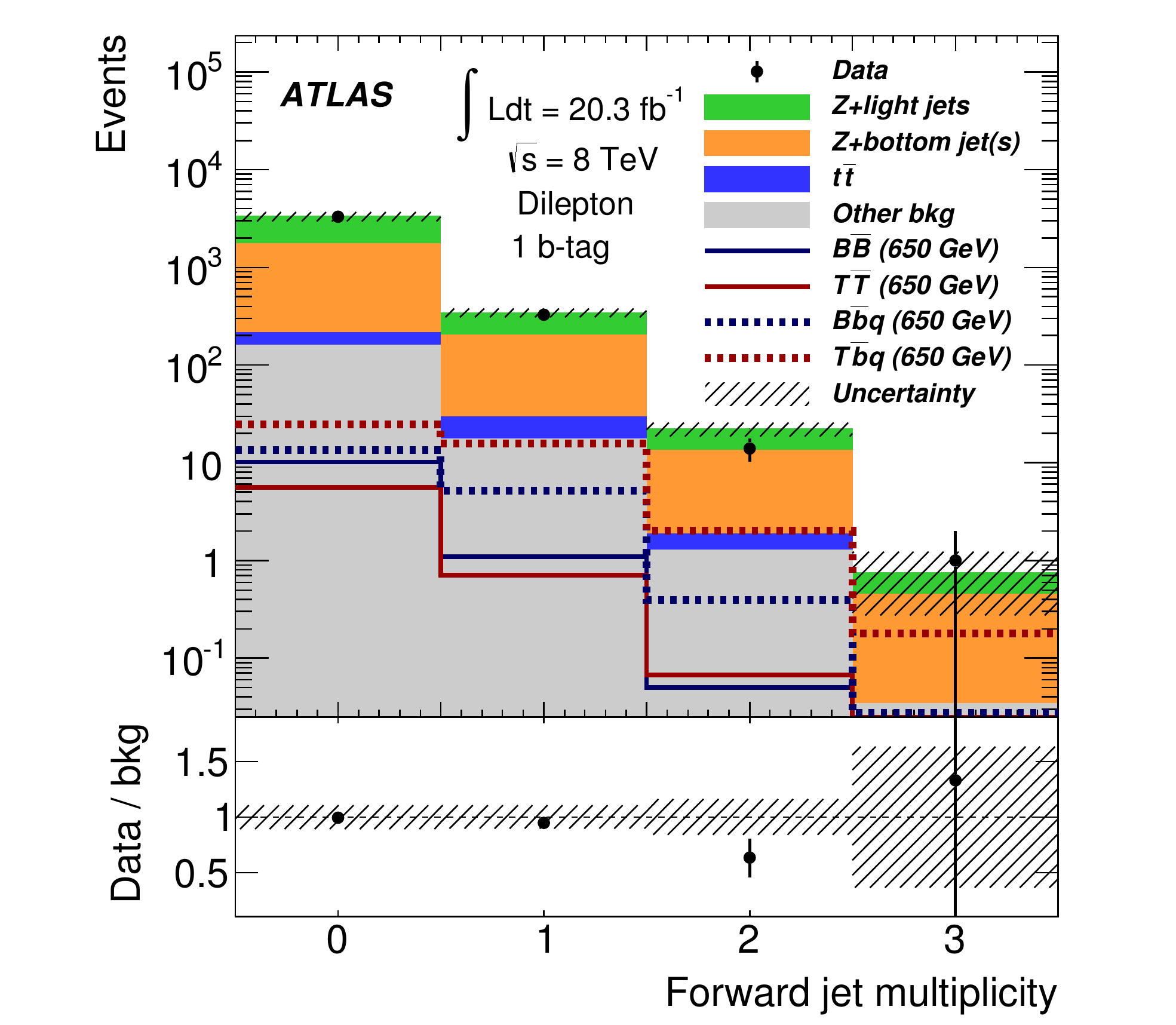}\label{fig:dilepsingle_nfjet1}}
\subfigure[]{\includegraphics[width=0.48\textwidth]{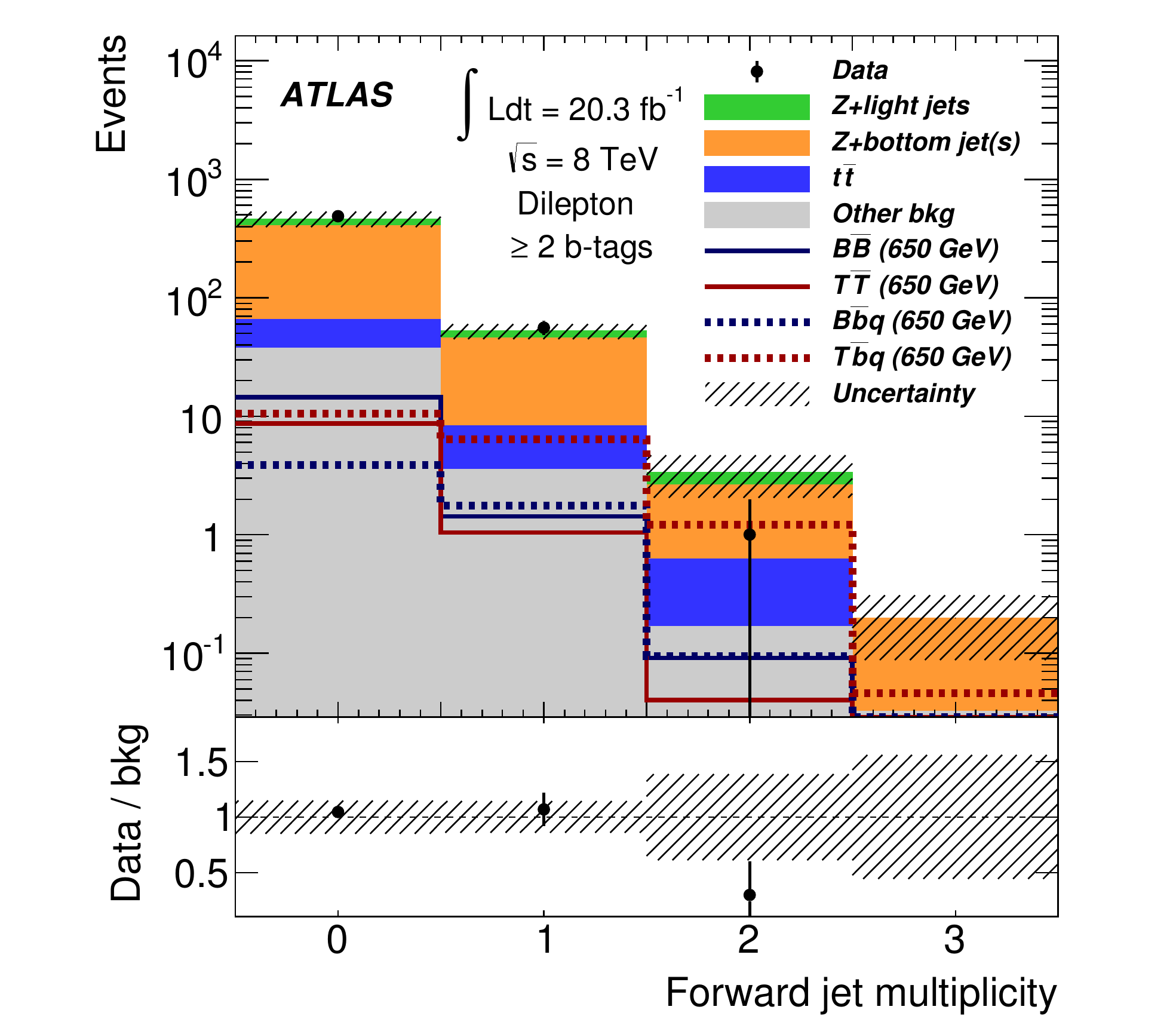}\label{fig:dilepsingle_nfjet2}}
\subfigure[]{\includegraphics[width=0.48\textwidth]{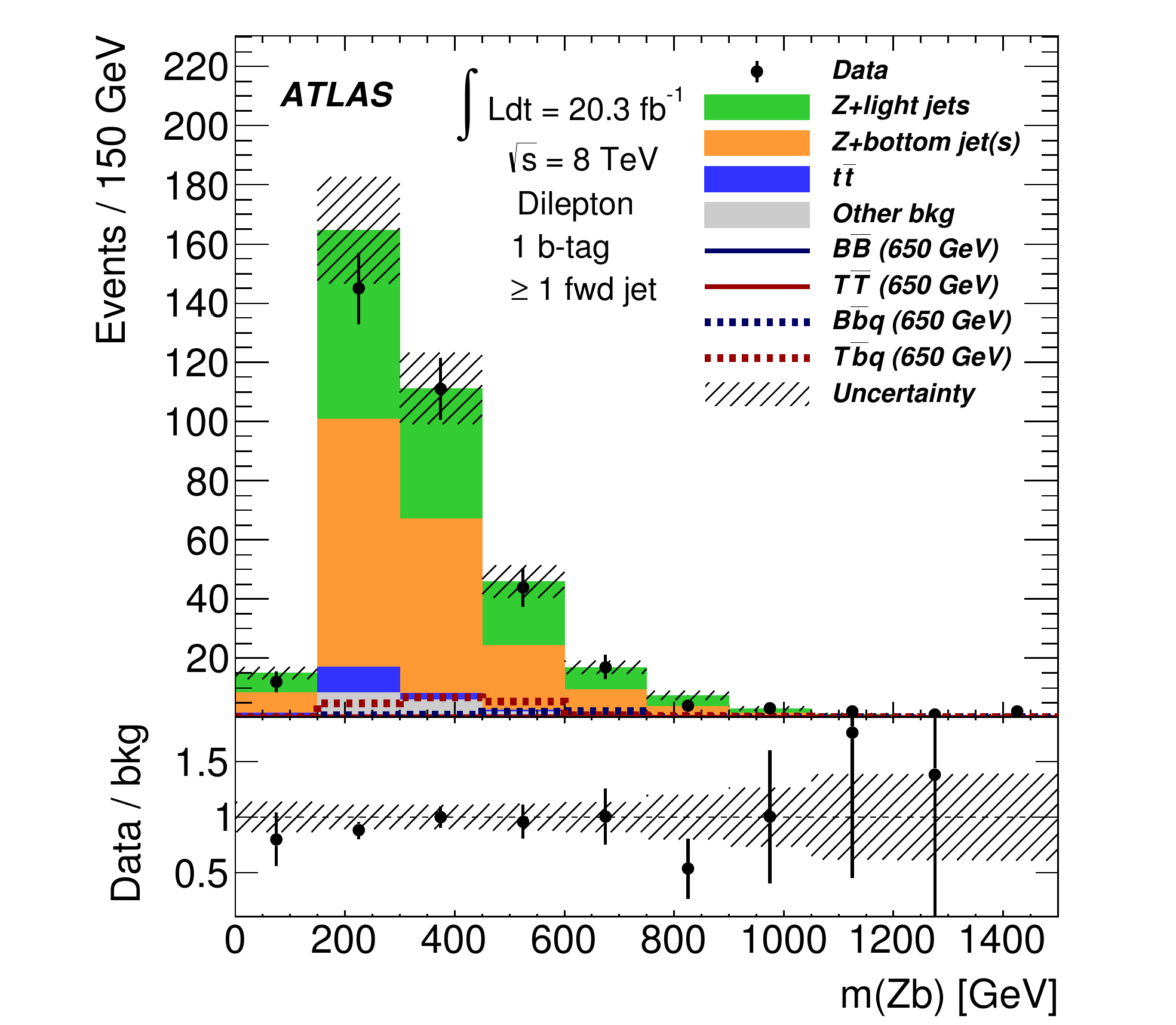}\label{fig:dilepsingle_mzb1}}
\subfigure[]{\includegraphics[width=0.48\textwidth]{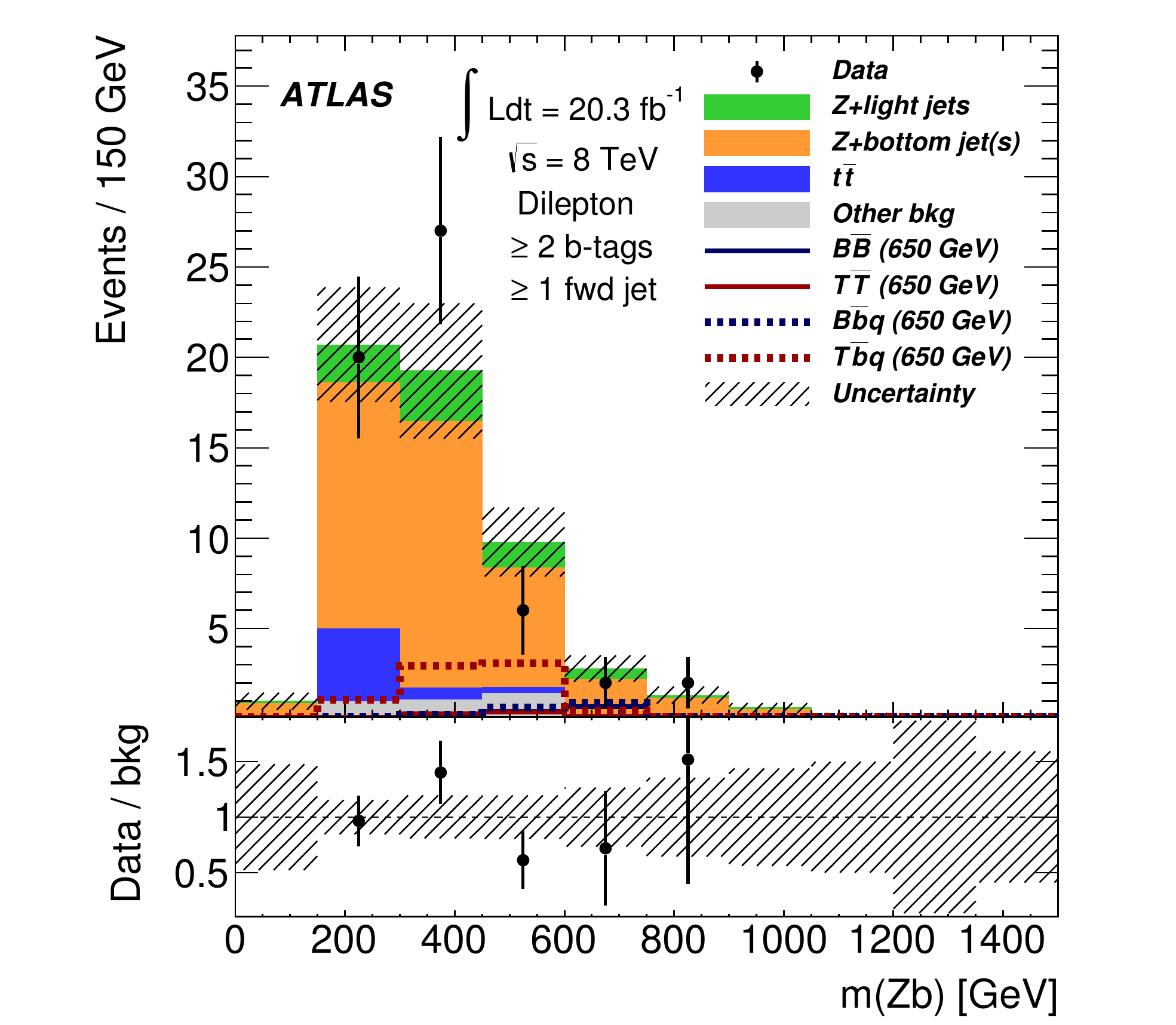}\label{fig:dilepsingle_mzb2}}
\caption{The forward-jet multiplicity distribution in dilepton channel events with $\geq2$ central jets, satisfying $p_{\rm T}(Z)>150$~GeV, and (a) $N_{\rm tag}=1$, or (b) $N_{\rm tag} \geq2$.  The $m(Zb)$ distribution following the final requirement of at least one forward jet in events with (c) $N_{\rm tag}=1$ or (d) $N_{\rm tag}\geq2$.  The predicted $T\bar{b}q$ signal assumes a mixing parameter value of $\lambda_{T}=2$, while the predicted $B\bar{b}q$ signal assumes a mixing parameter value of $X_{bB}=0.5$.}
\label{fig:dileptonsingleprod}
\end{figure*}

Figures~\ref{fig:dileptonsingleprod}(a,b) display the forward-jet multiplicity distribution in dilepton channel events after requiring at least two central jets, $p_{\rm T}(Z)>150$~GeV, and $N_{\rm tag}=1$ or $N_{\rm tag}\geq2$, respectively.  This is the same selection stage as that shown in the $H_{\rm T}({\rm jets})$ distributions of figures~\ref{fig:dilepton_htmzb}(c,d).  The predicted background is reduced by over an order of magnitude, and a large fraction of the single-production signal maintained, by restricting the sample to those events that contain at least one forward jet.  Figures~\ref{fig:dileptonsingleprod}(c,d) present the final $m(Zb)$ distributions in the control and signal regions, respectively, after applying the forward-jet requirement.

\begin{figure*}[tbp]
\centering
\subfigure[]{\includegraphics[width=0.49\textwidth]{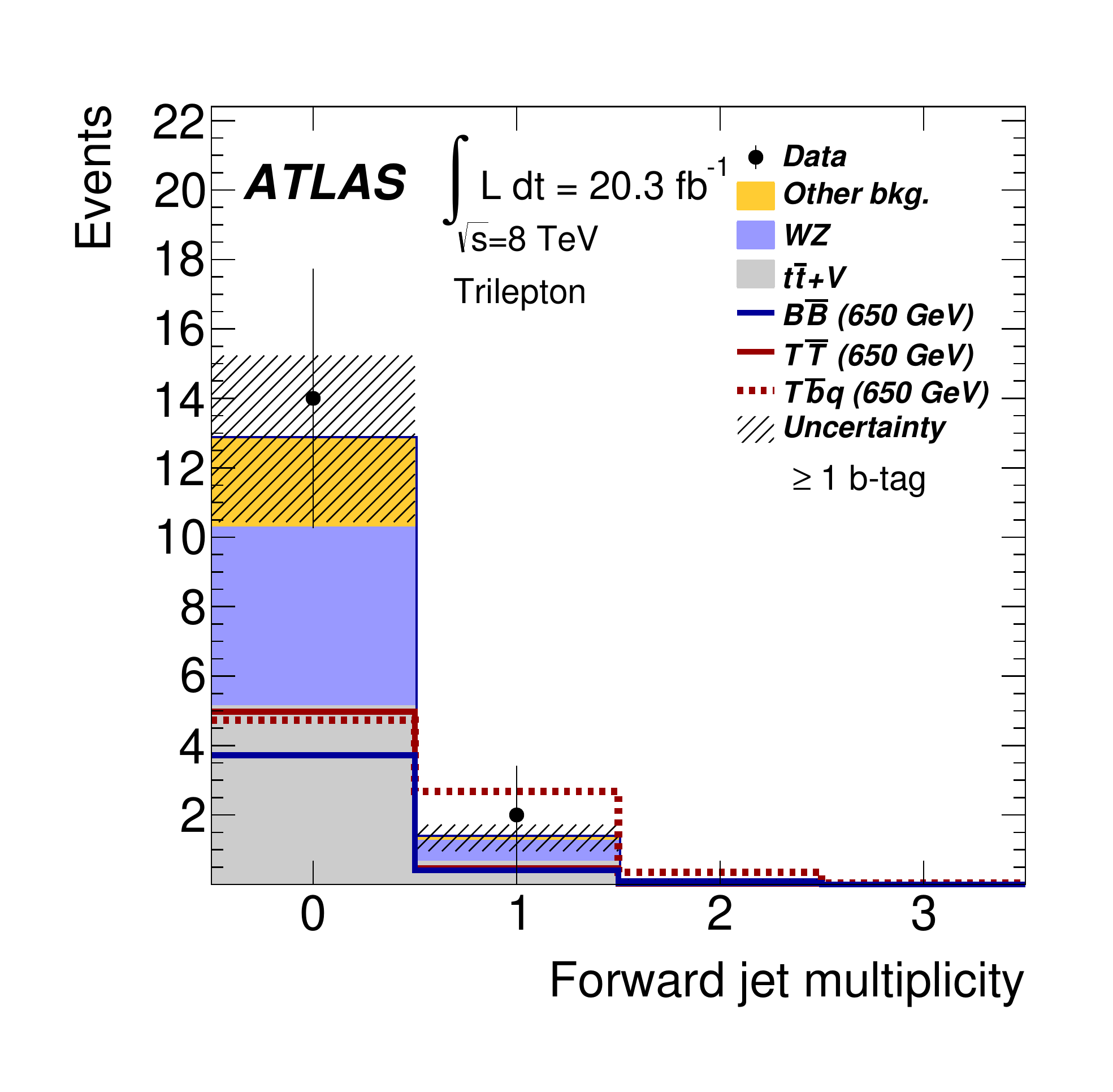}\label{fig:trilepton_7}}
\subfigure[]{\includegraphics[width=0.49\textwidth]{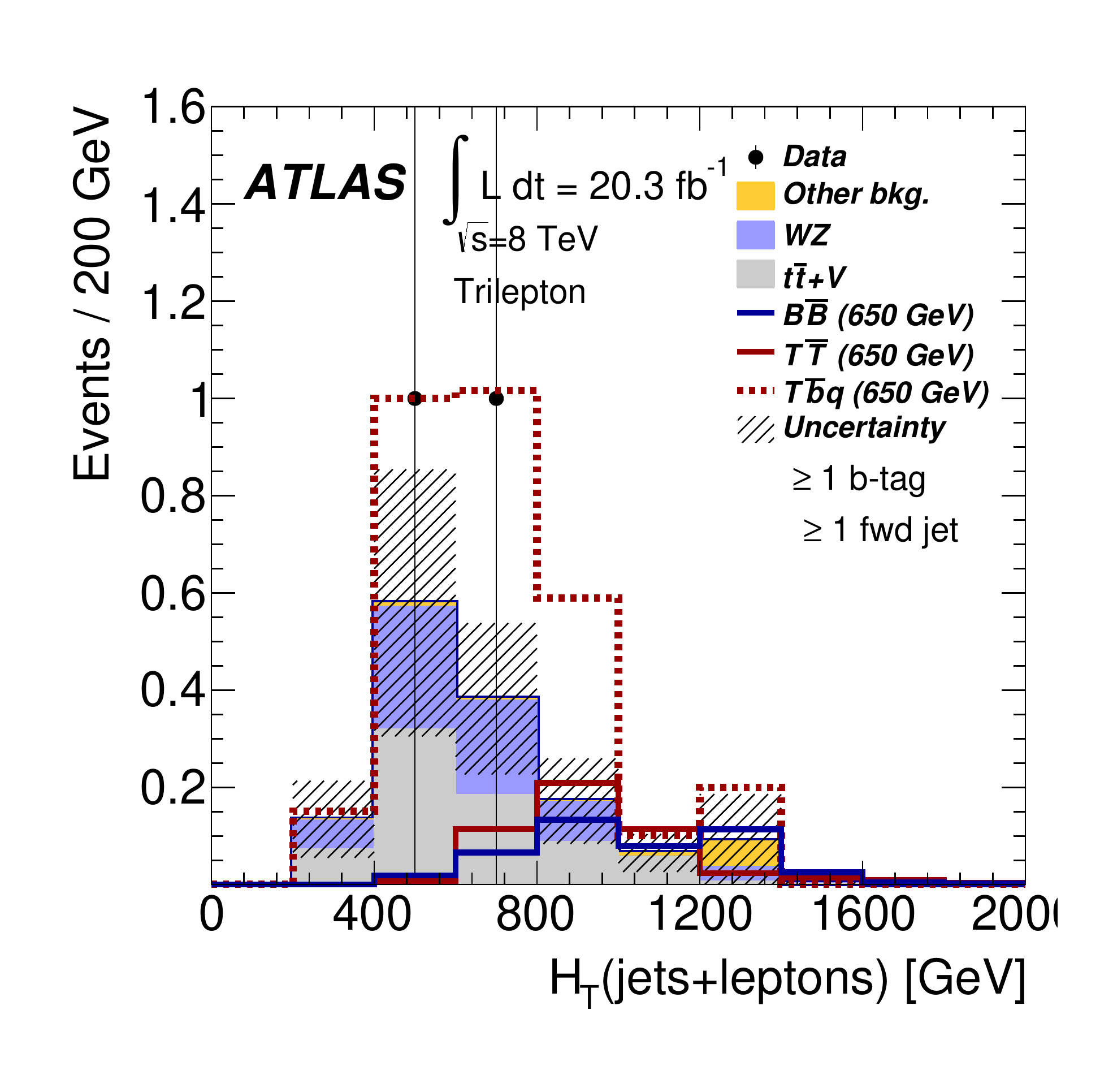}\label{fig:trilepton_8}}
\caption{The forward-jet multiplicity distribution (a) in trilepton channel events with $\geq2$~central jets, satisfying $p_{\rm T}(Z)>150$~GeV, and $N_{\rm tag} \geq 1$.  The $H_{\rm T}({\rm jets+leptons})$ distribution (b) following the requirement of at least one forward jet.  The predicted $T\bar{b}q$ signal assumes a mixing parameter value of $\lambda_{T}=2$.}
\label{fig:trileptonsingleprod}
\end{figure*}

Figure~\ref{fig:trileptonsingleprod}(a) shows the forward-jet multiplicity in the trilepton channel after requiring at least two central jets, $p_{\rm T}(Z)>150$~GeV, and $N_{\rm tag}\geq 1$.  These requirements constitute the final selection criteria for testing the pair-production hypotheses in the trilepton channel.  An increase in the sensitivity to the $T\bar{b}q$ process is achieved by restricting the sample to events with at least one forward jet.  Figure~\ref{fig:trileptonsingleprod}(b) shows the final $H_{\rm T}({\rm jets+leptons})$ distribution after the forward-jet requirement is applied.

Table~\ref{table:yields_singleprod} presents the observed data and predicted background events after the final event selection for testing the single-production hypotheses in both the dilepton and trilepton channels.  Predicted signal yields are shown for the $T\bar{b}q$ and $B\bar{b}q$ single-production processes, with reference coupling parameters of $\lambda_{T}=2$ and $X_{bB}=0.5$,\footnote{The maximum possible value of $X_{bB}$ is 0.5 in the case of an $SU(2)$ singlet $B$ quark.  This is a consequence of the relationship between $V_{tB}$ and $X_{bB}$, which can be found in tables 9 and 10 of ref.~\cite{JA_new}.} respectively, as well as the predicted contribution of pair-production signal events in the single-production signal regions.  In each case the heavy quark is an $SU(2)$ singlet with a mass of 650~GeV.

\begin{table}[tbp]
\begin{center}
\begin{tabular}{c|c||c|c}\hline\hline
\multicolumn{2}{c||}{Dilepton channel} & \multicolumn{2}{|c}{Trilepton channel}	\\ \hline
$Z+$light 		& $7.3\pm2.0$ 		& $WZ$  			& $0.62\pm0.27$ \\ 
$Z+$bottom 	& $40\pm10$ 		& $t\bar{t}+V$  		& $0.74\pm0.24$ \\ 
$t\bar{t}$ 	 	& $5.2\pm2.1$ 		&  				&  \\ 
Other SM 	 	& $3.8\pm1.3$ 		& Other SM 		&  $0.07\pm0.10$\\ \hline 
Total SM 	 	& $56\pm12$ 		& Total SM 		&  $1.4\pm0.4$	\\ \hline \hline
Data 	 	& 57 				& Data 		&  2\\ \hline \hline
$B\bar{b}q$ {\small ($m_{B}=650$~GeV, $X_{bB}=0.5$)} 	 	& $1.88\pm0.27$		&  	& \\
$T\bar{b}q$ {\small ($m_{T}=650$~GeV, $\lambda_{T}=2$)} 	 	& $7.7\pm1.0$		&  $T\bar{b}q$ {\small ($m_{T}=650$~GeV, $\lambda_{T}=2$)} 	&  $3.1\pm0.5$\\   \hline \hline
$B\bar{B}$ {\small ($m_{B}=650$~GeV)} & $1.53\pm0.24$ & $B\bar{B}$ {\small ($m_{B}=650$~GeV)} & $0.45\pm0.10$ \\
$T\bar{T}$ {\small ($m_{T}=650$~GeV)} & $1.08\pm0.15$ & $T\bar{T}$ {\small ($m_{T}=650$~GeV)} & $0.50\pm0.10$ \\ \hline \hline
\end{tabular}
\end{center}
\caption{Number of predicted and observed dilepton and trilepton channel events after the final selection for testing the single-production hypotheses, which includes a forward-jet requirement.  The expected yield of $T\bar{b}q$ and $B\bar{b}q$ events is listed for $SU(2)$ singlet $T$ and $B$ quarks with a mass of 650~GeV and for reference mixing parameters.  The predicted contribution of pair-production events in the single-production signal regions is also provided.  The uncertainties on the predicted yields include statistical and systematic sources.}
\label{table:yields_singleprod}          
\end{table}

\section{Systematic uncertainties}

Several sources of systematic uncertainty affect the predicted yield of SM background and signal events after the full selection criteria are applied, as well as the distribution of these events in the discriminating variables, $m(Zb)$ and $H_{\rm T}({\rm jets+leptons})$.  The sources of uncertainty described below are assumed to be uncorrelated.  The impact is evaluated by propagating each uncertainty through the full analysis chain for each signal or background source, and allowing the final predictions to vary accordingly during hypothesis testing.  Tables~\ref{table:dilep_sys} and \ref{table:trilep_sys} list the fractional uncertainty in the normalization of the final signal and background predictions for each category of systematic uncertainty in the dilepton and trilepton pair-production signal regions, respectively.  The characteristics of these categories are explained below.

\paragraph{Luminosity}

The uncertainty on the integrated luminosity is 2.8\%, resulting in a normalization uncertainty for processes estimated with simulated samples.  The uncertainty was derived following the same methodology as that detailed in ref.~\cite{Lumi}.  In addition, since the $Z+{\rm jets}$ background prediction is corrected to account for differences between data and all other backgrounds in a control region, the luminosity uncertainty also indirectly impacts the yield of this background source.

\paragraph{Signal and background cross sections}

Signal and background cross-section uncertainties influence the predicted yield of events from processes estimated with simulated samples.  As explained above in the case of the luminosity uncertainty, the SM background cross-section uncertainties~\cite{toppp1,xsec_DB,xsec_ST,xsec_TTV} also indirectly influence the $Z+{\rm jets}$ background prediction.  While the impact is small in the dilepton channel analysis, uncertainties in the cross sections of background processes constitute the dominant systematic uncertainty in the trilepton channel analysis.  The uncertainty on the $t\bar{t}+V$ processes is conservatively assessed to be $30\%$ using the results of ref.~\cite{xsec_TTV}.  The uncertainty on the $WZ+$~jets background is taken to be $50\% \times H_{\rm T}({\rm jets+leptons})/~1~{\rm TeV}$ following the methods described in ref.~\cite{dibosonerr}.

\begin{table}[tbp]
\begin{center}
\begin{tabular}{c|cccccc}
\hline\hline
\multicolumn{7}{c}{Fractional uncertainties (\%): dilepton channel}                                                \\\hline
                           		&        $Z$+jets &       $t\bar t$ &      Other bkg.	&        Total bkg.   &  $B\bar{B}$& $T\bar{T}$ \\\hline
               Luminosity  	&             1.4 	&             2.8 	&             2.8 	&               0.3   &        2.8 &             2.8		 \\\hline
            Cross section &             5.5 	&             6.4 	&            29 	&               0.7   &          - &               - 		\\\hline
   Jet reconstruction  	&              13 	&            10 	&            14 	&              11   &        	2.0 &             2.1 	\\\hline
              $b$-tagging &             9.1 	&            13 	&             9.9 	&               5.7   &        7.2 &             5.9 	\\\hline
$e$ reconstruction  &             2.9 	&            16 	&             5.9 	&               4.6   &        2.5 &             1.5 	\\\hline
$\mu$ reconstruction  &             3.8 	&             7.8 	&             7.2 	&               4.2   &        3.2 &             1.3 	\\\hline
$Z$+jets $p_{\rm T}(Z)$ correction&     9.0 	&               - 	&               - 	&               6.5   &          - &               - 	\\\hline
        $Z$+jets rate correction  &             6.9 	&               - 	&               - 	&               5.0   &          - &               - 	\\\hline\
MC statistics &             5.0 &            25 &            12 &             5.4       &      2.4 &             2.9 \\\hline \hline
\end{tabular}
\end{center}
\caption{The fractional uncertainties (\%) in the yields of signal and background events after the final dilepton channel selection for testing the pair production hypotheses.  The signals correspond to $SU(2)$ singlet $T$ and $B$ quarks with a mass of 650~GeV.  The uncertainties are grouped into categories that are explained in more detail in the text.}
\label{table:dilep_sys}
\end{table}

\begin{table}[tbp]
\begin{center}
\begin{tabular}{c|ccccccc}
\hline \hline
\multicolumn{7}{c}{Fractional uncertainties (\%): trilepton channel}                                                \\\hline
 			& $WZ$ 	& $t\bar{t}+V$ 	& Other bkg. 	&  Total bkg. 	& $B\bar{B}$ 	& $T\bar{T}$   \\ \hline 
Luminosity 	& 2.8 	& 2.8 		& 2.8 		& 2.8 		& 2.8 		& 2.8 		 \\ \hline
Cross section 	& 17 		& 30 			& 8.9 		& 21 			& - 			& - 		 \\ \hline
Jet reconstruction 		& 5.4 	& 1.2 		& 8.1 		& 3.1 		& 4.0 		& 1.8		 \\  \hline  
$b$-tagging 	& 13 		& 3.6 		& 13 			& 6.7 		& 5.6 		& 5.5 	 \\ \hline
$e$ reconstruction 		& 9.3 	& 3.9 		& 37 			& 11 			& 5.9 		& 12 		 \\ \hline
$\mu$ reconstruction 	& 14 		& 3.9 		& 18 			& 4.2 		& 6.2 		& 5.7 	 \\ \hline 
 MC statistics & 11 & 3.1 & 27 & 6.6 & 4.8 & 8.3 \\ \hline\hline
\end{tabular}
\end{center}
\caption{The fractional uncertainties (\%) in the yields of signal and background events after the final trilepton channel selection for testing the pair production hypotheses.  The signals correspond to $SU(2)$ singlet $T$ and $B$ quarks with a mass of 650~GeV.  The uncertainties are grouped into categories that are explained in more detail in the text.}
\label{table:trilep_sys}
\end{table}

\paragraph{Jet reconstruction}

The jet energy scale~\cite{JES} was determined using information from test-beam data, LHC collision data, and simulation.  The corresponding uncertainty varies between 0.8\% and 6\%, depending on the $p_{\rm T}$ and $\eta$ of selected jets in this analysis.  Additional uncertainties associated with other $pp$ interactions in the same bunch crossing (pile-up) can be as large as 5\%{}.  Likewise, an additional uncertainty of up to 2.5\%{}, depending on the $p_{\rm T}$ of the jet, is applied for $b$-tagged jets.  The energy resolution of jets was measured in dijet events and agrees with predictions from simulations within 10\%{}, and the corresponding uncertainty is evaluated by smearing the jet energy accordingly.  The jet reconstruction efficiency was estimated using minimum-bias and dijet events.  The inefficiency was found to be at most 2.7\% for low-$p_{\rm T}$ and at the per mil level for high-$p_{\rm T}$ jets.  This uncertainty is taken into account by randomly removing jets in simulated events.  A requirement is made on the tracks associated with central jets in order to reduce the contribution of jets that arise from pile-up.  The performance of this requirement was compared in data and simulation for \mbox{$Z(\to \ell^+\ell^-)+1$}-jet events, selecting separately events enriched in hard-scatter jets and events enriched in pile-up jets.  Simulation correction factors were determined separately for both types. For hard-scatter jets they decrease from $\sim 1.03$ at $p_{\rm T}=25$~GeV to $\sim 1.01$ at $p_{\rm T}>50$~GeV, while for pile-up jets they are consistent with unity.

\paragraph{$\boldmath{b}$-tagging}

Dedicated performance studies of the $b$-tagging algorithm have been performed and calibration factors determined~\cite{btag1,BTAG_NEW}.  Efficiencies for tagging $b$-jets ($c$-jets) in simulation are corrected by $p_{\rm T}$-dependent factors in the range 0.9--1.0 (0.9--1.1), whereas the light-jet efficiency is corrected by $p_{\rm T}$- and $\eta$-dependent factors in the range 1.2--1.5.  The uncertainties in these corrections are between 2--6\% for $b$-jets, 10--15\%{} for $c$-jets, and 20--40\% for light jets.

\paragraph{Lepton reconstruction and trigger}

The uncertainties on the identification and reconstruction efficiency of electrons and muons, as well as the efficiency of the single-lepton triggers used in the analysis, affect the nominal scale factors used to correct differences observed between data and simulation.  When combined, these lepton efficiency uncertainties contribute to an uncertainty on the final signal and background estimates at the level of 5\%.  Data events with leptonic decays of the $Z$ boson were used to measure the lepton momentum scale and resolution, and simulation correction factors with associated uncertainties were derived~\cite{EleRef,MuonRef1,MuonRef2}.  The effect of momentum scale uncertainties were evaluated by repeating the event selection with the electron and muon momentum varied according to the corresponding uncertainties.  The impact of the momentum resolution uncertainty was evaluated by smearing the lepton momentum in simulation accordingly.  The lepton momentum uncertainties contribute to an uncertainty on the final signal and background estimates at the level of 1\%.

\paragraph{Systematic uncertainties associated with data-driven $\boldmath{Z}+{\boldmath \rm jets}$ corrections}

The $Z+{\rm jets}$ scaling factor and the $p_{\rm T}(Z)$ shape correction are derived in control regions and applied to the signal region.  The rate correction was derived in both the \mbox{$p_{\rm T}(Z)<100$~GeV} and the \mbox{$50<p_{\rm T}(Z)<150$~GeV} regions and the difference between the resulting predictions was used to assess an uncertainty.  Similarly, the $p_{\rm T}(Z)$ spectrum correction was derived in both the $N_{\rm tag}=0$ and $N_{\rm tag}=1$ control regions, and the difference when applied to the $N_{\rm tag}\geq2$ signal region used to assign an uncertainty.  Dedicated {\sc sherpa} $Z+{\rm jets}$ samples were also produced with varied renormalization, factorization, and matching scales, and used to cross-check the uncertainties derived from the data-driven methods.

\section{Results}

A binned Poisson likelihood test is performed on the distributions of the final discriminating variables to assess the compatibility of the observed data with the background-only and signal-plus-background hypotheses.  The test employs a log-likelihood ratio function, $-2 \ln (L_{\rm s + b} / L_{\rm b} )$, where $L_{\rm s + b}$ ($L_{\rm b}$) is the Poisson probability to observe data under the signal-plus-background (background-only) hypothesis.  Poisson pseudo-experiments are generated for the two hypotheses using the predicted signal and background distributions and the impact of each systematic uncertainty.  The latter are evaluated for their impact on both the normalization and the shape of the final discriminating variables, and are varied during the generation of the pseudo-experiments assuming a Gaussian distribution as the prior probability distribution function.

For the pair-production hypotheses, the final discriminating variable in the dilepton channel is the $m(Zb)$ distribution shown in figure~\ref{fig:dilepton_htmzb}(d), while the final discriminating variable in the trilepton channel is the $H_{\rm T}({\rm jets+leptons})$ distribution shown in figure~\ref{fig:trilepton2}(b).  For the single-production hypotheses, the final discriminating variable in the dilepton channel is the $m(Zb)$ distribution shown in figure~\ref{fig:dileptonsingleprod}(d), while the final discriminating variable in the trilepton channel is the $H_{\rm T}({\rm jets+leptons})$ distribution shown in figure~\ref{fig:trileptonsingleprod}(b).

The data are found to be consistent with the background-only hypotheses in each of the four final distributions, and limits are subsequently derived according to the ${\rm CL}_{\rm s}$ prescription~\cite{mclimit,cls}.  Upper limits at the 95\% confidence level (CL) are set on the pair- and single-production cross sections of vector-like $T$ and $B$ quarks.  The cross-section limits are then used to set lower limits on the quark masses, as well as upper limits on electroweak coupling parameters.

\subsection{Limits on the pair-production hypotheses}

\begin{figure*}[tp]
\centering
\subfigure[]{\includegraphics[width=0.49\textwidth]{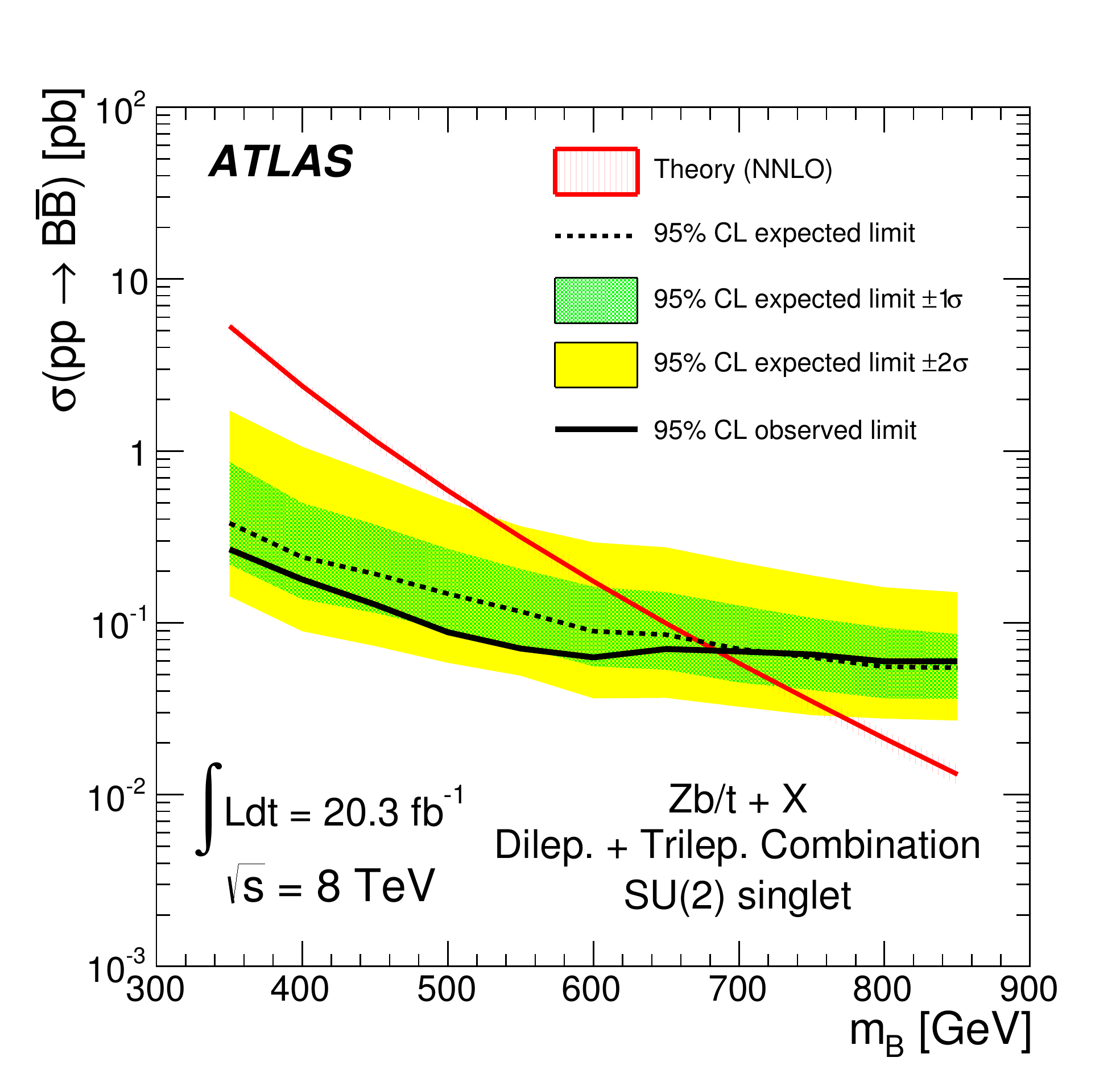}\label{fig:limitBBs}}
\subfigure[]{\includegraphics[width=0.49\textwidth]{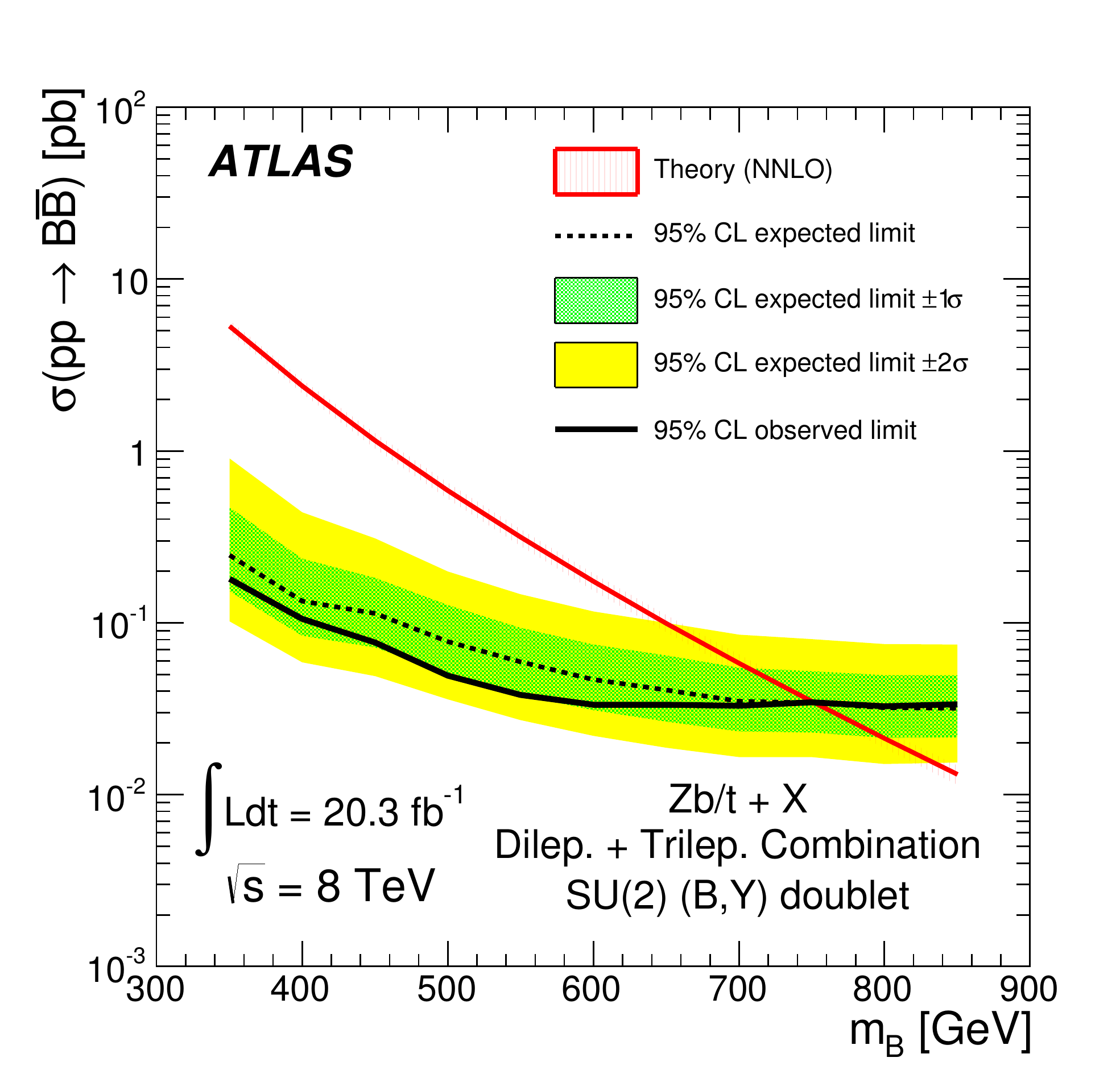}\label{fig:limitBBd}}
\subfigure[]{\includegraphics[width=0.49\textwidth]{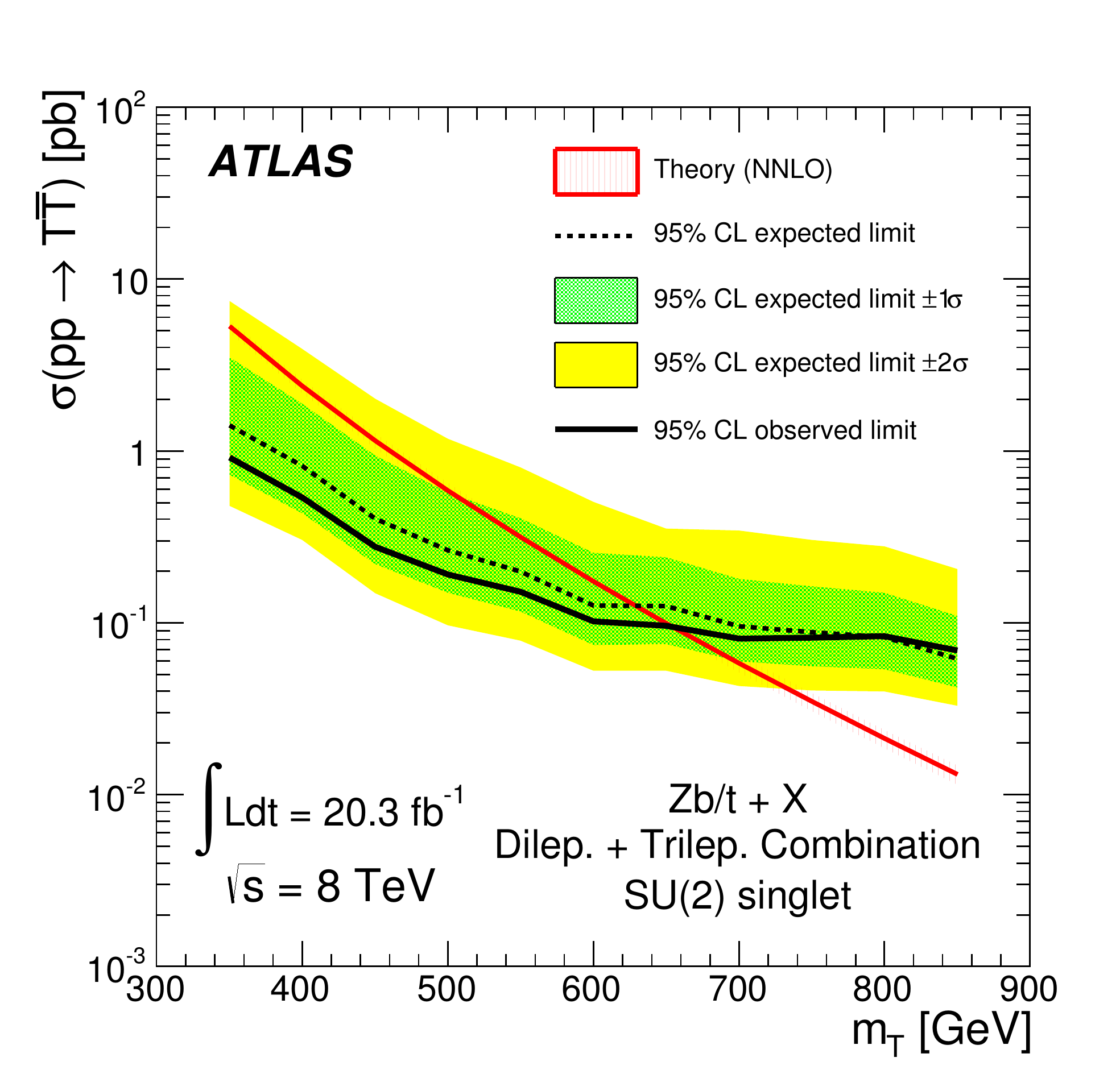}\label{fig:limitTTs}}
\subfigure[]{\includegraphics[width=0.49\textwidth]{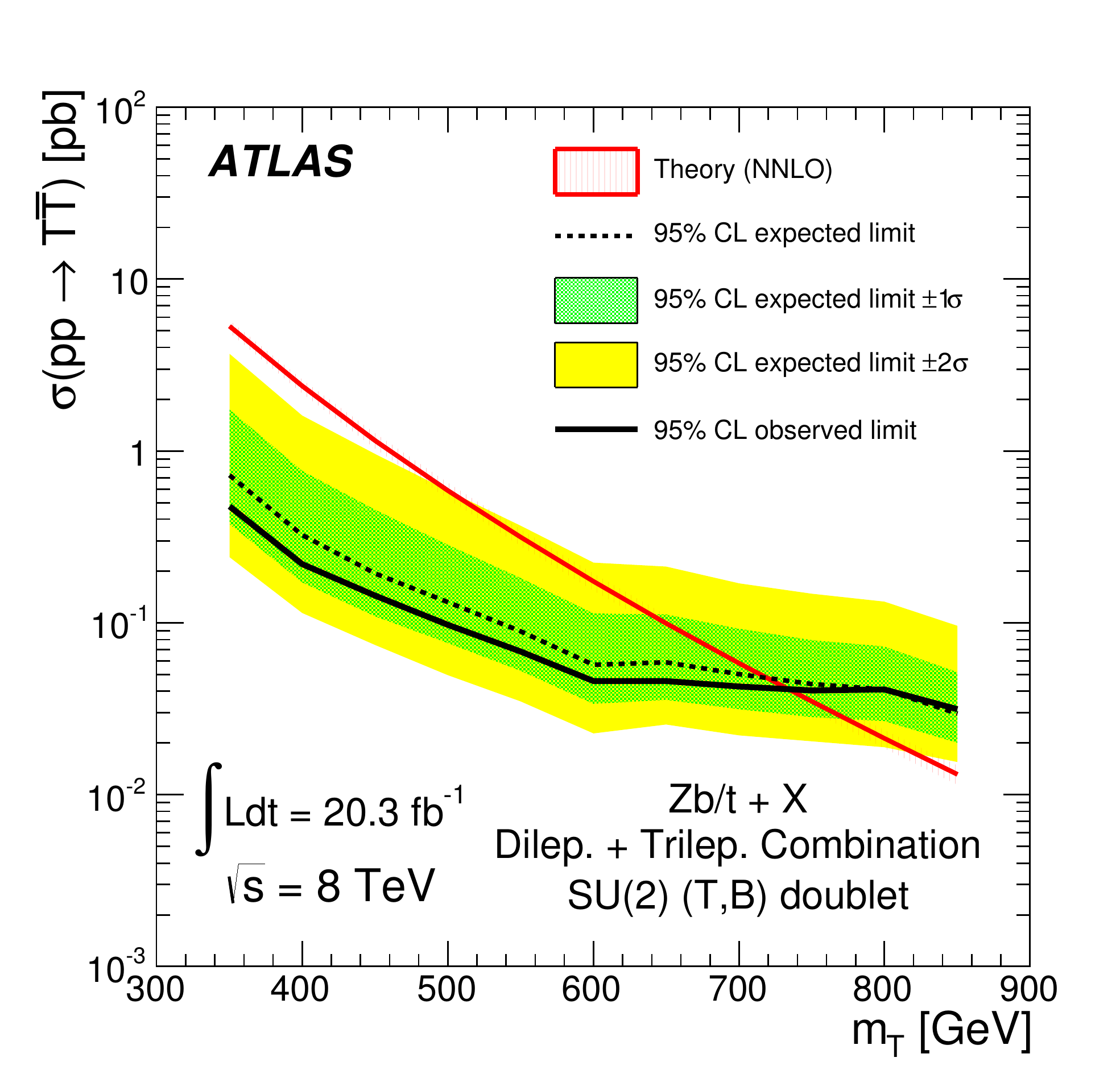}\label{fig:limitTTd}}
\caption{Predicted pair-production cross section as a function of the heavy quark mass and $95\%$~CL observed and expected upper limits for (a) an $SU(2)$ singlet $B$ quark, and (b) a $B$ quark forming an $SU(2)$ $(B,Y)$ doublet with a charge $-4/3$ $Y$ quark.  Likewise, the upper limit on the pair-production cross section as a function of the heavy quark mass for (c) an $SU(2)$ singlet $T$ quark, and (d) a $T$ quark forming an $SU(2)$ $(T,B)$ doublet with a charge $-1/3$ $B$ quark.}
\label{fig:pairprodlimits}
\end{figure*}

Figures~\ref{fig:pairprodlimits}(a,b) show the pair-production cross-section limit for $B$ quark masses in the interval 350--850~GeV, assuming the branching ratios of an $SU(2)$ singlet $B$ quark and a $B$ quark in a $(B,Y)$ doublet, respectively.  The theoretical curve represents the total pair-production cross section calculated with {\sc Top++}, and the width of the curve indicates the uncertainty on the prediction from PDF$+\alpha_{s}$ and scale uncertainties.  The observed (expected) limit on the mass of an $SU(2)$ singlet $B$ quark is 685~GeV (670~GeV), while the observed (expected) limit on the mass of a $B$ quark in a $(B,Y)$ doublet is 755~GeV (755~GeV).   These limits are derived by combining the dilepton and trilepton channels in a single likelihood function.  Table~\ref{table:masslimits} lists the combined $B$ quark mass limits along with the mass limits obtained from the dilepton and trilepton channels independently.  The dilepton channel provides the greater degree of sensitivity for both the singlet and doublet $B$ quark hypotheses.

\begin{table}[tp]
\begin{center}
\begin{tabular}{|c||c|c|c||c|c|c|} \hline \hline
                      & \multicolumn{3}{c||}{Singlet mass limit [GeV]}                     &  \multicolumn{3}{c|}{Doublet mass limit [GeV]}          \\ \hline 
Hypothesis & Dilepton 	& Trilepton 	& Comb. & Dilepton 		& Trilepton 	& Comb. \\ \hline
$B\bar{B}$  &	690 (665)	&  610 (610)	&  685 (670)     &  765 (750)	&   540 (530)	& 755 (755) \\
$T\bar{T}$  &	620 (585)	&  620 (620)	&  655 (625)     &  705 (665)	&   700 (700)	& 735 (720) \\ \hline \hline
\end{tabular}
\end{center}
\caption[Summary of mass limits from pair production]{Observed (expected) 95\% CL limits on the $T$ and $B$ quark mass (GeV) assuming pair production of $SU(2)$ singlet and doublet quarks, and using the dilepton and trilepton channels separately, as well as combined.}\label{table:mass_limits}
\label{table:masslimits}
\end{table}

Figures~\ref{fig:pairprodlimits}(c,d) show the pair-production cross-section limit for $T$ quark masses in the interval 350--850~GeV, assuming the branching ratios of an $SU(2)$ singlet $T$ quark and a $T$ quark in a $(T,B)$ doublet, respectively.  The observed (expected) limit on the mass of an $SU(2)$ singlet $T$ quark is 655~GeV (625~GeV), while the observed (expected) limit on the mass of a $T$ quark in a $(T,B)$ doublet is 735~GeV (720~GeV).  These limits are derived by combining the dilepton and trilepton channels in a single likelihood function.  Table~\ref{table:masslimits} lists the combined $T$ quark mass limits along with the mass limits obtained from the dilepton and trilepton channels independently.  The sensitivity of the two channels is similar, though the trilepton channel is more sensitive in both cases.  

\begin{figure*}[tp]
\centering
\subfigure[]{\includegraphics[width=0.80\textwidth]{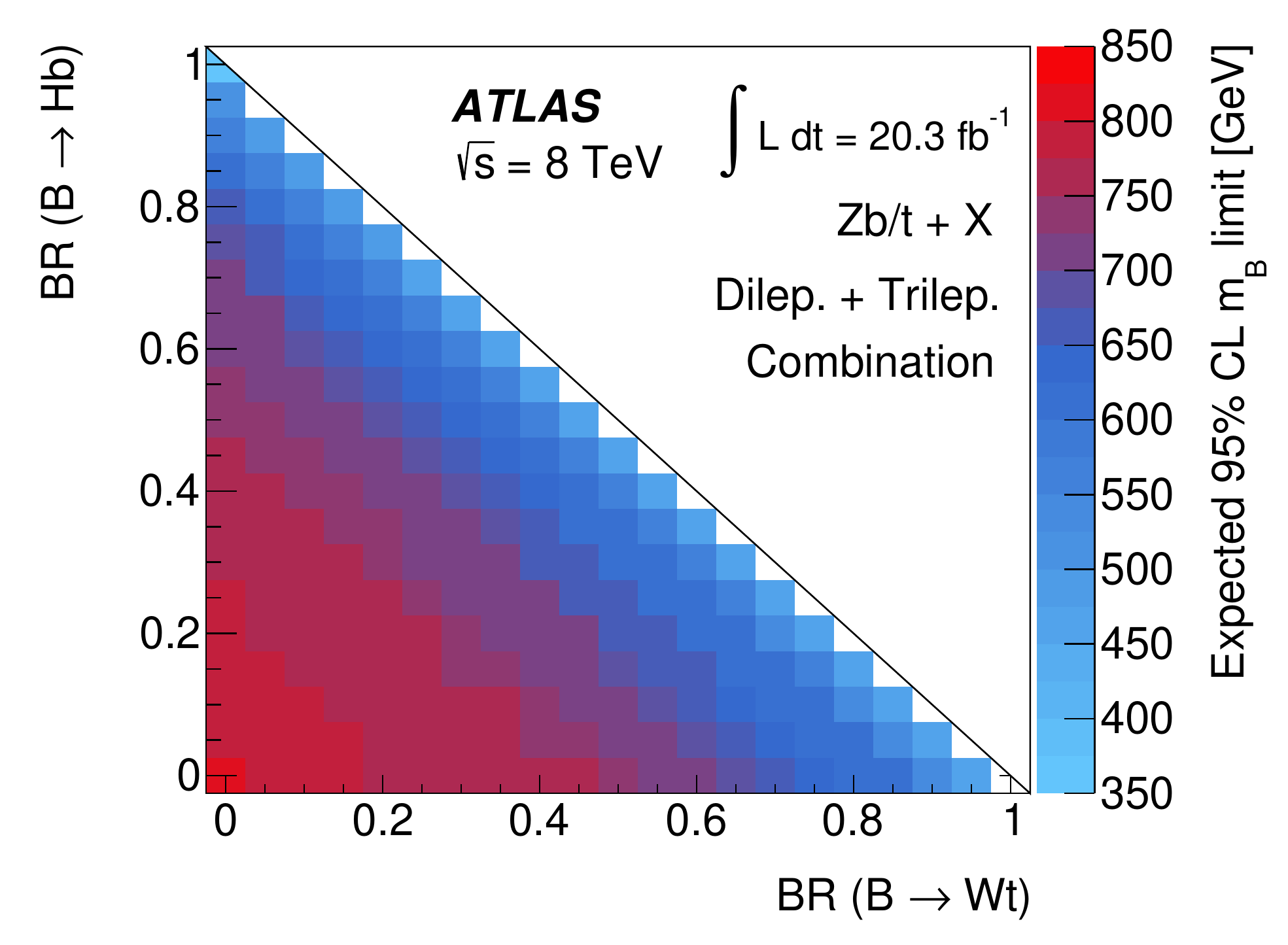}}
\subfigure[]{\includegraphics[width=0.80\textwidth]{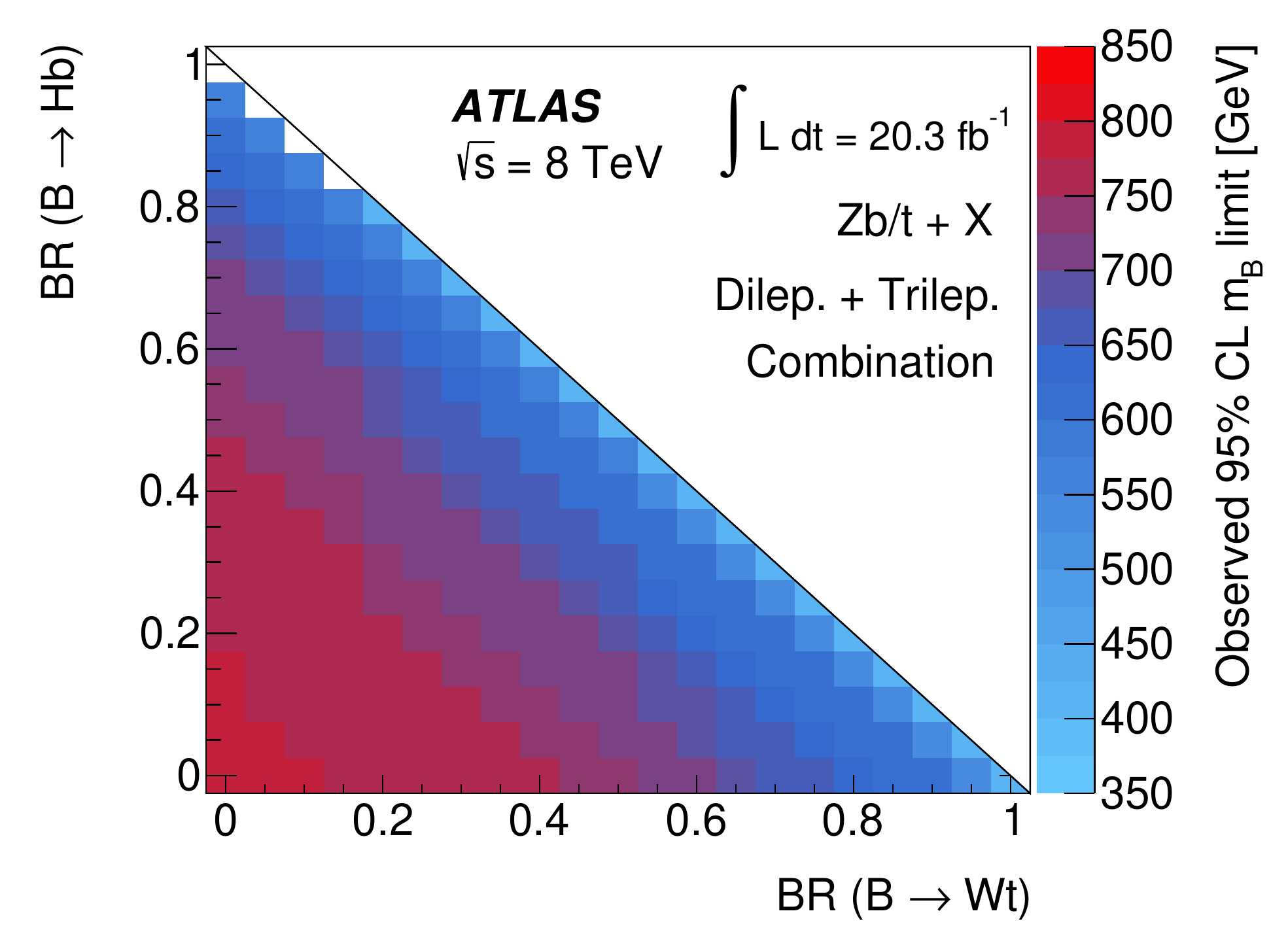}\label{fig:2dBBs}}
\caption{Expected (a) and observed (b) limit (95\%~CL) on the mass of the $B$ quark assuming the pair-production hypothesis and presented in the $(Wt,Hb)$ branching ratio plane.}
\label{fig:BBtemperature}
\end{figure*}

\begin{figure*}[tp]
\centering
\subfigure[]{\includegraphics[width=0.80\textwidth]{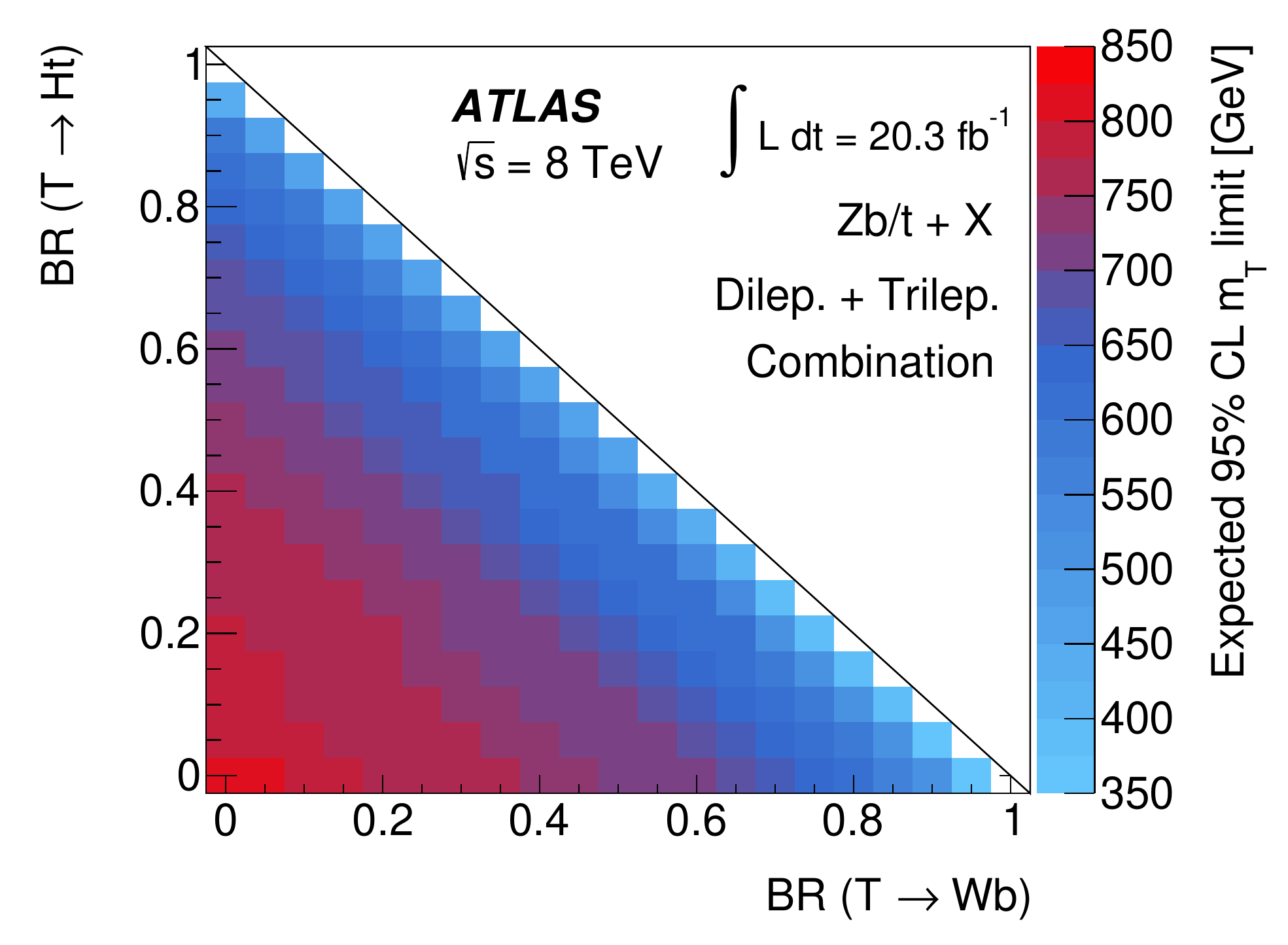}}
\subfigure[]{\includegraphics[width=0.80\textwidth]{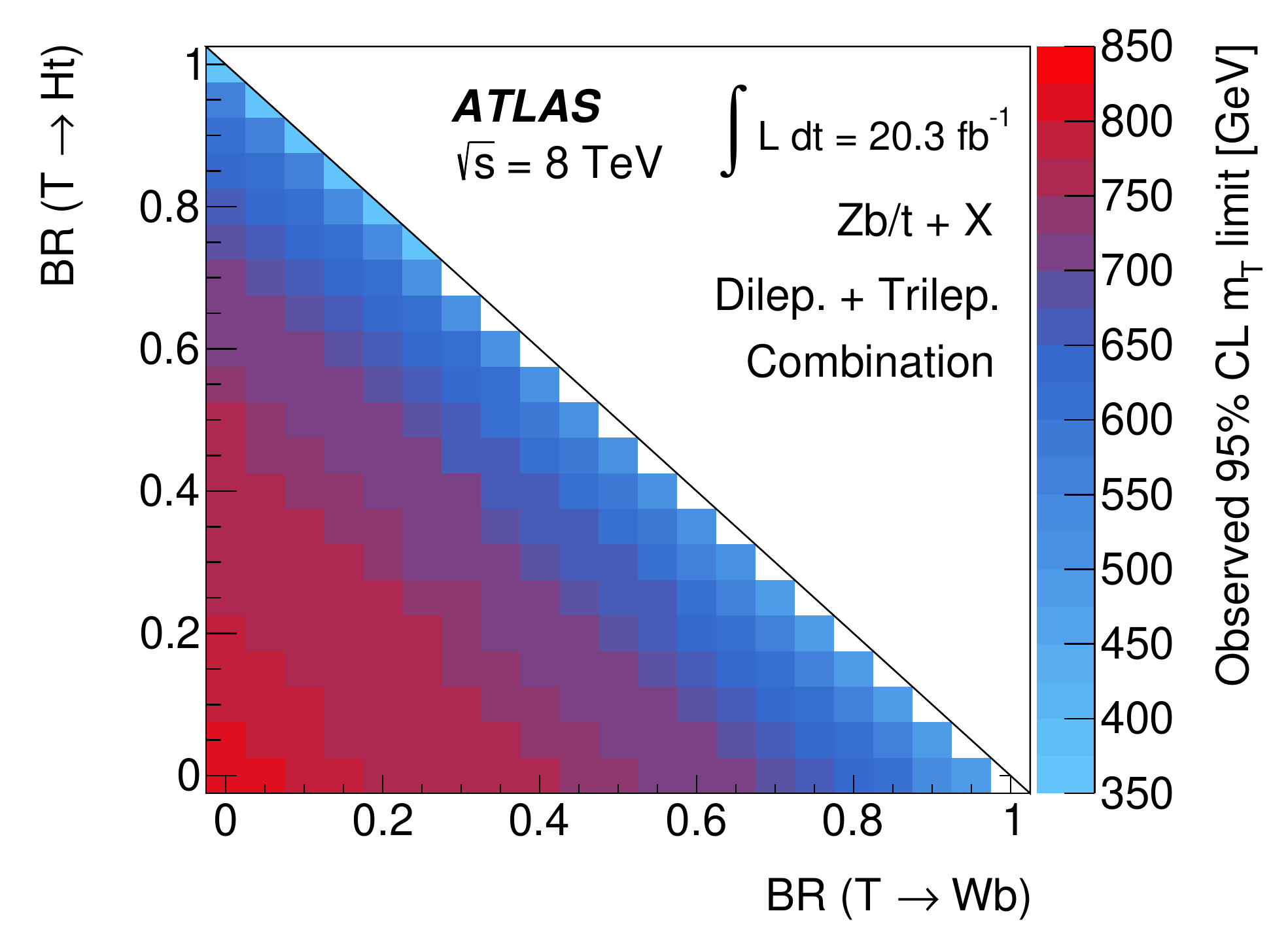}\label{fig:2dTTs}}
\caption{Expected (a) and observed (b) limit (95\%~CL) on the mass of the $T$ quark assuming the pair-production hypothesis and presented in the $(Wb,Ht)$ branching ratio plane.}
\label{fig:TTtemperature}
\end{figure*}

In addition to lower limits on the quark masses for these benchmark $SU(2)$ singlet and doublet scenarios, limits are also derived using the combination of the dilepton and trilepton channels for all sets of heavy quark branching ratios consistent with the three decay modes ($W$, $Z$, and $H$) summing to unity.  Figures~\ref{fig:BBtemperature}(a,b) present expected and observed $B$ quark mass limits, respectively, in a two-dimensional plane of branching ratios, with $BR(B \rightarrow Hb)$ plotted on the vertical axis and $BR(B \rightarrow Wt)$ on the horizontal axis.  The sensitivity is greatest in the lower-left corner where the branching ratio to the $ZbZb$ final state is 100\%.  In this case, the expected $B$ quark mass limit is 800~GeV and the observed limit is 790~GeV.  Likewise, figures~\ref{fig:TTtemperature}(a,b) present the expected and observed $T$ quark mass limits, respectively, in the $BR(T \rightarrow Ht)$ versus $BR(T \rightarrow Wb)$ plane of branching ratios.  In the case of a 100\% branching ratio to the $ZtZt$ final state, both the expected and observed $T$ quark mass limits are 810~GeV.

\subsection{Limits on the single-production hypotheses}

\begin{figure*}[tp]
\centering
\subfigure[]{\includegraphics[width=0.49\textwidth]{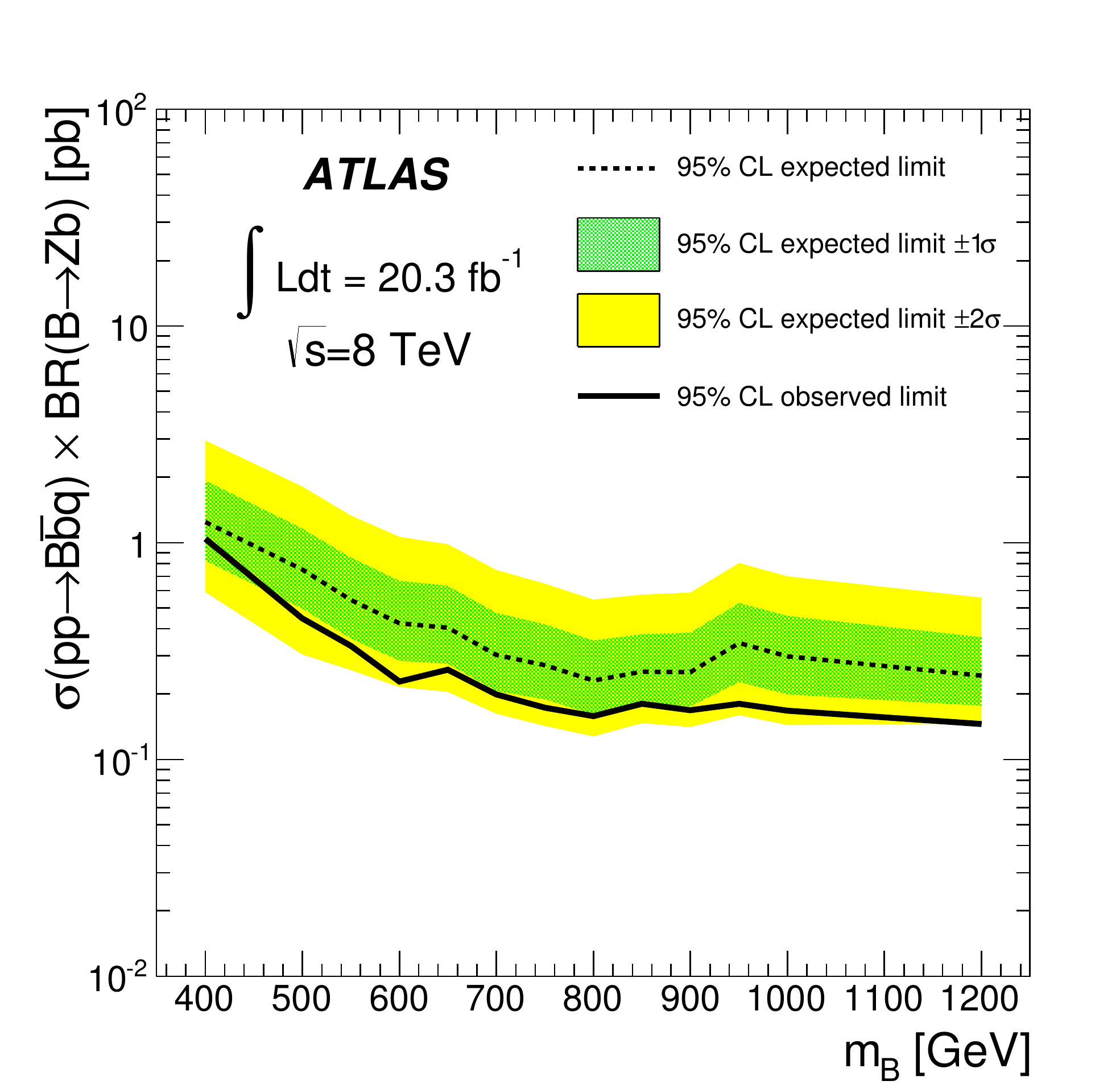}\label{fig:singleVLB}}
\subfigure[]{\includegraphics[width=0.49\textwidth]{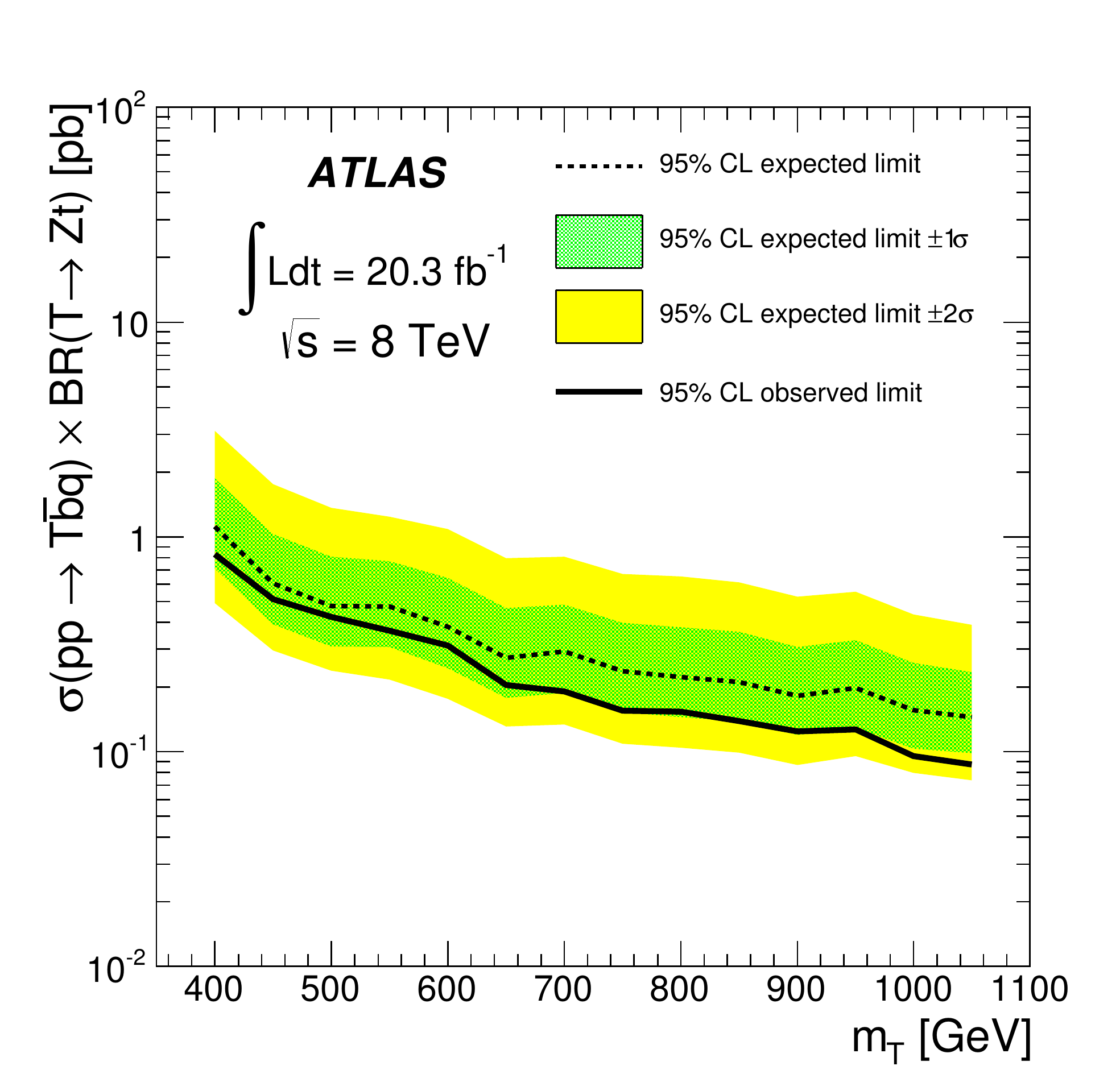}\label{fig:singleVLT}}
\caption{Upper limit ($95\%$ CL) on the single-production cross section times branching ratio as a function of the heavy quark mass: (a) $\sigma(pp \rightarrow B\bar{b}q) \times BR(B \rightarrow Zb)$, and (b) $\sigma(pp \rightarrow T\bar{b}q) \times BR(T \rightarrow Zt)$.}
\label{fig:singleprodlimits}
\end{figure*}

Figures~\ref{fig:singleprodlimits}(a,b) present upper limits on the cross section as a function of the heavy quark mass for the electroweak single-production processes $pp \rightarrow B\bar{b}q$ and $pp \rightarrow T\bar{b}q$, respectively, multiplied by the branching ratio to the $Z$ boson decay mode.  The cross-section limits on the $B\bar{b}q$ process are obtained using the dilepton channel only, as the trilepton channel is not sensitive to this process.  The cross-section limits on the $T\bar{b}q$ process are derived by combining the dilepton and trilepton channels.  The sensitivity of the two channels is comparable in the low-mass regime, while the dilepton channel provides the greater sensitivity in the high-mass regime.

Constraints on the $\lambda_{T}$ parameter of the reference composite Higgs model~\cite{CH_Natascia} are assessed using the limits on the $T\bar{b}q$ process.  The implementation of the model invokes the approximation described in footnote 2 of Ref.~\cite{ContinoServant}.  Values $\lambda_{T}<1.5$ are neither expected nor observed to be excluded at 95\%~CL for any $SU(2)$ singlet $T$ quark in the mass range considered.  The sensitivity to the $V_{Tb}$ and $X_{bB}$ mixing parameters is also assessed using the limits on the $T\bar{b}q$ and $B\bar{b}q$ processes, respectively.  In addition to accounting for the scaling of the cross section with the mixing parameters, the $SU(2)$ singlet branching ratios shown in Fig.~\ref{fig:BRs} are recalculated for large mixing values using the relationships tabulated in Appendix~A of Ref.~\cite{JA_new}.  Sensitivity to values of $V_{Tb}<1$ is not expected for any $T$ quark mass considered, but values as low as $0.7$ are observed to be excluded at 95\%~CL for some masses in the range 450--650~GeV due to a modest downward fluctuation of the data relative to the background prediction.  Values of $X_{bB}<0.5$ are neither expected nor observed to be excluded for any $B$ quark mass considered.

\section{Conclusions}
A search for heavy quarks that decay to a $Z$ boson and a third-generation quark has been performed with a dataset corresponding to $20.3~{\rm fb}^{-1}$ that was collected by the ATLAS detector at the LHC in $pp$ collisions at \mbox{$\sqrt{s}=8$~TeV}.  No evidence for a heavy quark signal is observed when selecting events with topologies sensitive to heavy quarks produced either in pairs via the strong interaction or singly via the electroweak interaction.  The results are used to set lower mass limits of 685~GeV and 755~GeV (at 95\% CL) on vector like $B$ quarks when assuming the $SU(2)$ singlet and doublet hypotheses, respectively.  Likewise, lower mass limits of 655~GeV and 735~GeV (at 95\% CL) are obtained for vector-like $T$ quarks when assuming the $SU(2)$ singlet and doublet hypotheses, respectively.  Lower mass limits are also derived for all sets of $B$ and $T$ quark branching ratios to third-generation quarks and $W$, $Z$, or $H$ bosons.  Using signal regions defined to test the single-production hypotheses, upper limits are derived on the cross sections for the $T\bar{b}q$ and $B\bar{b}q$ processes multiplied by the branching ratio of the heavy quark to decay to a $Z$ boson and a third-generation quark.  For example, limits of 190~fb and 200~fb are derived for the $T$ and $B$ quark processes, respectively, assuming a heavy quark mass of 700~GeV.  The results on electroweak single production are used to assess constraints on electroweak coupling parameters of the new quarks.

\acknowledgments




We thank CERN for the very successful operation of the LHC, as well as the
support staff from our institutions without whom ATLAS could not be
operated efficiently.

We acknowledge the support of ANPCyT, Argentina; YerPhI, Armenia; ARC,
Australia; BMWFW and FWF, Austria; ANAS, Azerbaijan; SSTC, Belarus; CNPq and FAPESP,
Brazil; NSERC, NRC and CFI, Canada; CERN; CONICYT, Chile; CAS, MOST and NSFC,
China; COLCIENCIAS, Colombia; MSMT CR, MPO CR and VSC CR, Czech Republic;
DNRF, DNSRC and Lundbeck Foundation, Denmark; EPLANET, ERC and NSRF, European Union;
IN2P3-CNRS, CEA-DSM/IRFU, France; GNSF, Georgia; BMBF, DFG, HGF, MPG and AvH
Foundation, Germany; GSRT and NSRF, Greece; ISF, MINERVA, GIF, I-CORE and Benoziyo Center,
Israel; INFN, Italy; MEXT and JSPS, Japan; CNRST, Morocco; FOM and NWO,
Netherlands; BRF and RCN, Norway; MNiSW and NCN, Poland; GRICES and FCT, Portugal; MNE/IFA, Romania; MES of Russia and ROSATOM, Russian Federation; JINR; MSTD,
Serbia; MSSR, Slovakia; ARRS and MIZ\v{S}, Slovenia; DST/NRF, South Africa;
MINECO, Spain; SRC and Wallenberg Foundation, Sweden; SER, SNSF and Cantons of
Bern and Geneva, Switzerland; NSC, Taiwan; TAEK, Turkey; STFC, the Royal
Society and Leverhulme Trust, United Kingdom; DOE and NSF, United States of
America.

The crucial computing support from all WLCG partners is acknowledged
gratefully, in particular from CERN and the ATLAS Tier-1 facilities at
TRIUMF (Canada), NDGF (Denmark, Norway, Sweden), CC-IN2P3 (France),
KIT/GridKA (Germany), INFN-CNAF (Italy), NL-T1 (Netherlands), PIC (Spain),
ASGC (Taiwan), RAL (UK) and BNL (USA) and in the Tier-2 facilities
worldwide.

\label{app:References}
\bibliographystyle{JHEP}
\bibliography{Ztag}

\clearpage
\begin{flushleft}
{\Large The ATLAS Collaboration}

\bigskip

G.~Aad$^{\rm 84}$,
B.~Abbott$^{\rm 112}$,
J.~Abdallah$^{\rm 152}$,
S.~Abdel~Khalek$^{\rm 116}$,
O.~Abdinov$^{\rm 11}$,
R.~Aben$^{\rm 106}$,
B.~Abi$^{\rm 113}$,
M.~Abolins$^{\rm 89}$,
O.S.~AbouZeid$^{\rm 159}$,
H.~Abramowicz$^{\rm 154}$
% ...
\bigskip
\\
% ...
$^{11}$ Institute of Physics, Azerbaijan Academy of Sciences, Baku, Azerbaijan\\
% ...
$^{84}$ CPPM, Aix-Marseille Universit{\'e} and CNRS/IN2P3, Marseille, France\\
% ...
$^{89}$ Department of Physics and Astronomy, Michigan State University, East Lansing MI, United States of America\\
% ...
$^{106}$ Nikhef National Institute for Subatomic Physics and University of Amsterdam, Amsterdam, Netherlands\\
% ...
$^{112}$ Homer L. Dodge Department of Physics and Astronomy, University of Oklahoma, Norman OK, United States of America\\
$^{113}$ Department of Physics, Oklahoma State University, Stillwater OK, United States of America\\
% ...
$^{116}$ LAL, Universit{\'e} Paris-Sud and CNRS/IN2P3, Orsay, France\\
% ...
$^{152}$ Institute of Physics, Academia Sinica, Taipei, Taiwan\\
% ...
$^{154}$ Raymond and Beverly Sackler School of Physics and Astronomy, Tel Aviv University, Tel Aviv, Israel\\
% ...
$^{159}$ Department of Physics, University of Toronto, Toronto ON, Canada\\
% ...
\end{flushleft}


\end{document}